\documentclass[11pt,a4paper]{article}
\pdfoutput=1

\usepackage[T1]{fontenc}
\usepackage[utf8]{inputenc}
\usepackage[greek,english]{babel}
\usepackage{jcappub}
\usepackage{bm}
\usepackage{lmodern}
\usepackage[usenames,dvipsnames,svgnames,table]{xcolor}
\usepackage{tcolorbox}
\usepackage{etoolbox}

\tcbuselibrary{skins}
\tcbuselibrary{breakable}

\usepackage{ctable}
\usepackage{array}
\usepackage{relsize}
\usepackage{mathrsfs}
\usepackage{tablefootnote}

\newenvironment{grouppanel}{\begin{tcolorbox}[enhanced,colback=gray!10,frame hidden]}{\end{tcolorbox}}

\RequirePackage{ifpdf}
\ifpdf
\else 
\fi

\makeatletter
\@ifundefined{c@rownum}{%
  \let\c@rownum\rownum
}{}
\@ifundefined{therownum}{%
  \def\therownum{\@arabic\rownum}%
}{}
\makeatother

\definecolor{ERCorange}{HTML}{F26623}
\definecolor{SussexCobaltBlue}{HTML}{1E428A}
\definecolor{SussexDeepAquamarine}{HTML}{487A7B}
\hypersetup{colorlinks=true,citecolor=SussexDeepAquamarine,linkcolor=SussexCobaltBlue,urlcolor=SussexCobaltBlue}

\newcommand{\StandardTable}{\small
    	\heavyrulewidth=.08em
    	\lightrulewidth=.05em
    	\cmidrulewidth=.03em
    	\belowrulesep=.65ex
    	\belowbottomsep=0pt
    	\aboverulesep=.4ex
    	\abovetopsep=0pt
    	\cmidrulesep=\doublerulesep
    	\cmidrulekern=.5em
    	\defaultaddspace=.5em
    	\renewcommand{\arraystretch}{1.5}
}

\setupctable{botcap,mincapwidth=\textwidth,doinside={\StandardTable\rowcolors{2}{gray!25}{white}},footerwidth}

\newcommand{\grad}{\nabla}

\renewcommand{\d}{\mathrm{d}}
\newcommand{\im}{\mathrm{i}}
\newcommand{\e}[1]{\mathrm{e}^{{#1}}}

\newcommand{\Mp}{M_{\mathrm{P}}}

\newcommand{\vect}[1]{\bm{\mathrm{{#1}}}}

\newcommand{\LCDM}{$\Lambda$CDM}
\newcommand{\MSbar}{$\overline{\text{MS}}$}

\newcommand{\nonlinear}{\text{\textsc{nl}}}

\newcommand{\Mpc}{\text{Mpc}}

\newcommand{\cs}{c_s}
\newcommand{\csdot}{\dot{c}_s}

\newcommand{\vphys}{v_{\text{phys}}}
\newcommand{\vflow}{v_{\text{H}}}
\newcommand{\rhoflow}{\rho_{\text{f}}}
\newcommand{\rhocondensation}{\rho_{\text{c}}}

\newcommand{\tfreefall}{t_{\text{ff}}}
\newcommand{\tR}{t_R}

\newcommand{\deltarsd}{\delta_s}
\newcommand{\deltarsdmu}[1]{\delta_{s,{#1}}}

\newcommand{\DA}{D_A}
\newcommand{\DB}{D_B}
\newcommand{\DD}{D_D}
\newcommand{\DE}{D_E}
\newcommand{\DF}{D_F}
\newcommand{\DG}{D_G}
\renewcommand{\DJ}{D_J}
\newcommand{\DK}{D_K}
\newcommand{\DL}{D_L}
\newcommand{\DM}{D_M}
\newcommand{\DN}{D_N}
\newcommand{\DP}{D_P}
\newcommand{\DQ}{D_Q}
\newcommand{\DR}{D_R}
\newcommand{\DS}{D_S}

\newcommand{\fA}{f_A}
\newcommand{\fB}{f_B}
\newcommand{\fD}{f_D}
\newcommand{\fE}{f_E}
\newcommand{\fF}{f_F}
\newcommand{\fG}{f_G}
\newcommand{\fJ}{f_J}

\newcommand{\looplevel}[1]{{#1}\ell}
\newcommand{\attreelevel}{=\looplevel{0}}
\newcommand{\atoneloop}{=\looplevel{1}}
\newcommand{\uptooneloop}{\leq\looplevel{1}}

\newcommand{\VSCF}{\text{\textsc{vscf}}}
\newcommand{\SPT}{\text{\textsc{spt}}}

\newcommand{\wiggle}{\text{w}}
\newcommand{\nowiggle}{\text{nw}}

\newcommand{\Pinit}{P^\ast}
\newcommand{\Pinitnw}{P^\ast_{\nowiggle}}
\newcommand{\Pinitw}{P^\ast_{\wiggle}}

\newcommand{\Pref}{P_{\text{ref}}}

\newcommand{\Prsd}{P_s}
\newcommand{\Prsdnw}{P_{s,\nowiggle}}
\newcommand{\Prsdw}{P_{s,\wiggle}}
\newcommand{\Prsdren}{P_s^{\renormalizedtag}}
\newcommand{\PrsdrenVSCF}{P_{s,\VSCF}^{\renormalizedtag}}

\newcommand{\Ploopnw}{P^{\uptooneloop}_{\nowiggle}}
\newcommand{\Ploopw}{P^{\uptooneloop}_{\wiggle}}
\newcommand{\Patloopw}{P^{\atoneloop}_{\wiggle}}
\newcommand{\Ptree}{P^{\attreelevel}}
\newcommand{\Ptreenw}{P^{\attreelevel}_{\nowiggle}}
\newcommand{\Ptreew}{P^{\attreelevel}_{\wiggle}}
\newcommand{\PloopVSCF}{P^{\uptooneloop}_{\VSCF}}

\newcommand{\Prsdloopnw}{P^{\uptooneloop}_{s,\nowiggle}}
\newcommand{\Prsdloopw}{P^{\uptooneloop}_{s,\wiggle}}

\newcommand{\Prsdtreew}{P^{\attreelevel}_{s,\wiggle}}
\newcommand{\PrsdloopVSCF}{P^{\uptooneloop}_{s,\VSCF}}

\newcommand{\PSPT}{P_{\SPT}}
\newcommand{\PNL}{P_{\nonlinear}}

\newcommand{\qmin}{q_{\text{min}}}
\newcommand{\qmax}{q_{\text{max}}}

\newcommand{\vrec}{\vect{v}_{\text{r}}}

\newcommand{\UVcontains}{\overset{\text{UV}}{\supseteq}}

\newcommand{\ctrterm}[2]{c_{{#1}|{#2}}}
\newcommand{\ctrconst}[2]{Z_{{#1}|{#2}}}
\newcommand{\ctrfree}[2]{\zeta_{{#1}|{#2}}}

\newcommand{\muctrterm}[2]{c_{{#1}|\deltarsd,{#2}}}
\newcommand{\ellctrterm}[2]{d_{{#1}|\deltarsd,{#2}}}

\newcommand{\renormalizedtag}{\text{\textsc{r}}}
\newcommand{\renormalized}[1]{{#1}^{\renormalizedtag}}
\newcommand{\renormalizedc}[1]{({#1})^{\renormalizedtag}}
\newcommand{\renormalizedft}[1]{[{#1}]^{\renormalizedtag}}

\newcommand{\kNL}{k_{\nonlinear}}
\newcommand{\kpiv}{k_{\text{piv}}}
\newcommand{\kren}{k_{\renormalizedtag}}
\newcommand{\kdamp}{k_{\text{damp}}}

\newcommand{\kmin}{0.1 h/\Mpc}
\newcommand{\kmax}{0.4 h/\Mpc}
\newcommand{\kminreal}{0.15 h/\Mpc}
\newcommand{\kmaxreal}{0.4 h/\Mpc}

\newcommand{\sigmav}{\sigma_{\text{v}}}

\newcommand{\Legendre}[2]{\mathscr{P}_{#1}(#2)}
\newcommand{\FabrikantThree}[3]{\mathscr{J}^{#1}_{{#2}{#3}}}

\newcommand{\FabrikantOne}[1]{\mathscr{J}_{#1}}

\newcommand{\IRtag}{\text{IR}}
\newcommand{\KIR}{K_{\IRtag}}
\newcommand{\AIR}{A^{\IRtag}}
\newcommand{\PiIR}{\Pi_{\IRtag}}
\newcommand{\DiracD}{\delta_{\text{D}}}

\newcommand{\semibold}[1]{{\fontseries{b}\selectfont{#1}}}
\newcommand{\para}[1]{\par\vspace{2mm}\noindent\semibold{{#1.}---}\ignorespaces}

\DeclareMathOperator{\newRe}{Re}

\renewcommand{\Re}{\newRe}

\newcommand{\pochhammer}[2]{({#1})^{#2}}

\renewcommand{\geq}{\geqslant}
\renewcommand{\leq}{\leqslant}

\newcommand{\llangle}{\langle\kern-2\nulldelimiterspace\langle}
\newcommand{\rrangle}{\rangle\kern-2\nulldelimiterspace \rangle}

\DeclareMathOperator{\sgn}{sgn}
\DeclareMathOperator{\erf}{erf}
\DeclareMathOperator{\BigO}{\mathcal{O}}

\newcommand{\packagefont}{\sffamily\fontseries{sbc}\selectfont}
\newcommand{\GitRevision}[2]{{\small\sffamily \packagefont\href{#2}{{#1}}}}
\newcommand{\gevolution}{{\packagefont gevolution}}
\newcommand{\git}{{\packagefont git}}
\newcommand{\CppTransport}{{\packagefont CppTransport}}
\newcommand{\Halofit}{{\packagefont HALOFIT}}
\newcommand{\CAMB}{{\packagefont CAMB}}
\newcommand{\SQLite}{{\packagefont SQLite}}

\newcommand\CC{C\nolinebreak\hspace{-.05em}\raisebox{.4ex}{\relsize{-3}{\textbf{+}}}\nolinebreak\hspace{-.10em}\raisebox{.4ex}{\relsize{-3}{\textbf{+}}}}

\usepackage{color}

\graphicspath{{./}}

\begin{document}

\title{\fontseries{s}\selectfont The matter power spectrum in redshift space
using effective field theory
\vspace{5mm}\hrule}
\author{\fontseries{s}\selectfont\large Luc\'{\i}a Fonseca de la Bella$^1$, Donough Regan$^1$, David Seery$^1$ and Shaun Hotchkiss$^2$}

\affiliation{\vspace{2mm}\small
$^1$ Astronomy Centre, University of Sussex, Falmer, Brighton BN1 9QH, UK \\
$^2$ Department of Physics, The University of Auckland, Private Bag 92019, Auckland, New Zealand
}

\emailAdd{L.Fonseca-De-La-Bella@sussex.ac.uk}
\emailAdd{D.Regan@sussex.ac.uk}
\emailAdd{D.Seery@sussex.ac.uk}
\emailAdd{s.hotchkiss@auckland.ac.nz}

\abstract{The use of Eulerian `standard perturbation theory'
to describe mass assembly
in the early universe has traditionally been limited to modes with
$k \lesssim 0.1 h/\Mpc$ at $z=0$. At larger $k$ the SPT
power spectrum deviates from measurements made using $N$-body simulations.
Recently, there has been progress in extending the reach of perturbation theory to
larger $k$ using ideas borrowed from effective field theory.
We revisit the computation of the
redshift-space matter power spectrum within this framework, including for the first time
the full one-loop time dependence.
We use a resummation scheme proposed by Vlah et al.
to account for damping of baryonic acoustic oscillations due to
large-scale random motions and show that this has a significant effect on
the multipole power spectra.
We renormalize by
comparison to a suite of custom $N$-body simulations
matching the MultiDark MDR1 cosmology.
At $z=0$ and for
scales $k \lesssim 0.4 h/\Mpc$ we find that the
EFT furnishes a description of the real-space power
spectrum up to $\sim 2\%$,
for the $\ell = 0$ mode up to $\sim 5\%$, and for
the $\ell = 2, 4$ modes up to $\sim 25\%$.
We argue that, in the MDR1 cosmology,
positivity of the $\ell=0$ mode gives a firm
upper limit of $k \approx 0.74 h/\Mpc$ for the
validity of the one-loop
EFT prediction in redshift space using only the lowest-order
counterterm.
We show that replacing the one-loop growth factors by their Einstein-de Sitter
counterparts is a good approximation for the $\ell=0$ mode, but
can induce deviations as large as $2\%$ for the $\ell=2, 4$ modes.
An accompanying software bundle, distributed under
open source licenses, includes Mathematica notebooks
describing the calculation, together with parallel pipelines
capable of computing both the necessary one-loop SPT integrals
and the effective field theory counterterms.
\par\vspace{6mm}\mbox{}\hfill
\raisebox{-0.5\height}{\includegraphics[scale=0.2]{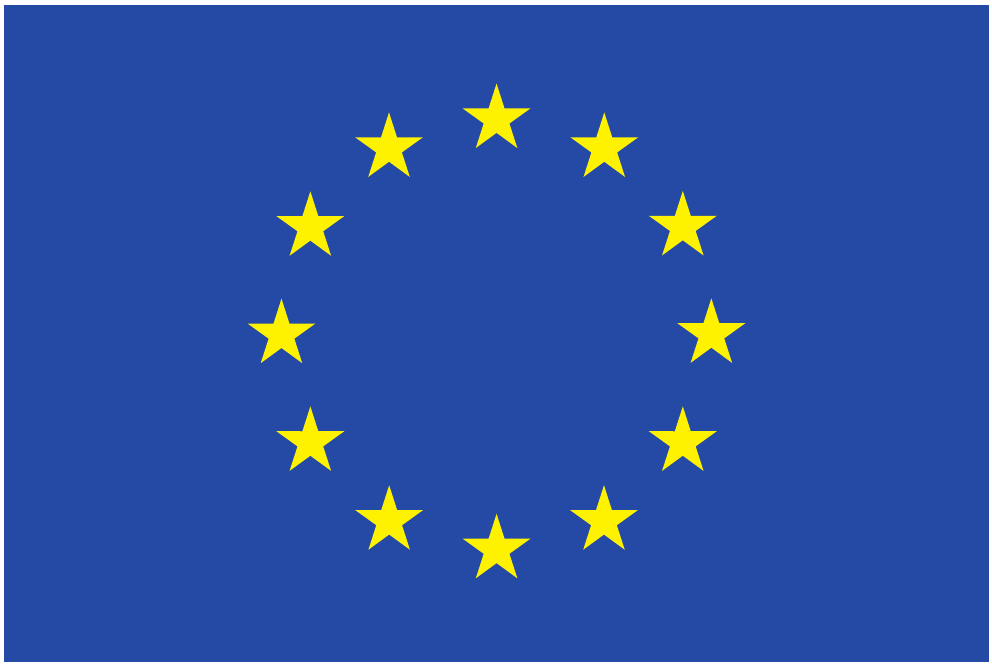}}
\;\textcolor{ERCorange}{\vrule width 1pt}\;
\raisebox{-0.5\height}{\includegraphics[scale=0.1]{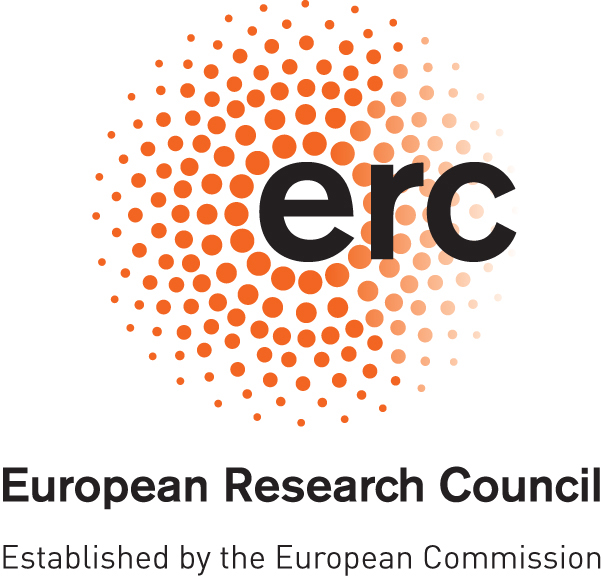}}
\quad
\raisebox{-0.5\height}{\includegraphics[scale=0.5]{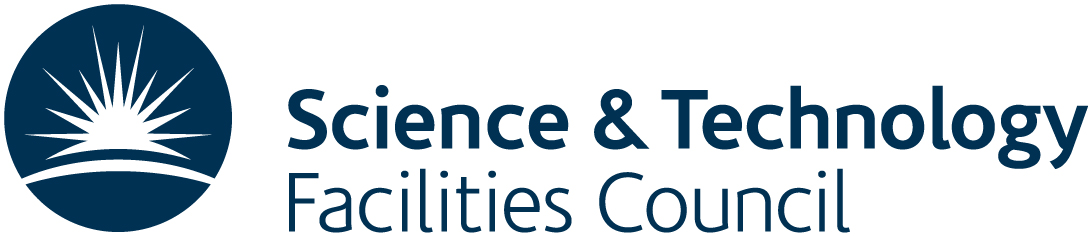}}}

\maketitle
\newpage

\section{Introduction}
\label{sec:introduction}
The long dominance of the cosmic microwave background (CMB) as our
principal source of information regarding the early universe
will soon come to an end,
displaced by
new datasets from large galaxy redshift surveys.
In addition to present-day surveys such
as the Dark Energy Survey,
the list will expand over the next decade to include
at least Euclid,
the Dark Energy Spectroscopic Instrument,
the Large Synoptic Survey Telescope,
the Square Kilometer Array,
and the 4-metre Multi Object Spectroscopic Telescope.
The ensemble of Fourier modes visible to each of these instruments
carries information about \emph{both} (i) the gravitational
potentials---presumably generated by inflation---that seeded
structure formation, and (ii) the effective force laws that
operated while matter was drawn into these potential wells
and condensed into halos.
This sensitivity to a rich range of physical processes
means that
the imminent era of large galaxy surveys should drive a step
change in our understanding of the standard cosmological model---%
and especially
its poorly-understood early- and late-time accelerating phases.

The price to be paid for access to this information is an obligation
to connect our theoretical description
with observation
by carrying out
sophisticated modelling of both gravitational potentials and force
laws.
Analytic control has traditionally come from the use of perturbation
theory~\cite{1981MNRAS.197..931J,1983MNRAS.203..345V,Fry:1983cj,
Goroff:1986ep,Bertschinger:1993zv,Suto:1990wf,Makino:1991rp,
Bernardeau:2001qr},
but its reach is limited in scale
to $k \lesssim 0.1 h/\Mpc$ at $z=0$
and therefore
excludes a significant fraction of the modes visible
to the surveys listed above.
Large $N$-body simulations
provide an alternative,
but
although Moore's Law has significantly reduced their time cost
they are still expensive---certainly too expensive to be
considered routine for extensions of the standard cosmological model
that entail a significant increase in the parameter space.
These pressures have produced a large literature
based on enhancements of standard perturbation theory (`SPT')
that extend its reach to moderate $k$ in the approximate range
$0.1 h/\Mpc$ to $0.5 h/\Mpc$.
One such approach is based on modern ideas from
effective field theory~\cite{Baumann:2010tm,Carrasco:2012cv,
Carrasco:2013mua,Porto:2013qua,Senatore:2014via,Senatore:2014vja,
Vlah:2015sea,Vlah:2015zda},
leading to the so-called `effective field theory of large-scale structure'.
This has yielded encouraging results for the matter power spectrum and bispectrum,
at the cost of adjustable counterterms
that must be estimated from observation or from $N$-body simulations.

\para{Redshift space effects}
In this paper we revisit the application of these ideas to
the \emph{redshift-space} power spectrum.
Real surveys must estimate the radial distance to a source from its redshift,
and therefore do not measure the galaxies' true spatial configuration.
Unknown peculiar velocities associated with each source
bias our distance estimate,
introducing a systematic `redshift space distortion'
that must be modelled appropriately if we are to
extract reliable results~\cite{Jackson:2008yv}.
This is both a challenge and an opportunity. While redshift-space effects
complicate the analysis, they enable us to measure correlations between
densities and velocities that carry
information about the effective gravitational
force law on cosmological scales.
In Einstein gravity,
for the non-relativistic regime applicable to large-scale
structure, this effective force is composed
of an attractive $1/r^2$ component
that is offset by a repulsive contribution from the
cosmological expansion.
In non-Einstein gravities the competition between these effects
may be altered,
or the scale-dependence of the force law may itself be modified
due to processes involving exchange
of new force-carrying particles.

These non-Einstein gravities could be constrained by precise
measurements of the effective force law on cosmological scales,
but only if
its scale-dependence can be separated from uncertainties in our
computation of its behaviour in the standard cosmological model.
For this purpose
effective field theory
should be a helpful tool,
enabling us to extend the range of wavenumbers
that can be reached analytically and used for comparison.
Senatore \& Zaldarriaga~\cite{Senatore:2014vja}
and later Lewandowski et al.~\cite{Lewandowski:2015ziq}
provided an analysis of the
redshift-space matter power spectrum
within such an effective description.
More recently, Perko et al. extended this analysis to include
biased tracers of the dark matter distribution~\cite{Perko:2016puo}
(see also Refs.~\cite{Assassi:2014fva,Gleyzes:2016tdh}).
By itself, the dark matter
can be measured only through its impact on
cosmological weak lensing.

In this paper we revisit the redshift-space analysis for the
pure matter power spectrum.
Our computation is similar to that of Lewandowski et al.,
with which it shares a common language and point of departure.
However, it differs in certain technical details
such as construction of the counterterms,
our procedure for estimating their numerical values,
and our procedure to resum large loop-level terms
involving integrals over the infrared part of the power spectrum.
Moreover, we compute all time dependent terms exactly.
Because these time dependences are known to be relatively insensitive to
cosmology they are often approximated as powers of the
Einstein--de Sitter growth function $D(z)$.
Since we retain the full time dependence we are able to
assess the accuracy of the Einstein--de Sitter approximation.%
    \footnote{Bose \& Koyama introduced a software tool that can numerically integrate
    the one-loop power spectrum in redshift space for a variety
    of models~\cite{Bose:2016qun,Bose:2017dtl}.
    Exact time dependence is sometimes
    considered in analytic calculations; eg., for recent
    examples, see Refs.~\cite{Fasiello:2016qpn,Lewandowski:2017kes}.
    Fasiello \& Vlah quoted exact results for cosmologies
    more general than {\LCDM}, but because they did not commit to a specific scenario
    their results were expressed as quadratures~\cite{Fasiello:2016qpn}.
    Closed-form analytic expressions
    for the one-loop {\LCDM}
    power spectrum in redshift space have not previously
    been given.}

This enables us to assess the accuracy of the Einstein--de Sitter
approximation.
We renormalize to a suite of custom $N$-body simulations
performed using the
\href{https://github.com/gevolution-code/gevolution-1.1}{\gevolution}
numerical relativity code.

As part of our analysis we describe some computational
innovations that we believe to be improvements over
the traditional methods used by Matsubara to compute the
redshift-space power spectrum in standard perturbation theory~\cite{Matsubara:2007wj}.
One such innovation is an algorithm to extract the explicit $\mu$-dependence%
    \footnote{Here, $\mu = \hat{\vect{k}} \cdot \hat{\vect{n}}$ is the
    orientation of a $\vect{k}$ mode contributing to the matter density
    field relative to the line of sight $\hat{\vect{n}}$ from Earth.}
of the redshift-space power spectrum
using the Rayleigh plane-wave expansion and analytic
formulae for weighted integrals over products of two or three
spherical Bessel functions.
A procedure to compute these three-Bessel integrals
was described by Gervois \& Navelet~\cite{doi:10.1137/0520067},
and more recently by Fabrikant~\cite{fabrikant2013elementary}.
However, their results do not yet seem to have entered the cosmological
literature.%
    \footnote{Certain three-Bessel integrals were computed as long ago
    as 1936 by
    Bailey~\cite{bailey1936some}.
    However, Bailey's method (and its descendents) required a triangle
    inequality to be satisfied by the arguments of the Bessel functions.
    To be effective our algorithm requires knowledge of the integral
    for any values of the arguments and not just those that satisfy
    the triangle inequality.
    It is for this extension that we require the more advanced
    methods of Refs.~\cite{doi:10.1137/0520067,fabrikant2013elementary}.}

\para{Code availability}
To assist those who wish to replicate or extend our analysis,
we have made our computer codes
and supporting datasets
available as part of a software
bundle accompanying this paper.
These include the parameter files needed to reconstruct our
initial linear power spectra, the settings files
required to reproduce our
{\gevolution} simulations,
and databases containing the loop integrals and
one-loop power spectra evaluated using the EFT.
Each of these products can be downloaded by following the links given
in Appendix~\ref{appendix:software}.

\para{Summary}
Our presentation is organized as follows.
In~\S\ref{sec:one-loop-real} we fix notation
by summarizing the construction of the
renormalized real-space matter power spectrum,
originally described
by Carrasco et al.~\cite{Carrasco:2012cv,Carrasco:2013mua}.
In~\S\ref{sec:matter-equations}
we collect the equations of structure formation during the
matter era and describe their non-relativistic limit.
In~\S\ref{sec:eulerian-perturbation-theory}
we construct Eulerian perturbation theory based on these
equations and compute the one-loop correction to the
power spectrum of the density contrast $\delta = \delta\rho / \rho$.
The time-dependent factors $D_A, \ldots, D_J$
and the loop integrals
$P_{AA}, \ldots, P_{BB}, P_D, \ldots, P_{J_1}, P_{J_2}$
are the key results from this section.
They are re-used extensively in~\S\ref{sec:one-loop-rsd}.

In~\S\ref{sec:ultraviolet-sensitivity}
we briefly summarize the use of effective field-theory methods
to parametrize the unknown
ultraviolet parts of these loop integrals.
In~\S\ref{sec:renormalized-operators}
we describe
renormalization of the velocity field,
and explain how to relate the perspective used in this
paper to the `smoothing' prescription for renormalized
operators used in Refs.~\cite{Baumann:2010tm,Carrasco:2012cv,Carrasco:2013mua,Mercolli:2013bsa}
and elsewhere.
In~\S\ref{sec:resummation-methods}
we introduce a scheme proposed by
Vlah, Seljak, Chu \& Feng
to resum the damping effect of displacements
on large scales
and assess its impact on the real-space power spectrum.
We conclude this section by
describing the renormalization of the power spectrum at
redshift $z=0$ (\S\ref{sec:real-space-results}), and compare our results with those already
reported in the literature.

This section can be read as a mini-primer on the use of effective field-theory
methods.
Readers already familiar with their application to large-scale structure
may wish to focus on~\S\ref{sec:eulerian-perturbation-theory}---%
which introduces our notation for time-dependent factors,
the SPT kernels, and loop integrals---%
and~\S\ref{sec:resummation-methods},
which describes our resummation prescription.
These summarize the principal technical differences
between our formalism and the existing literature.

In~\S\ref{sec:one-loop-rsd}
we describe the renormalization of the redshift-space power spectrum.
In~\S\ref{sec:redshift-space}
we write down an expression suitable for computing the redshift-space
density contrast $\deltarsd$ up to one-loop
and discuss the counterterms needed to renormalize it.
In~\S\ref{sec:evaluate-rsd-twopf}
we describe the calculation of the $\deltarsd$ power
spectrum up to one-loop,
introducing a new method to simplify evaluation of the
tensor integrals that appear at this order.
We extend the Vlah et al. resummation scheme to redshift space
in~\S\ref{sec:rsd-resummation}
and comment on its relation to empirical schemes
for capturing the suppression of power on small scales
due to randomized virial motions within halos.
In~\S\ref{sec:multipoles}
we describe the construction of the Legendre multipoles.
A significant advantage of the Vlah et al. resummation scheme is that
this can be done \emph{analytically}, reducing the requirement
for expensive numerical computation.
The $N$-body
simulations needed to obtain non-linear measurements of these multipoles
are described in~\S\ref{sec:simulations}.
We comment on a number of difficulties encountered when
extracting reliable estimates of the redshift-space
multipoles.
Finally, in~\S\ref{sec:results}
we fit for the counterterms of the effective description
and discuss the resulting power spectra.
We assess the accuracy of the Einstein--de Sitter approximation
and comment on the time-dependence of the EFT counterterms.
We conclude in~\S\ref{sec:conclusions}.
A number of Appendices extend the discussion presented
in the main text.

\para{Notation}
We use units in which $c = \hbar = 1$ and define the reduced
Planck mass to be $\Mp = (8\pi G)^{-1/2}$.
Our Fourier convention
is $f(\vect{x}) = \int \d^3 k \, (2\pi)^{-3} \,
f(\vect{k}) \e{\im \vect{k}\cdot\vect{x}}$.

Latin indices $a$, $b$, \ldots, from the beginning of the
alphabet range over
spacetime coordinates $(t, x, y, z)$ or $(0, 1, 2, 3)$.
Latin indices $i$, $j$, \ldots, from the middle of the alphabet
range over spatial indices only.
Repeated spacetime indices are taken to be contracted with the
metric $g_{ab}$.
Repeated spatial indices all in the `up' or `down' position are contracted with the
three-dimensional Euclidean metric $\delta_{ij}$, so that
(for example) $v^2 = v^i v^i = \delta_{ij} v^i v^j = \sum_{i} (v^i)^2$,
and likewise for $v_i v_i = \delta^{ij} v_i v_j$.

\section{One-loop renormalization of the matter power spectrum in real space}
\label{sec:one-loop-real}

In this section we briefly recapitulate the construction of the
one-loop matter power spectrum, neglecting the complexities of
redshift-space distortions.
The material presented here is a review of the
theory developed by Baumann et al.~\cite{Baumann:2010tm},
Carrasco et al.~\cite{Carrasco:2012cv,Carrasco:2013mua}
and Mercolli \& Pajer~\cite{Mercolli:2013bsa},
although some results are new (including renormalization of
the velocity accounting for its full time dependence),
and parts of our presentation are
different to discussions
that have already appeared in the literature.
We develop the formalism in detail
because we will rely on the notation and methodology
developed here when we study the power spectrum in redshift space.

\subsection{Matter equations of motion}
\label{sec:matter-equations}

Initially we work in a non-linear Newtonian gauge for which the metric can be
written
\begin{equation}
    \d s^2 = - \e{2\Psi} \, \d t^2 + a^2 \, \e{2\Phi} \, \d \vect{x}^2 .    
\end{equation}
The comoving dark matter velocity satisfies $u^a = \e{-\Psi} \gamma ( 1, \vect{v} )$,
where $\gamma = (1-\vphys^2)^{-1/2}$ is the special-relativistic Lorentz
factor and $\vphys = a \e{\Phi - \Psi} v$ is the physical peculiar 3-velocity.
To obtain the true physical velocity
for a source at distance $d$
we should add $\vphys$ to the Hubble flow $\vflow = H d$.
In this metric,
the continuity equation for a perfect fluid with pressure $p$ and density
$\rho$ can be written
\begin{equation}
    \partial_t \Big( \gamma^2 (p + \rho) - p \Big)
    + \grad \cdot \Big( \gamma^2 (p + \rho) \vect{v} \Big)
    + \gamma^2 ( p + \rho )
    \Big(
        \vect{v} \cdot \grad (\Psi + 3 \Phi)
        + (H + \dot{\Phi})(4 - \gamma^{-2})
    \Big)
    = 0 ,
    \label{eq:relativistic-continuity}
\end{equation}
and the Euler equation is
\begin{equation}
    \begin{split}
        \partial_t \Big( \gamma^2 \vect{v} (p + \rho) \Big)
        & + (\vect{v} \cdot \grad) \Big( \gamma^2 \vect{v} (p + \rho) \Big)
        + \frac{1}{a^2} \e{2\Psi - 2\Phi} \grad p
        \\
        & \mbox{} +
        \gamma^2 \vect{v} (p + \rho)
        \Big(
            \grad \cdot \vect{v} + 5 (H + \dot{\Phi}) + 5 \vect{v} \cdot \grad \Phi
            - \vect{v} \cdot \grad \Psi - \dot{\Psi}
        \Big)
        \\
        & \mbox{} +
        \frac{1}{a^2} \e{2\Psi - 2\Phi} ( p + \rho )
        \Big(
            \gamma^2 \grad \Psi
            - (\gamma^2 - 1) \grad \Phi
        \Big)
        = 0
        .
    \end{split}
    \label{eq:relativistic-euler}
\end{equation}
An overdot denotes a derivative with respect to time $t$.
The gravitational potentials satisfy the Poisson constraint,
\begin{equation}
    \begin{split}
        \frac{1}{a^2} \grad^2 \Phi + \frac{1}{2a^2} \grad \Phi \cdot \grad \Phi
        =
        &
        \mbox{}
        - \frac{\e{2\Phi}}{2\Mp^2}
        \Big(
            \gamma^2 ( p + \rho ) - p
            - 3H a^2 \e{-2\Psi} \grad^{-2}
                \grad \cdot [\gamma^2 \e{2\Phi} (p + \rho) \vect{v}]
        \Big)
        \\ & \mbox{}
        + \e{2\Phi - 2\Psi}
        \Big(
            \frac{3}{2} H^2 + 3 H \grad^{-2}
            [
                (H + \dot{\Phi}) \grad^2 \Psi
                + \grad \dot{\Phi} \cdot \grad \Psi
            ]
            + \frac{3}{2} \dot{\Phi}^2
        \Big) .
    \end{split}
    \label{eq:relativistic-poisson}
\end{equation}
Finally, for Einstein gravity coupled to
a perfect fluid, the gravitational potentials
will be related by the no-slip condition $\Psi = - \Phi$.
All these equations are exact.
In particular, they do not assume that the density and pressure
are perturbatively small or that velocities are non-relativistic.

\para{Non-relativistic limit}
Up to this point we have retained all terms in order to make clear
what is entailed by our approximations.
We wish to use these equations to
describe deposition of matter by a gravitationally-driven flow
within the potential wells associated with $\Phi$ and $\Psi$.
Assume that the flow carries density $\rhoflow$ which is deposited onto
a condensation of density $\rhocondensation$.
Therefore
the density contrast $\delta$ is approximately $\rhocondensation/\rhoflow$.
In a static Newtonian universe
the flow velocity at distance $R$ from the condensation
is roughly
\begin{equation}
    \label{eq:velocity-scaling}
    v \sim \frac{R}{\tfreefall} \frac{\rhocondensation}{\rhoflow}
    \sim \frac{R}{\tfreefall} \delta ,
\end{equation}
where $\tfreefall \approx (G \rhoflow)^{-1/2}$ is the free-fall time associated with
the flow.
This correlation between $v$ and $\delta$ is characteristic of an
inverse-square-law force.
It continues to apply in an expanding universe described by Einstein
gravity, adjusted by a scale-independent constant of order
unity that accounts for competition between Newtonian attraction
and cosmological repulsion.
In a non-Einstein gravity
we should expect its $R$ dependence or the overall constant of
proportionality to receive corrections.
Ultimately,
it is these corrections that we wish to explore
using redshift-space distortions.

Returning to Einstein gravity
and temporarily restoring factors of $c$
we conclude that $v/c$ scales like $\tR / \tfreefall$,
where $\tR = R/c$ is the light-crossing time at distance $R$.
In the case of cosmological structure formation the flow density $\rhoflow$
is the background matter density and $\tfreefall$ is of order a Hubble time.
Therefore $\tR / \tfreefall \ll 1$ on any scale well inside the Hubble
radius, making $v/c \ll 1$.
On these scales
it follows that
relativistic corrections $\sim \BigO(\gamma)$
will be negligible.
A similar discussion was given by Fry~\cite{Fry:1983cj}.

On the other hand, terms of order
$\grad \vect{v} \sim \tfreefall^{-1} \rhocondensation / \rhoflow$
need not be suppressed.
In combination with a time derivative such as $\dot{\rho}$ or $\dot{\vect{v}}$
the relative importance of such terms
will be of order $\tfreefall \grad \vect{v} \sim
\rhocondensation / \rhoflow$, which need not be especially small.
Therefore it is meaningful to develop a series expansion in $\grad \vect{v}$
while neglecting relativistic corrections from
terms of order $v^2$ and higher.
This is standard perturbation theory or `SPT'.
Specializing to matter domination, in which the gravitational potentials
are determined by the matter density fluctuation,
and keeping only terms linear in $\Phi = -\Psi$,
Eqs.~\eqref{eq:relativistic-continuity}--\eqref{eq:relativistic-poisson}
for pressureless cold dark matter
reduce to
\begin{subequations}
    \begin{align}
        \label{eq:SPT-continuity}
        \dot{\delta} + \grad \cdot \Big( (1 + \delta) \vect{v} \Big) & = 0 \\
        \label{eq:SPT-Euler}
        \dot{\vect{v}} + (\vect{v} \cdot \grad) \vect{v}
        + 2 H \vect{v} - \frac{1}{a^2} \grad \Phi & = 0 \\
        \label{eq:SPT-Poisson}
        \frac{1}{a^2} \grad^2 \Phi & = - \frac{3 H^2}{2} \Omega_m \delta ,
    \end{align}
\end{subequations}
where we have decomposed the density as $\rho = \rho_0 + \delta \rho$,
with $\rho_0$ the homogeneous background, and $\delta = \delta \rho / \rho$
is the density contrast.
The quantity $\Omega_m = \rho_m / (3 H^2 \Mp^2)$ is the redshift-dependent
matter density parameter.

\para{Radial inflow approximation}
On large scales the flow $\vect{v}$ will be oriented
nearly radially into
a nearby potential well and the vorticity $\vect{\omega} =
\grad \times \vect{v}$ will be very small.
In this `potential flow' region the velocity can be written as a gradient
$\vect{v} = \grad \grad^{-2} \theta$, where
$\theta = \grad \cdot \vect{v}$
is the velocity divergence.
In this approximation, after translation to Fourier space,
Eqs.~\eqref{eq:SPT-continuity}--\eqref{eq:SPT-Poisson}
become
\begin{subequations}
    \begin{align}
        \label{eq:alpha-eq}
        \dot{\delta}_{\vect{k}} + \theta_{\vect{k}} & =
            - \int \frac{\d^3 q \, \d^3 s}{(2\pi)^6} \, (2\pi)^3
            \delta(\vect{k} - \vect{q} - \vect{s}) \alpha(\vect{q}, \vect{s})
            \theta_{\vect{q}} \delta_{\vect{s}} , \\
        \label{eq:beta-eq}
        \dot{\theta}_{\vect{k}} - 2H \dot{\theta}_{\vect{k}}
            + \frac{3 H^2}{2} \Omega_m \delta_{\vect{k}}
            & = - \int \frac{\d^3 q \, \d^3 s}{(2\pi)^6} \, (2\pi)^3
            \delta(\vect{k} - \vect{q} - \vect{s}) \beta(\vect{q}, \vect{s})
            \theta_{\vect{q}} \theta_{\vect{s}} ,
    \end{align}
\end{subequations}
where the dimensionless kernels $\alpha(\vect{q}, \vect{s})$ and
$\beta(\vect{q}, \vect{s})$ satisfy
\begin{subequations}
    \begin{align}
        \label{eq:alpha-def}
        \alpha(\vect{q}, \vect{s}) & =
            \frac{\vect{q} \cdot (\vect{q} + \vect{s})}{q^2} , \\
        \label{eq:beta-def}
        \beta(\vect{q}, \vect{s}) & =
            \frac{\vect{q} \cdot \vect{s}}{2 q^2 s^2}(\vect{q} + \vect{s})^2 .
    \end{align}
\end{subequations}
Notice that $\beta$ is symmetric but $\alpha$ is not.
For future use it is helpful to define a symmetrized version
of weight unity,
\begin{equation}
    \label{eq:alpha-sym-def}
    \bar{\alpha}(\vect{q}, \vect{s}) = \frac{1}{2} \alpha(\vect{q}, \vect{s}) +
    \frac{1}{2} \alpha(\vect{s}, \vect{q}) .
\end{equation}
We also define a third kernel $\gamma(\vect{q}, \vect{s})$ to be a sum
of the $\alpha$ and $\beta$ kernels,
\begin{equation}
    \label{eq:gamma-def}
    \gamma(\vect{q}, \vect{s}) = \alpha(\vect{q}, \vect{s}) + \beta(\vect{q}, \vect{s}) .    
\end{equation}
Like $\alpha$, it can be symmetrized to give $\bar{\gamma}(\vect{q}, \vect{s})$.
Observe that the linear part of Eq.~\eqref{eq:alpha-eq} reads $\theta_{\vect{k}} = -
\dot{\delta}_{\vect{k}}$, which replicates our conclusion above that
$\grad \vect{v} \sim \tfreefall^{-1} \rhocondensation/\rhoflow \sim H \delta$.

Combining Eqs.~\eqref{eq:alpha-eq}--\eqref{eq:beta-eq}
to eliminate $\theta_{\vect{k}}$
and obtain a single second-order equation for $\delta_{\vect{k}}$,
and exchanging cosmic time $t$
for redshift $z$, defined by
\begin{equation}
    1 + z = \frac{a_0}{a} ,    
\end{equation}
where $a_0 = a(t_0)$ is the present-day value of the scale factor,
we find
\begin{equation}
    \begin{split}
        \delta_{\vect{k}}''
        - \frac{1-\epsilon}{1 + z} \delta_{\vect{k}}'    
        & \mbox{} - \frac{3}{2} \frac{\Omega_m}{(1+z)^2} \delta_{\vect{k}}
        \\ &
        = - \int \frac{\d^3 q \, \d^3 s}{(2\pi)^6} (2\pi)^3 \delta(\vect{k} - \vect{q} - \vect{s})
        S_2(\vect{q}, \vect{s})
        \\ &
        \quad - \int \frac{\d^3 q \, \d^3 s}{(2\pi)^6} (2\pi)^3
            \delta(\vect{k} - \vect{q} - \vect{s})
            \int \frac{\d^3 t \, \d^3 u}{(2\pi)^6} (2\pi)^3
            \delta(\vect{s} - \vect{t} - \vect{u})
            S_3(\vect{q}, \vect{s}, \vect{t}, \vect{u})
        \\ &
        \quad + \BigO(\delta^4) .
    \end{split}
    \label{eq:total-delta-eq}
\end{equation}
We have retained terms only
up to $\BigO(\delta^3)$;
those of higher order do not contribute to the one-loop power spectrum.
A prime $'$ denotes a derivative with respect to $z$.
The quantity $\epsilon$ is defined by $\epsilon = - \dot{H}/H^2$
and can be related to the deceleration parameter.
The source terms $S_2$ and $S_3$ satisfy
\begin{equation}
    \label{eq:S2-def}
    S_2(\vect{q}, \vect{s}) =
    \bar{\gamma}(\vect{q}, \vect{s}) \delta'_{\vect{q}} \delta'_{\vect{s}}
    + \frac{3}{2} \frac{\Omega_m}{(1+z)^2} \bar{\alpha}(\vect{q}, \vect{s})
    \delta_{\vect{q}} \delta_{\vect{s}} ,
\end{equation}
and
\begin{equation}
    \label{eq:S3-def}
    S_3(\vect{q}, \vect{s}, \vect{t}, \vect{u}) =
    \alpha(\vect{s}, \vect{q}) \beta(\vect{t}, \vect{u})
    \delta_{\vect{q}} \delta'_{\vect{t}} \delta'_{\vect{u}}
    + 2 \beta(\vect{q}, \vect{s}) \alpha(\vect{t}, \vect{u})
    \delta'_{\vect{q}} \delta'_{\vect{t}} \delta_{\vect{u}}
    + \alpha(\vect{s}, \vect{q}) \alpha(\vect{t}, \vect{u})
    \delta'_{\vect{q}} \delta'_{\vect{t}} \delta_{\vect{u}} .    
\end{equation}

\subsection{Eulerian perturbation theory}
\label{sec:eulerian-perturbation-theory}
The most straightforward approach to solution of Eq.~\eqref{eq:total-delta-eq}
is via an expansion in powers of $\delta$.
The outcome of this procedure is described as Eulerian perturbation theory.

\para{Linear solution}
First consider the linear term, which does not require the sources $S_2$ and $S_3$.
Because Eq.~\eqref{eq:total-delta-eq} applies only during matter domination
we should suppose the initial condition $\delta_{\vect{k}} = \delta_{\vect{k}}^\ast$
to be set at some redshift $z = z^\ast$
that is well within the matter era, but still early enough that
terms of order $(\delta_{\vect{k}}^\ast)^2$ or higher can be neglected.
For practical calculations we normally set $z^\ast \approx 50$.

The linear solution is $\delta_{\vect{k}}(z) = D(z) \delta_{\vect{k}}^\ast$,
where the growth function $D(z)$ satisfies
\begin{equation}
    D'' - \frac{1-\epsilon}{1+z} D' - \frac{3}{2} \frac{\Omega_m}{(1+z)^2} D = 0     
    \label{eq:meszaros-eq} .
\end{equation}
If the initial time $z^\ast$ is chosen sufficiently early
then the initial condition requires that $D(z)$
is approximately given by the matter-dominated
solution $D(z) \approx a(z)/a(z^\ast)$.
Notice that $D^\ast = D(z^\ast) = 1$.
Solutions to this equation
were studied by M\'{e}sz\'{a}ros~\cite{Meszaros:1974tb}
and Groth \& Peebles~\cite{1975A&A....41..143G}.
The velocity can be determined from the linear part of Eq.~\eqref{eq:alpha-eq},
yielding
\begin{equation}
    \theta_{\vect{k}} = - f H \delta_{\vect{k}}
    \label{eq:theta-linear}
\end{equation}
where the growth factor $f(z)$ is defined to be
\begin{equation}
    f \equiv - (1+z) \frac{D'}{D} = \frac{\d \ln D}{\d \ln a} .
    \label{eq:growth-factor}
\end{equation}
Eqs.~\eqref{eq:theta-linear}--\eqref{eq:growth-factor}
are nothing more than the estimate~\eqref{eq:velocity-scaling}
in this model, with $\tfreefall = 1/H$ and $R \sim 1/k$,
and $f$ representing a scale-independent damping of the
gravitational force due to cosmological
expansion.
In the matter-only Einstein--de Sitter model
we have $f=1$ and there is no damping
of the correlation between
$\vect{v}$ and $\delta$; the effect of the expansion
is only to soften exponential growth of $\delta$ into a power-law.
For $\Omega_m < 1$ there is extra suppression
which can be estimated in Einstein gravity by
$f \approx \Omega_m^{5/9}$~\cite{Linder:2005in}.

\para{Second-order solution}
To distinguish the different contributions to $\delta_{\vect{k}}$ and
$\theta_{\vect{k}}$ we attach a label $n$ indicating the order
in perturbation theory. The linear solution described above gives the
first-order component
$\delta_{\vect{k},1}$.
The second-order component
$\delta_{\vect{k},2}$
is generated by insertion of linear solutions
into the quadratic source $S_2$.
It gives
\begin{equation}
    \delta_{\vect{k},2} =
    \int \frac{\d^3 q \, \d^3 s}{(2\pi)^6}
    \, (2\pi)^3 \delta(\vect{k} - \vect{q} - \vect{s})
    \delta_{\vect{q}}^\ast \delta_{\vect{s}}^\ast
    \Big(
        \DA(z) \bar{\alpha}(\vect{q}, \vect{s})
        + \DB(z) \bar{\gamma}(\vect{q}, \vect{s})
    \Big)
    ,
    \label{eq:delta-2}
\end{equation}
for which the spatial average $\vect{k}=0$ mode vanishes
because $\alpha(\vect{q},-\vect{q}) =
\beta(\vect{q},-\vect{q}) = \gamma(\vect{q},-\vect{q}) = 0$.
The time-dependent growth functions $\DA(z)$ and $\DB(z)$ are
analogues of the linear growth function $D(z)$.
They are
solutions to the equations
\begin{subequations}
    \begin{align}
        \label{eq:DA-def}
        \DA'' - \frac{1-\epsilon}{1+z} \DA' - \frac{3}{2} \frac{\Omega_m}{(1+z)^2} \DA
            & = \frac{3}{2} \frac{\Omega_m}{(1+z)^2} D^2 \\
        \label{eq:DB-def}
        \DB'' - \frac{1-\epsilon}{1+z} \DB' - \frac{3}{2} \frac{\Omega_m}{(1+z)^2} \DB
            & = (D')^2 .
    \end{align}
\end{subequations}
We choose initial equations so that $D_A$ and $D_B$ match the corresponding growth functions
in a matter-only model at the initial redshift $z = z^\ast$.
This makes our results practically independent of the choice of $z^\ast$, provided
it is taken to be sufficiently early.

\para{Third-order solution}
The third-order solution is sourced by insertion of linear
solutions into the cubic term $S_3$ together with insertion of
one linear and one 
second-order solution in the quadratic term $S_2$.
It can be written
\begin{equation}
    \begin{split}
        \delta_{\vect{k},3} =
        \int & \frac{\d^3 q \, \d^3 s \, \d^3 t}{(2\pi)^9}
        \, (2\pi)^3 \delta(\vect{k} - \vect{q} - \vect{s} - \vect{t})
        \delta_{\vect{q}}^\ast \delta_{\vect{s}}^\ast \delta_{\vect{t}}^\ast
        \\ &
        \mbox{} \times
        \Bigg(
            2 [\DD(z) - \DJ(z)] \bar{\gamma}(\vect{s} + \vect{t}, \vect{q}) \bar{\alpha}(\vect{s}, \vect{t})
            + 2 \DE(z) \bar{\gamma}(\vect{s} + \vect{t}, \vect{q}) \bar{\gamma}(\vect{s}, \vect{t})
        \\ &
        \qquad
            + 2 [\DF(z) + \DJ(z)] \bar{\alpha}(\vect{s} + \vect{t}, \vect{q}) \bar{\alpha}(\vect{s}, \vect{t})
            + 2 \DG(z) \bar{\alpha}(\vect{s} + \vect{t}, \vect{q}) \bar{\gamma}(\vect{s}, \vect{t})
        \\ &
        \qquad
            + \DJ(z)
            \Big[
                \alpha(\vect{s} + \vect{t}, \vect{q}) \bar{\gamma}(\vect{s}, \vect{t})
                - 2 \alpha(\vect{s} + \vect{t}, \vect{q}) \bar{\alpha}(\vect{s}, \vect{t})
            \Big]
        \Bigg) .
    \end{split}
    \label{eq:delta-3}
\end{equation}
The new growth functions $\DD$, $\DE$, $\DF$, $\DG$ and $\DJ$ satisfy
\begin{subequations}
    \begin{align}
        \DD'' - \frac{1-\epsilon}{1+z} \DD' - \frac{3}{2} \frac{\Omega_m}{(1+z)^2} \DD & = D' \DA' \\        
        \DE'' - \frac{1-\epsilon}{1+z} \DE' - \frac{3}{2} \frac{\Omega_m}{(1+z)^2} \DE & = D' \DB' \\        
        \DF'' - \frac{1-\epsilon}{1+z} \DF' - \frac{3}{2} \frac{\Omega_m}{(1+z)^2} \DF & = \frac{3}{2} \frac{\Omega_m}{(1+z)^2} D \DA \\        
        \DG'' - \frac{1-\epsilon}{1+z} \DG' - \frac{3}{2} \frac{\Omega_m}{(1+z)^2} \DG & = \frac{3}{2} \frac{\Omega_m}{(1+z)^2} D \DB \\        
        \DJ'' - \frac{1-\epsilon}{1+z} \DJ' - \frac{3}{2} \frac{\Omega_m}{(1+z)^2} \DJ & = (D')^2 D .
    \end{align}    
\end{subequations}
As above,
each $D_i$ should be solved subject to the
boundary condition that it matches a matter-only model
at $z = z^\ast$.

\begin{figure}
    \begin{center}
        \includegraphics{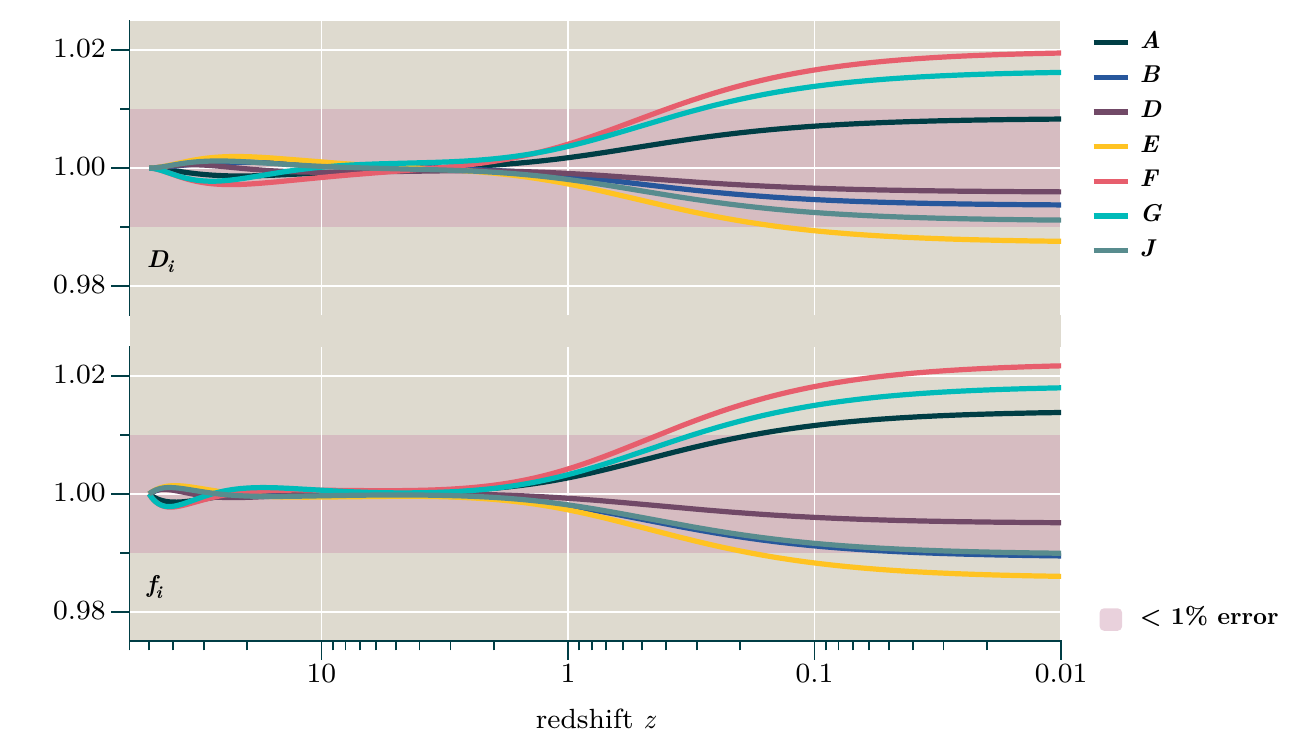}
    \end{center}
    \caption{\label{fig:EdS-growth}Time evolution of the growth functions
    $D_i$ and growth factors $f_i$
    for Planck2015 parameter values~\cite{Ade:2015xua}
    relative to the Einstein-de Sitter
    approximations of Table~\ref{table:EdS-growth}.
    The light pink shaded region shows where the Einstein--de Sitter approximation
    is accurate to better than $1\%$.
    Some jitter is visible near the initial redshift $z^\ast = 50$,
    which is caused by slight inaccuracies in our initial conditions.
    These are set assuming matter domination and neglect the
    radiation component. The effect is negligible for $z < 1$.}
\end{figure}

\para{Einstein--de Sitter approximation}
It is common to simplify Eqs.~\eqref{eq:delta-2} and~\eqref{eq:delta-3}
by exchanging the non-linear growth functions $D_i$
for powers of the linear growth function $D$.
(See Appendix B.3 of Scoccimarro et al.~\cite{Scoccimarro:1997st}.)
This procedure is exact for the $\Omega_m = 1$ Einstein--de Sitter model.
If we define a growth factor $f_i$ for each $D_i$
by analogy with Eq.~\eqref{eq:growth-factor},
\begin{equation}
    \label{eq:growth-factor-def}
    f_i \equiv - (1+z) \frac{D_i'}{D_i} ,
\end{equation}
then the solutions for $D_i$ and $f_i$ in an
Einstein--de Sitter model are given in Table~\ref{table:EdS-growth}.
With these choices the combination
$\DA \bar{\alpha}(\vect{q},\vect{s}) + \DB \bar{\gamma}(\vect{q},\vect{s})$
in Eq.~\eqref{eq:delta-2}
becomes the standard kernel $D^2 F_2(\vect{q},\vect{s})$
and the kernel in Eq.~\eqref{eq:delta-3}
becomes
$D^3 F_3(\vect{q},\vect{s},\vect{t})$~\cite{1981MNRAS.197..931J,1983MNRAS.203..345V,
Goroff:1986ep,Wise:1988kua,Makino:1991rp,Scoccimarro:1995if,Scoccimarro:1996se}.
\ctable[
    caption = {Relation between the non-linear growth functions
    $D_i$ and their Einstein--de Sitter counterparts, which can be expressed as powers
    of the linear growth function $D$.},
    label = table:EdS-growth
]{rrrrrrrr}{}{
    \toprule
        & $A$
        & $B$
        & $D$
        & $E$
        & $F$
        & $G$
        & $J$ \NN
        growth function $D_i$
            & $\frac{3}{7} D^2$
            & $\frac{2}{7} D^2$
            & $\frac{2}{21} D^3$
            & $\frac{4}{63} D^3$
            & $\frac{1}{14} D^3$
            & $\frac{1}{21} D^3$
            & $\frac{1}{9} D^3$ \NN
        growth factor $f_i$
            & $2f$
            & $2f$
            & $3f$
            & $3f$
            & $3f$
            & $3f$
            & $3f$ \NN
    \bottomrule
}
In Fig.~\ref{fig:EdS-growth} we plot the time evolution of the $D_i$ and $f_i$,
calculated for a Planck2015 cosmology~\cite{Ade:2015xua},
relative to the `Einstein--de Sitter approximation'
computed using Table~\ref{table:EdS-growth}.%
    \footnote{To be clear, note that what we describe as the
    \emph{Einstein--de Sitter approximation} consists of taking the $D_i$
    and $f_i$ to satisfy the relations of Table~\ref{table:EdS-growth}
    using the appropriate linear $D(z)$ for the cosmology under discussion.
    We do not use the specific $D(z)$ corresponding to an
    $\Omega_m=1$, $\Omega_\Lambda=0$ Einstein--de Sitter model.}
At large $z$ the growth functions match the Einstein--de Sitter
values rather closely~\cite{Scoccimarro:1997st}.
At $z \sim 2$, where the vacuum energy becomes significant,
they begin to deviate from the Einstein--de Sitter prediction.
At low redshift $z \sim 0$ the largest discrepancies
are roughly $2\%$,
implying that the full time dependence may be required
for very accurate calculations.

In this paper we retain the distinction between the different
growth functions,
and in~\S\ref{sec:results}
we will quantify the error incurred by the Einstein--de Sitter
approximation.

\para{Power spectra}
The two-point function following from
Eqs.~\eqref{eq:meszaros-eq},
\eqref{eq:delta-2} and~\eqref{eq:delta-3}
was computed by Suto \& Sasaki~\cite{Suto:1990wf},
and later for the velocity power spectrum
by Makino, Sasaki \& Suto~\cite{Makino:1991rp};
see also Scoccimarro \& Frieman~\cite{Scoccimarro:1995if,Scoccimarro:1996se}
and Scoccimarro~\cite{Scoccimarro:2004tg}.
Assuming $\delta^\ast_{\vect{k}}$ to be a Gaussian
random field there are three contributions,
conventionally labelled $P_{11}$, $P_{22}$ and $P_{13}$,
\begin{subequations}
    \begin{align}
        \langle \delta_{\vect{k}_1,1} \delta_{\vect{k}_2,1} \rangle & = (2\pi)^3 \delta(\vect{k}_1 + \vect{k}_2) P_{11}(k) \\
        \langle \delta_{\vect{k}_1,2} \delta_{\vect{k}_2,2} \rangle & = (2\pi)^3 \delta(\vect{k}_1 + \vect{k}_2) P_{22}(k) \\
        \langle \delta_{\vect{k}_1,1} \delta_{\vect{k}_2,3} + \delta_{\vect{k}_1,3} \delta_{\vect{k}_2,1} \rangle & = (2\pi)^3 \delta(\vect{k}_1 + \vect{k}_2) P_{13}(k) ,
    \end{align}
\end{subequations}
where $k$ is the common magnitude of the wavevectors $\vect{k}_1$ and $\vect{k}_2$,
and to prevent clutter we have suppressed the $z$-dependence of each quantity.
The linear contribution $P_{11}$ is described as the tree-level power spectrum,
and the sum $P_{22} + P_{13}$ is the one-loop contribution.
Defining
the initial power spectrum $\Pinit(k)$ to satisfy
$\langle \delta_{\vect{k}_1}^\ast \delta_{\vect{k_2}}^\ast \rangle = (2\pi)^3 \delta(\vect{k}_1 + \vect{k}_2) P^\ast(k)$,
these different contributions can be written
\begin{subequations}
\begin{align}
    \label{eq:P11}
    P_{11}(k) & = D^2 \Pinit(k) \\
    \label{eq:P22}
    P_{22}(k) & = \DA^2 P_{AA}(k) + \DA \DB P_{AB}(k) + \DB^2 P_{BB}(k)
\end{align}
and
\begin{equation}
    \label{eq:P13}
    \begin{split}
        P_{13}(k) = D P^\ast(k) \Big[ & (\DD-\DJ) P_D(k) + \DE P_E(k) + (\DF+\DJ) P_F(k) \\ & \mbox{} + \DG P_G(k) + \frac{\DJ}{2} \big[ P_{J2}(k) - 2 P_{J1}(k) \big] \Big] .
    \end{split}
\end{equation}
\end{subequations}
The quantities $P_i$ appearing in these expressions are defined by
\begin{subequations}
    \begin{align}
        \label{eq:PAA}
        P_{AA} & \equiv 2 \int \frac{\d^3 q}{(2\pi)^3} \bar{\alpha}(\vect{k}-\vect{q}, \vect{q})^2 \Pinit(\vect{q}) \Pinit(\vect{k}-\vect{q}) \\
        \label{eq:PAB}
        P_{AB} & \equiv 4 \int \frac{\d^3 q}{(2\pi)^3} \bar{\alpha}(\vect{k}-\vect{q}, \vect{q}) \bar{\gamma}(\vect{k}-\vect{q}, \vect{q}) \Pinit(\vect{q}) \Pinit(\vect{k}-\vect{q}) \\
        \label{eq:PBB}
        P_{BB} & \equiv 2 \int \frac{\d^3 q}{(2\pi)^3} \bar{\gamma}(\vect{k}-\vect{q}, \vect{q})^2 \Pinit(\vect{q}) \Pinit(\vect{k}-\vect{q}) \\
        \label{eq:PD}
        P_D & \equiv 8 \int \frac{\d^3 q}{(2\pi)^3} \bar{\gamma}(\vect{k}-\vect{q}, \vect{q}) \bar{\alpha}(\vect{k}, -\vect{q}) \Pinit(\vect{q}) \\
        \label{eq:PE}
        P_E & \equiv 8 \int \frac{\d^3 q}{(2\pi)^3} \bar{\gamma}(\vect{k}-\vect{q}, \vect{q}) \bar{\gamma}(\vect{k}, -\vect{q}) \Pinit(\vect{q}) \\
        \label{eq:PF}
        P_F & \equiv 8 \int \frac{\d^3 q}{(2\pi)^3} \bar{\alpha}(\vect{k}-\vect{q}, \vect{q}) \bar{\alpha}(\vect{k}, -\vect{q}) \Pinit(\vect{q}) \\
        \label{eq:PG}
        P_G & \equiv 8 \int \frac{\d^3 q}{(2\pi)^3} \bar{\alpha}(\vect{k}-\vect{q}, \vect{q}) \bar{\gamma}(\vect{k}, -\vect{q}) \Pinit(\vect{q}) \\
        \label{eq:PJ1}
        P_{J1} & \equiv 8 \int \frac{\d^3 q}{(2\pi)^3} \alpha(\vect{k}-\vect{q}, \vect{q}) \bar{\alpha}(\vect{k}, -\vect{q}) \Pinit(\vect{q}) \\
        \label{eq:PJ2}
        P_{J2} & \equiv 8 \int \frac{\d^3 q}{(2\pi)^3} \alpha(\vect{k}-\vect{q}, \vect{q}) \bar{\gamma}(\vect{k}, -\vect{q}) \Pinit(\vect{q})
    \end{align}
\end{subequations}
If we replace the growth functions $D_i$ by their
Einstein--de Sitter counterparts
of Table~\ref{table:EdS-growth}
then Eqs.~\eqref{eq:P11}--\eqref{eq:PJ2}
reproduce the one-loop $\delta$ power spectrum
reported by Suto et al.~\cite{Suto:1990wf}.

\para{Infrared safety}
Each of these integrals converges individually
in the infrared region
$q \ll k$
provided $\Pinit(k)$ is no more divergent than $1/k$
at small $k$,
which is amply satisfied for realistic power spectra.
We discuss the ultraviolet region
$q \gg k$ in detail in~\S\ref{sec:ultraviolet-sensitivity} below.

If $\Pinit(k)$ diverges in the infrared
more strongly than $1/k$ but less than $1/k^3$,
Scoccimarro \& Frieman~\cite{Scoccimarro:1995if}
demonstrated that
any low-$q$ divergences would cancel
between the $22$ and $13$ terms
in Galilean-invariant correlation
functions.
This is part of a more general
cancellation of the low-$q$
contribution~\cite{Sugiyama:2013gza,Pajer:2013jj,Carrasco:2013sva}.
Assuming an Einstein--de Sitter background
and focusing on the low-$q$ region we have
\begin{subequations}
    \begin{align}
        \label{eq:P13-IR}
        P_{13} & = - \frac{2}{3} D^2 k^2 \Pinit(k)
            \int \frac{\d q}{(2\pi)^3}
            \bigg[
                1 + \BigO\Big(\frac{q^2}{k^2}\Big)
            \bigg]
            \Pinit(\vect{q}) \\
        \label{eq:P22-IR}
        P_{22} & = \frac{1}{3} D^2 k^2
            \int \frac{\d q}{(2\pi)^3}
            \bigg[
                1 + \BigO\Big(\frac{q^2}{k^2}\Big)
            \bigg]
            \Pinit(\vect{q})\Pinit(\vect{k}-\vect{q}) .
    \end{align}
\end{subequations}
The leading part of~\eqref{eq:P22-IR} comes from regions centred on
$\vect{q}=0$ and $\vect{q}=\vect{k}$
which each give a contribution of the same form as~\eqref{eq:P13-IR},
and therefore we have
cancellation between these terms.
The cancellation between the $\BigO(q^2/k^2)$ corrections
is not exact, so
the total one-loop term will diverge in the low-$q$ region
if $\Pinit(k)$ is more divergent than $1/k^3$.

This cancellation means that it is necessary to compute the
integrals~\eqref{eq:PAA}--\eqref{eq:PJ1}
with sufficient accuracy that we retain a good estimate of the
remainder after cancellation has occurred.
Alternatively, they can be grouped in a form in which
cancellation is explicit, as described in Ref.~\cite{Carrasco:2013sva}.
In practice we do not find it is onerous to achieve the required
accuracy for realistic input power spectra $\Pinit(k)$.

\subsection{Ultraviolet sensitivity and renormalization}
\label{sec:ultraviolet-sensitivity}
Each $P_i$ defined in Eqs.~\eqref{eq:PAA}--\eqref{eq:PJ2}
involves a weighted integral over the power spectrum $\Pinit(\vect{q})$
(or the convolution $\Pinit(\vect{q}) \Pinit(\vect{k}-\vect{q})$ in the case
of integrals contributing to $P_{22}$),
with weighting function given by a combination of the kernels $\alpha$ and $\gamma$.
The terminology `one-loop' is borrowed from the diagrammatic expansion of
quantum field theory
in which similar integrals are encountered.
In either case we can regard the loop as an estimate of the average
influence of fluctuations over the range $\vect{q}$
on the single mode of wavenumber $\vect{k}$.

In a free quantum field theory, the typical amplitude
of quantum fluctuations
of four-momentum $q^a$ decays like $1/q$ for large $|q|$, and therefore
the influence of individual high-momentum fluctuations decreases.
However, because the number of such fluctuations grows like $q^4$
their aggregated influence can be very large---indeed,
in perturbation theory,
the prediction may be unboundedly large.
The same behaviour can occur in
Eqs.~\eqref{eq:PAA}--\eqref{eq:PJ2},
in which the typical amplitude of fluctuations
on scale $q$ decreases like $\Pinit(q)^{1/2}$.
The corresponding contribution to the average
may be suppressed or enhanced depending on the details of the weighting
function,
but
since the number of modes grows like $q^3$ the aggregated effect
of high-momentum modes may again be significant or unbounded.

The resolution of this difficulty is to recognize that our predictions
for the typical amplitude of high-wavenumber fluctuations are unreliable.%
    \footnote{In this case, Eqs.~\eqref{eq:PAA}--\eqref{eq:PJ2}
    are also unreliable at very low wavenumbers,
    for which the relativistic corrections in
    Eqs.~\eqref{eq:relativistic-continuity}--\eqref{eq:relativistic-poisson}
    are no longer small.
    However, this is an artefact of our gauge choice
    and may be neglected
    provided there are no large contributions to the loop from
    Hubble-scale modes.}
In quantum field theory this is true because of our ignorance of the details
of very high energy physics.
In applications to structure formation we would (in principle) encounter the same
fundamental uncertainty at high enough energies, but in practice our ability
to accurately model amplitudes is already compromised at much lower wavenumber
because we cannot adequately
describe the details of non-linear halo and galaxy
formation, gas dynamics, feedback from active galactic nuclei, and so on.
Therefore our estimates of the aggregate influence on some low wavenumber
$\vect{k}$ from much higher wavenumbers $\vect{q}$ are not trustworthy
even if they are finite.

Although we cannot trust Eqs.~\eqref{eq:PAA}--\eqref{eq:PJ2}
as they stand, we can break them into two parts:
first, an integral that aggregates the influence of wavenumbers in a range
for which we believe that our estimate of typical
amplitudes is adequate;
and second, an integral over the remaining $\vect{q}$.
We cannot evaluate this second integral, but we can parametrize it.
Once suitable parameters have been determined, by comparison with observation
or simulation,
the theory is as predictive as if we had a reliable \emph{ab initio} estimate
of the typical amplitude for high-energy fluctuations.
This parametrization of unknown high-$\vect{q}$
effects is the content of the renormalization programme.
 
\para{Large $q$ contributions from $P_{13}$ terms}
The first step is to find a suitable parametrization
for the ultraviolet part of each integral.
The procedure is much the same as for conventional quantum
field theory,
although complicated by the presence of a time-dependent background.

First consider the large $|\vect{q}|$ contributions
to $P_D$, \ldots, $P_{J2}$, given by
Eqs.~\eqref{eq:PD}--\eqref{eq:PJ2}.
These contribute to the $P_{13}$ part of the one-loop power spectrum.
If there were no time dependence to accommodate,
we would express the dimensionless weighting functions
in these integrals as a Taylor series in $k^2/q^2$.
Using rotational invariance,
it follows immediately that the $q \gg k$
part of each integral can be parametrized as
\begin{equation}
    P \UVcontains \sum_{n=0}^\infty \frac{k^{2n}}{M_n^{2n}} ,
    \label{eq:P-parametrization}
\end{equation}
for some mass scales $M_n$.
(This parametrization may miss effects, associated with the remainder of the Taylor expansion,
that vanish for small $k$
faster than any finite power of $k$.
Such effects are not captured by the effective description.%
    \footnote{In quantum field theory the combination
    $k^2/q^2$ is typically replaced by $k^2/M^2$
    for some hard scale $M$.
    The remainder term captures effects that
    are not visible at any finite order in perturbation theory
    such as $\exp(-M^2/k^2)$.})
Notice that it does not matter how we divide the
$\vect{q}$ integral
and define its untrustworthy $q \gg k$ region,
because
any change in the division can be absorbed into
a redefinition of the mass scales $M_n$.

The low-energy region $q \lesssim k$ may also generate positive
powers of $k^2$.
If so, these are degenerate with the unknown ultraviolet
contributions.
But unlike the ultraviolet region, the low-energy region
may generate terms that are not analytic in $k^2$.
These non-analytic contributions cannot be modified by ultraviolet effects
and are unambiguous predictions of the low-energy theory
(see, eg., Refs.~\cite{Donoghue:1994dn,Donoghue:1995cz}).

In this picture it would be sufficient to measure six
independent mass scales (one for each of $P_D$, \ldots, $P_{J2}$)
for each power of $k$ included in the parametrization.
Unfortunately, if our description of the high-wavenumber modes is
inadequate to predict their amplitudes, it will also be
inadequate to predict their time dependence.
Therefore we cannot rely on these modes evolving in the way
prescribed by perturbation theory.
The result is that, rather than requiring just six numbers to fix the
relative size of each contribution to~\eqref{eq:P13},
we must allow the coefficient of each power of $k$
to become an arbitrary undetermined function of redshift.%
    \footnote{If we wish, we can apply this statement to each
    combination such as $D_D(z) P_D(k)$, but all these
    undetermined functions of time will assemble
    to give a single undetermined function of time
    for each term of the form $k^{2n} \Pinit(k)$ in
    $P_{13}$.
    It is only this single undetermined function that can be
    constrained.
    The division of the $P_{13}$ time dependence into
    $D$, $E$, \ldots, $J$
    components is part of the structure of low-energy
    perturbation theory and need not be respected by the
    ultraviolet terms.}
This procedure becomes predictive once we have
made enough measurements to constrain this function
over the redshift range of interest.
Depending on the range required this could
entail many more than six independent numbers.
We will return to this issue in~\S\ref{sec:redshift-dependence}.

The final result must still be independent of how we divide
the $\vect{q}$ integrals.
For this reason the unknown time-dependent function must contain
a component with the same redshift dependence as the $q \gg k$
region of each loop integral.
This enables it to subtract any unphysical dependence on
the arbitrary upper limit of this region.
If we cut off each integral at the same scale $\Lambda$,
then up to $\BigO(k^2)$ the 
$q \gg k$ region of $P_{13}$ behaves like
\begin{equation}
    P_{13} \supseteq - D \Pinit(k)
    \Big(
        18 \DD + 28 \DE - 7 \DF - 2 \DG - 13 \DJ
    \Big)
    \frac{k^2}{15\pi^2}
    \!\!\!\int\limits_{k \ll q \lesssim \Lambda}\!\!\! \d q \, \Pinit(q)
    + \cdots .
    \label{eq:P13-UV-divergence}
\end{equation}
Notice that the $k^0$ term is
absent~\cite{Goroff:1986ep,Wise:1988kua},
which
is a consequence of conservation of energy and momentum.%
    \footnote{See, eg., Peebles~\cite{1974A&A....32..391P}.
    The argument in this reference amounts to the observation that
    the large-scale matter distribution feels only tidal effects
    from small scales.
    Mercolli \& Pajer gave an explicit demonstration
    of this property for $\delta$ and (under certain circumstances)
    also the velocity $\vect{v}$~\cite{Mercolli:2013bsa}.
    The connexion to tidal forces
    was made explicitly in~{\S}5.2 of
    Baumann et al.~\cite{Baumann:2010tm}.\label{footnote:no-kzero-in-delta}}
Therefore up to $\BigO(k^2)$
the unknown ultraviolet dependence must take the form
\begin{equation}
    \begin{split}
        P_{13} & \UVcontains 2 D^2 k^2 \Pinit(k)
        \bigg\{
            \frac{18 \DD + 28 \DE - 7\DF - 2\DG - 13\DJ}{D} \ctrconst{2}{\delta}
            + \ctrfree{2}{\delta}(z)
        \bigg\}
        \\
        & \equiv - 2 D^2 \frac{k^2}{\kNL^2} \ctrterm{2}{\delta}(z) \Pinit(k)
    \end{split}
    \label{eq:P13-UV-parametrization}
\end{equation}
where $\ctrconst{2}{\delta}$ is
a fixed number of dimension $[\text{mass}]^{-2}$
that effectively takes the place of the mass scale $M_n$ in
Eq.~\eqref{eq:P-parametrization},
and
$\ctrfree{2}{\delta}(z)$ is an arbitrary function of $z$
representing any time dependence of the ultraviolet modes that cannot
be predicted from perturbation theory at low $k$.
For example, retarded memory effects that are nonlocal
in time may contribute to this function~\cite{Carroll:2013oxa,Fuhrer:2015cia}.
Eq.~\eqref{eq:P13-UV-parametrization}
is the \emph{counterterm} needed to renormalize the $P_{13}$
part of the one-loop
$\delta\delta$ power spectrum
up to $k^2$.

The quantities $\kNL$
and
$\ctrterm{2}{\delta}(z)$
are defined by the second equality in~\eqref{eq:P13-UV-parametrization}.
Only the combination $\ctrterm{2}{\delta}/\kNL^2$ can be constrained by
fitting to data,
but the separation of
$\kNL$ is conceptually useful
if all higher-order powers of $k$
are controlled by the same scale.
In this case the parametrization orders itself as an
expansion in $k/\kNL$ with coefficients such as $\ctrterm{2}{\delta}$ that are not
too different from unity.
Provided we are satisfied with fixed accuracy, we need only retain
sufficiently many terms to make $(k/ \kNL)^{2n}$ suitably small.
In this paper we retain only terms up to $\BigO(k^2)$.
We discuss the procedure to fix $\ctrterm{2}{\delta}$ in~\S\ref{sec:real-space-results}.

In principle we can carry this parametrization to as many powers of $k$
as we wish,
in which case we would encounter
further counterterms
involving
$k^4$, $k^6$, \ldots,
as in Eq.~\eqref{eq:P-parametrization},
all multiplying the combination $D^2 \Pinit(k)$.
The time dependence of each
term would be analogous to~\eqref{eq:P13-UV-parametrization}:
a term matching the redshift dependence from the $q \gg k$ part
of each integral,
and a second arbitrary time-dependent term $\ctrfree{4}{\delta}$, $\ctrfree{6}{\delta}$, \ldots,
representing unknown time dependence that cannot be
predicted from perturbation theory.

\para{Large $q$ contributions from $P_{22}$ terms}
Now consider
the analogous contributions
to~\eqref{eq:PAA}--\eqref{eq:PBB}.
These contribute to the $P_{22}$ part of the one-loop
power spectrum.
Much of the discussion of $P_{13}$ terms also
applies to these integrals, with the exception
that they do not enter $P_{22}$ in proportion to the
input power spectrum $\Pinit(k)$ as in Eq.~\eqref{eq:P13-UV-parametrization}.
Instead, their contribution to $P_{22}$ is simply a power
series in $k^2$.
After recovery of the correlation function $\xi(\vect{r})$
from the Fourier transform of $P(\vect{k})$,
such powers
generate terms
proportional to the $\delta$-function
$\delta(\vect{r})$
and its derivatives.
The same applies for $22$-type contributions
to the power spectrum of any operator, not just $\delta$.

Because these ultraviolet
contributions do not enter the power spectrum
in combination with $\Pinit(k)$
they must describe
fluctuations that are stochastically independent of $\delta$.
To interpret them
we should return to the division
between
a known low-energy sector
$q < \Lambda$
and an unknown high-energy sector $q > \Lambda$
described above.
These sectors are coupled by processes
in which
low-energy fluctuations
interact to produce high-energy fluctuations
or vice-versa.
When
energy is carried into the high-energy sector
by such processes it must be removed from our
description, but can later be returned.
Because this return of energy is mediated by
high-energy interactions
it falls below the effective resolution $\sim 1/\Lambda$
of the low-energy description and appears
nearly local.
In the correlation function its contribution
is therefore proportional to
$\delta(\vect{r})$ and its derivatives, in exactly the
manner described above.
The presence of such \emph{noise} and \emph{dissipation}
effects is well-understood
in applications of field theory to condensed
matter~\cite{Feynman:1963fq,Caldeira:1981rx,Caldeira:1982iu,Caldeira:1982uj};
for a textbook description, see
Kamenev~\cite{kamenev2011field}.
The application to effective field theories was emphasized by
Calzetta \& Hu~\cite{Calzetta:1996sy,Calzetta:1999xh}.

The conclusion is that we should add
extra counterterms that account for fluctuations
that are stochastically independent of the long-wavelength
part of $\delta$.
Baumann et al. called these
\emph{stochastic counterterms}~\cite{Baumann:2010tm}.
For the $\delta\delta$ power spectrum,
the $P_{22}$ contribution for $q \gg k$
begins at
$\BigO(k^4)$.
Therefore, in this paper, we assume these stochastic counterterms
to be unnecessary at the level of accuracy to which we are working.

\subsection{Renormalized operators}
\label{sec:renormalized-operators}

\para{Renormalized $\delta$ operator}
The analysis of~\S\ref{sec:ultraviolet-sensitivity}
can be rephrased in the language of renormalized operators.
By doing so we will able to unify our
treatment of the renormalized redshift-space
power spectrum with the discussion given here.

The outcome of \S\ref{sec:ultraviolet-sensitivity}
was a prescription for computing correlation functions
by cutting off each $\vect{q}$ integral and parametrizing the ultraviolet region
by counterterms.
This yields results that are the same as
would have been obtained from a modified
$\delta$ operator that mixes with a $\partial^2 \delta$ term,
\begin{equation}
    \renormalized{\delta}(\vect{x})
    =
    \delta_\Lambda(\vect{x})
    +
    \frac{\ctrterm{2}{\delta}(z)}{\kNL^2}
    \partial^2 \delta_\Lambda(\vect{x}) ,
    \label{eq:delta-renormalized}
\end{equation}
in which $\ctrterm{2}{\delta}$
should be treated as one-loop level
and therefore any diagram containing $\ctrterm{2}{\delta}$ need be
computed only to tree-level.
The subscript $\Lambda$ is a reminder
that loops involving $\delta_\Lambda$ should be cut off for $q \gtrsim \Lambda$.
As explained above, the arbitrariness in our choice of $\Lambda$ can be
compensated by a redefinition of the counterterm,
but to keep the notation simple we do not write this
dependence explicitly.
We describe $\renormalized{\delta}$ as the \emph{renormalized} density contrast.
If we had retained higher powers $k^4$, $k^6$, \ldots,
in the parametrization of the ultraviolet region
then these would appear as
mixing with further
operators
$\partial^4 \delta$, $\partial^6 \delta$, and so on.
In Eq.~\eqref{eq:P13-UV-parametrization}
the $k^0$ term is absent,
but
if present it would represent a multiplicative adjustment
of the normalization of $\delta$ on the right-hand side
of~\eqref{eq:delta-renormalized};
we shall see an example
for the velocity power spectrum below.
Finally, any stochastic counterterms would appear as
additive contributions
to $\renormalized{\delta}$ that are uncorrelated with $\delta_\Lambda$.

\para{Renormalized $\theta$ operator}
A similar analysis can be given for the velocity.
In the potential flow approximation
this yields
$\vect{v} = \im \vect{k} ( \phi_{\vect{k},1} + \phi_{\vect{k},2} + \phi_{\vect{k},3} )$,
where
\begin{subequations}
\begin{equation}
    \phi_{\vect{k},1} = \frac{H}{k^2} f D \delta^\ast_{\vect{k}} ,
    \label{eq:v-1}
\end{equation}
\begin{equation}
    \phi_{\vect{k},2} = \frac{H}{k^2} \int \frac{\d^3 q \, \d^3 s}{(2\pi)^6}
    \; (2\pi)^3 \delta(\vect{k}-\vect{q}-\vect{s})
    \delta^\ast_{\vect{q}} \delta^\ast_{\vect{s}}
    \Big(
        \DK(z) \bar{\alpha}(\vect{q},\vect{r})
        +
        \DL(z) \bar{\gamma}(\vect{q},\vect{r})
    \Big) ,
    \label{eq:v-2}
\end{equation}
and
\begin{equation}
\begin{split}
    \phi_{\vect{k},3} = \frac{H}{k^2} \int & \frac{\d^3 q \, \d^3 s \, \d^3 t}{(2\pi)^9}
    (2\pi)^3 \delta(\vect{k}-\vect{q}-\vect{s}-\vect{t})
    \delta^\ast_{\vect{q}} \delta^\ast_{\vect{s}} \delta^\ast_{\vect{t}}
    \\ &
    \mbox{} \times
    \Bigg(
        2 \DM(z) \bar{\gamma}(\vect{s}+\vect{t},\vect{q}) \bar{\alpha}(\vect{s},\vect{t})
        +
        2 \DN(z) \bar{\gamma}(\vect{s}+\vect{t},\vect{q}) \bar{\gamma}(\vect{s},\vect{t})
    \\ &
    \qquad
        +
        2 \DP(z) \bar{\alpha}(\vect{s}+\vect{t},\vect{q}) \bar{\alpha}(\vect{s},\vect{t})
        +
        2 \DQ(z) \bar{\alpha}(\vect{s}+\vect{t},\vect{q}) \bar{\gamma}(\vect{s},\vect{t})
    \\ &
    \qquad
        +
        \DR(z) \alpha(\vect{s}+\vect{t},\vect{q}) \bar{\alpha}(\vect{s},\vect{t})
        +
        \DS(z) \alpha(\vect{s}+\vect{t},\vect{q}) \bar{\gamma}(\vect{s},\vect{t})
    \Bigg) .
\end{split}
\label{eq:v-3}
\end{equation}
\end{subequations}
The growth functions $\DK$, \ldots, $\DS$ are defined by
\begin{subequations}
\begin{align}
    \DK & \equiv \fA \DA - f D^2 , \\
    \DL & \equiv \fB \DB , \\
    \DM & \equiv \fD \DD - \fJ \DJ , \\
    \DN & \equiv \fE \DE , \\
    \DP & \equiv \fF \DF + \fJ \DJ - f D \DA , \\
    \DQ & \equiv \fG \DG - f D \DB , \\
    \label{eq:DR-def}
    \DR & \equiv f D^3 + (f - \fA) D \DA - 2 \fJ \DJ , \\
    \label{eq:DS-def}
    \DS & \equiv \fJ \DJ + (f - \fB) D \DB .
\end{align}
\end{subequations}
When these functions are
replaced by their Einstein--de Sitter counterparts
using Table~\ref{table:EdS-growth},
the kernels in Eqs.~\eqref{eq:v-2} and~\eqref{eq:v-3}
become the standard expressions
$D^2 G_2(\vect{q}, \vect{s})$
and
$D^3 G_3(\vect{q}, \vect{s}, \vect{t})$~\cite{1981MNRAS.197..931J,1983MNRAS.203..345V,
Goroff:1986ep,Wise:1988kua,Makino:1991rp,Scoccimarro:1995if,Scoccimarro:1996se}.

The one-loop two-point function $\langle v_i v_j \rangle$
can be computed in analogy with~\S\ref{sec:eulerian-perturbation-theory},
yielding tree, 13 and 22 contributions whose properties match those discussed
above. As for $\langle \delta \delta \rangle$, the ultraviolet $q \gg k$
region of the one-loop integrals must be replaced with a parametrization.
This is equivalent to replacing $\vect{v}$ with a renormalized velocity,%
    \footnote{Notice that each composite operator may have its own, independent
    counterterms. Formally we couple each composite operator to the Lagrangian with
    an independent source, and obtain Green's functions for the composite operator
    by functional differentiation with respect to it.
    Finally the source is set to zero~\cite{Itzykson:1980rh}.
    Although there is only one operator
    of the form
    $\partial^2 \delta$, $\partial^4 \delta$, $\partial^6 \delta$, etc.,
    in the Lagrangian, its coefficient becomes a polynomial in the sources, and this
    allows the different counterterms to be separated.}
\begin{equation}
    \renormalized{\vect{v}}(\vect{x}) =
    (
        1 + \ctrterm{0}{\vect{v}}
    )
    \vect{v}_\Lambda(\vect{x})
    +
    \ctrterm{2}{\vect{v}}
    \frac{H}{\kNL^2}
    \nabla \delta_\Lambda(\vect{x})
    .
    \label{eq:velocity-renormalized}
\end{equation}
As in Eq.~\eqref{eq:delta-renormalized} we should treat
$\ctrterm{0}{\vect{v}}$
and
$\ctrterm{2}{\vect{v}}$
as one-loop terms,
and therefore
it does not matter whether we
take $\vect{v}$ to mix with
$\nabla \delta_\Lambda$
or
$\partial^2 \vect{v}$
because
these are related by the tree-level continuity equation~\eqref{eq:alpha-def}.
The coefficients $\ctrterm{0}{\vect{v}}$ and $\ctrterm{2}{\vect{v}}$
must each contain a component matching the loop-level redshift dependence, and
a free function representing the unknown redshift dependence of the ultraviolet modes,
\begin{subequations}
\begin{align}
    \ctrterm{0}{\vect{v}}
    & = \frac{\DR + 2 \DS}{fD} \ctrconst{0}{\vect{v}} + \ctrfree{0}{\vect{v}}(z)
    \label{eq:c0v}
    \\
    \nonumber
    \ctrterm{2}{\vect{v}}
    & = \bigg(
        (5f - 12 \fA) \DA
        + (10f - 12 \fB) \DB
        + 12 f D^2
        \\
    & \qquad \mbox{}
        - \frac{18 f_D \DD + 28 f_E \DE - 7 \fF \DF - 2 \fG \DG - 13 \fJ \DJ}{D}
    \bigg) \ctrconst{2}{\vect{v}} + \ctrfree{2}{\vect{v}}(z) .
    \label{eq:c2v}
\end{align}
\end{subequations}
(Part of the perturbative time dependence in $\ctrterm{2}{\vect{v}}$ is fixed
by the $t$ derivative of the time dependence from $\ctrterm{2}{\delta}$,
but it cannot be expressed as
$\d \ctrterm{2}{\delta} / \d t$ because the coefficient
$\ctrconst{2}{\vect{v}}$ may be different.)

These counterterms are independent of $\ctrterm{2}{\delta}$.
Therfore, as emphasized by Mercolli \& Pajer~\cite{Mercolli:2013bsa},
the velocity requires
extra counterterms beyond those required to renormalize correlation functions
of the density. 

\para{Multiplicative renormalization of velocity}
Eq.~\eqref{eq:velocity-renormalized} differs from the renormalized
density constrast $\renormalized{\delta}$ because $\renormalized{\vect{v}}$
mixes not only with the higher-derivative
operator $\partial^2 \delta$ but also adjusts the normalization of the bare field
$\vect{v}$ through $\ctrterm{0}{\vect{v}}$.
This adjustment is the analogue of field-strength renormalization in quantum field
theory, but its appearance here is unexpected because it is known to be absent
in Einstein--de Sitter~\cite{Goroff:1986ep,Mercolli:2013bsa}.%
    \footnote{The absence of a $k^0$ term
    in Einstein--de Sitter has been known
    empirically for a long time.
    Mercolli \& Pajer showed that this could be justified, without making
    explicit use of the Einstein--de Sitter background,
    for a certain microscopic realization of the short-distance velocity field.
    Although we have not attempted to match our calculation to their
    microscopic model we believe that our results are not in conflict, since
    we make different assumptions.}
Therefore one might suspect that the combination $\DR + 2\DS$ that controls
the perturbative time-dependence of $\ctrterm{0}{\vect{v}}$ could be zero.
Although this is not true in general, it \emph{is} always a decaying mode.
One can show from Eqs.~\eqref{eq:DR-def}--\eqref{eq:DS-def}
and~\eqref{eq:DA-def}--\eqref{eq:DB-def} that
\begin{equation}
    \DR + 2\DS = f^\ast (D^\ast)^3 \left( \frac{1+z}{1+z^\ast} \right)^{1/2} ,
\end{equation}
where as above a superscript `$\ast$' indicates evaluation at the time when initial
conditions are set for the non-linear evolution.
In $\renormalized{\vect{v}}$ this part of the counterterm therefore decays like
$(D^\ast/D)(1+z)^{1/2}/(1+z^\ast)^{1/2}$, and
is identically zero for Einstein--de Sitter
in which $z^\ast \rightarrow \infty$.
Hence, it is projected out by our choice of initial conditions
for the $D_i$.

In practice, all multiplicative counterterms of this type
cancel out of the redshift-space density
contrast. Therefore even if we do not adopt Einstein--de Sitter
values for the growth factors
at the initial time,
it is not necessary to introduce an explicit renormalization
condition for $\ctrterm{0}{\vect{v}}$.

\para{Renormalized equations of motion}
Similar renormalized counterparts can be defined for each operator
appearing in the equations of motion~\eqref{eq:SPT-continuity}--\eqref{eq:SPT-Poisson}.
Beyond linear order this includes the composite operators
$\vect{v} \delta$ and
$(\vect{v} \cdot \grad) \vect{v}$.
In general, composite operators require extra
counterterms to produce finite correlation functions, even when their
constituents such as $\delta$ and
$\vect{v}$ have been renormalized~\cite{Collins:1984xc,Itzykson:1980rh}.
Once renormalized versions have been defined, they may be inserted into
Eqs.~\eqref{eq:SPT-continuity} and~\eqref{eq:SPT-Euler}
to obtain evolution equations.
The form of these equations was studied by
Mercolli \& Pajer, and depends on what relations we take to exist
among the counterterms~\cite{Mercolli:2013bsa}.

First consider the continuity equation.
For the renormalized operators this reads
\begin{equation}
    \frac{\d \renormalized{\delta}}{\d t}
    +
    \grad \cdot \renormalized{\vect{v}}
    +
    \grad \cdot \renormalizedc{\vect{v} \delta}
    =
    \bigg(
        \frac{\d \ctrterm{2}{\delta}}{\d t}
        +
        H
        \big[
            \ctrterm{2}{\vect{v}} + \ctrterm{2}{\vect{v} \delta}
        \big]
    \bigg)
    \frac{1}{\kNL^2}
    \partial^2 \delta ,
    \label{eq:renormalized-continuity}
\end{equation}
where we have defined
\begin{equation}
    \renormalizedc{\vect{v} \delta} =
    \vect{v}\delta
    + \ctrterm{2}{\vect{v}\delta} \frac{H}{\kNL^2}
    \partial^2 \delta .
    \label{eq:vdelta-renormalized}
\end{equation}
There is a possible multiplicative renormalization for $\vect{v} \delta$,
but as for $\vect{v}$ it is a decaying mode.
Therefore we have omitted it in~\eqref{eq:vdelta-renormalized}.
Whether the ordinary continuity equation applies to the renormalized operators
depends on whether we take the right-hand side of Eq.~\eqref{eq:renormalized-continuity}
to vanish.

In general there is no obligation to do so,
because we are free to choose the counterterms
$\ctrterm{2}{\delta}$,
$\ctrterm{2}{\vect{v}}$
and $\ctrterm{2}{\vect{v}\delta}$ independently.
A range of possible choices were surveyed by Mercolli \& Pajer~\cite{Mercolli:2013bsa}.
For example, we could use observational data or simulations to measure
a velocity correlation function such as
$\langle \delta(\vect{k}_1) \vect{v}(\vect{k}_2) \rangle$
or
$\langle \vect{v}(\vect{k}_1) \vect{v}(\vect{k}_2) \rangle$, and adjust
$\ctrterm{2}{\vect{v}}$ to fit the data over some range of $k$.
This is the analogue of an on-shell renormalization scheme.
Alternatively we could impose an arbitrary condition, such as fixing
$\langle \vect{v} \vect{v} \rangle$ to a specific value at some wavenumber
$\kren$. This would be an analogue of an off-shell scheme such as
minimal subtraction. (In an off-shell scheme we require an extra, finite renormalization
to express the observable $\vect{v}$ in terms of the renormalized
operator $\renormalized{\vect{v}}$.
We discuss these issues more carefully in~\S\ref{sec:real-space-results}.)
Depending on our choices, the right-hand side of~\eqref{eq:renormalized-continuity}
may not be zero.

Second, consider the Euler equation~\eqref{eq:SPT-Euler}.
This will become
\begin{equation}
    \frac{\d \renormalized{\vect{v}}}{\d t}
    +
    \renormalized{\big[(\vect{v}\cdot\nabla)\vect{v}\big]}
    +
    2H \renormalized{\vect{v}}
    -
    \frac{1}{a^2} \nabla \renormalized{\Phi}
    =
    \frac{\cs^2(z)}{\kNL^2} \partial^2 \delta ,
    \label{eq:renormalized-Euler}
\end{equation}
where $\cs^2$ is a redshift-dependent function built from the counterterms
for each of the operators used in Eqs.~\eqref{eq:SPT-continuity}
and~\eqref{eq:SPT-Euler}.
By analogy with the Navier--Stokes equations we can interpret
the net counterterm
as a viscosity. Its coefficient $\cs^2$ has dimensions of velocity-squared
which justifies the notation,
here chosen to match that used in
Refs.~\cite{Baumann:2010tm,Carrasco:2012cv,Carrasco:2013mua}.

Finally, the Poisson constraint~\eqref{eq:SPT-Poisson} is a linear relation
between $\nabla^2 \Phi$ and $\delta$ and is therefore preserved under
renormalization. We conclude that renormalization of $\Phi$ does not require
introduction of any new counterterms.

In Refs.~\cite{Baumann:2010tm,Carrasco:2012cv,Carrasco:2013mua,Mercolli:2013bsa},
analogues of Eqs.~\eqref{eq:renormalized-continuity}
and~\eqref{eq:renormalized-Euler}
were obtained
starting from the bare SPT equations~\eqref{eq:SPT-continuity}
and~\eqref{eq:SPT-Euler}
and smoothing them at some arbitrary scale.
The smoothed equations parametrize the influence of short-scale modes
on those of longer wavelength, and therefore must give the
same result as parametrizing the large-$\vect{q}$ part of the loop integrals.
We should therefore regard equations for renormalized operators,
such as Eqs.~\eqref{eq:renormalized-continuity}
and~\eqref{eq:renormalized-Euler},
as equivalent to the smoothed equations used in
Refs.~\cite{Baumann:2010tm,Carrasco:2012cv,Carrasco:2013mua,Mercolli:2013bsa}.

In Refs.~\cite{Senatore:2014vja,Lewandowski:2015ziq,Perko:2016puo},
a smoothing argument was used to obtain the counterterm for $\delta$
but composite operators were used to renormalize $\deltarsd$.
Consequently,
it was not immediately clear how these procedures were related.
When we discuss the redshift-space density contrast in~\S\ref{sec:one-loop-rsd}
we will employ the methods described in this section,
which makes clear that exactly the same procedure is being applied to
$\delta$ and $\deltarsd$.

\subsection{Resummation schemes}
\label{sec:resummation-methods}
Renormalized operators such as
$\renormalized{\delta}$ and $\renormalized{\vect{v}}$
correctly parametrize the effect of unknown short-scale modes, but this
does not mean that fixed-order perturbation theory in
these operators
(meaning that we calculate to a fixed
order in the loop expansion)
will provide an adequate description.
Eqs.~\eqref{eq:PAA}--\eqref{eq:PJ2}
show that
the typical magnitude of a loop-level term
is set by a weighted integral over the
initial power spectrum $\Pinit$.
For example, Eqs.~\eqref{eq:P13-IR}--\eqref{eq:P22-IR}
show that after making a Taylor expansion in $q$,
each integral can be regarded as a sum of weighted integrals
of the form
\begin{equation}
	\int_0^\Lambda \frac{\d q}{(2\pi)^3} q^{2n} \Pinit(q)	
\end{equation}
for integer $n \geq 0$.
In the full power spectrum these terms are
enhanced by powers of the growth functions $D$ or $D_i$.

Contributions with strong ultraviolet weighting $n \gg 0$
will be dominated by the region near the cutoff
and can be absorbed by counterterms.
But contributions with small $n$ may generate significant
contributions from all wavenumbers.
Porto, Senatore \& Zaldarriaga~\cite{Porto:2013qua}
and Senatore \& Zaldarriaga~\cite{Senatore:2014via}
introduced parameters $\epsilon_{s<}$,
$\epsilon_{s>}$
and
$\epsilon_{\delta<}$
to describe the size of these integrals
over different ranges of $q$,
\begin{subequations}
\begin{align}
	\label{eq:epsilon-sless}
	\epsilon_{s<}(z) & = k^2 D(z)^2 \int_0^k \frac{\d q}{2\pi^2} \Pinit(q) \\
	\label{eq:epsilon-sgtr}
	\epsilon_{s>}(z) & = k^2 D(z)^2 \int_k^\Lambda \frac{\d q}{2\pi^2} \Pinit(q) \\
	\label{eq:epsilon-deltaless}
	\epsilon_{\delta<}(z) & = k^2 D(z)^2 \int_0^\Lambda \frac{\d q}{2\pi^2} q^2 \Pinit(q) .
\end{align}
\end{subequations}
It was shown in Refs.~\cite{Porto:2013qua,Senatore:2014via}
that these parameters could become
order unity.
Therefore,
if they provide an accurate estimate of the size of
high-order terms,
fixed-order perturbation theory will cease to be
a good approximation.
Similar difficulties are frequently encountered in field theory.
In some cases it is possible to obtain a more statisfactory answer by
retaining an infinite subset of terms extending to all orders in the loop
expansion. The different strategies for doing so are called
resummation schemes.

In practice we will see that although $\epsilon_{s<}$,
$\epsilon_{s>}$ and $\epsilon_{\delta<}$ may become individually of
order unity, the loop expansion is better behaved because of cancellations.
For the real-space density power spectrum to be considered in this section,
the effect of resummation is modest---roughly a $2\%$ effect.
However, for the redshift-space density power spectrum studied in~\S\ref{sec:one-loop-rsd}
its effects are more significant.

\subsubsection{Vlah--Seljak--Chu--Feng resummation}
In any practical resummation scheme we require a template
that governs the form of some subset of loop corrections to arbitrary order.
If the template is sufficiently rigid then it will determine the sum
of all terms in the subset from matching to just the lowest few terms of
fixed-order perturbation theory.
For standard renormalization group flow the template is provided by the
criterion of renormalizability. In other cases, such as factorization
in QCD, rigorous theorems control the structure of the high-order terms.
For large-scale structure there are not yet any rigorous theorems of this kind
but we can still obtain suitable templates from models.
The situation is comparable to the use of approximate models to derive
properties of correlation functions in QCD~\cite{Polkinghorne:1980mk}.

\para{Lagrangian perturbation theory as a model}
The key observation,
suggested by Matsubara,
is that Lagrangian perturbation theory provides
a model from which templates can be derived~\cite{Matsubara:2007wj}.
In the Lagrangian approach one tracks the displacement
$\vect{\Psi}$ of a particle
from some initial comoving location $\vect{q}$ to a final location
$\vect{r}$,
\begin{equation}
	\vect{r}(\vect{q},t) = \vect{q} + \vect{\Psi}(\vect{q},t) .	
\end{equation}
This notation is conventional; note that in this section $\vect{q}$ is
position-space quantity, and should not be confused with the loop
momentum used in Eqs.~\eqref{eq:PAA}--\eqref{eq:PJ2}.
The density power spectrum
is given in terms of the
displacement correlation functions
by~\cite{1988grra.conf..385B,1993cvf..conf..585T,1996MNRAS.282..767T}
\begin{equation}
	P(k) = \int \d^3 q \; \e{-\im \vect{q}\cdot\vect{k}}
	\Big(
		\langle \e{-\im \vect{k} \cdot \Delta \vect{\Psi}} \rangle
		- 1
	\Big)
	,
\end{equation}
where $\Delta\vect{\Psi} \equiv \vect{\Psi}(\vect{q},t) - \vect{\Psi}(\vect{0},t)$.
The `$-1$' produces a $\delta$-function that can be dropped at finite wavenumber,
while
the cumulant expansion theorem can be used to rewrite the expectation of the
exponential,
\begin{equation}
	P(k) = \int \d^3 q \; \e{-\im \vect{q}\cdot\vect{k}}
	\exp\bigg(
		\sum_{n=1}^\infty \frac{(-\im)^n}{n!}
		\langle (\vect{k} \cdot \Delta\vect{\Psi})^n \rangle_c
	\bigg)
	,
	\label{eq:lagrangian-power-spectrum}
\end{equation}
where we have used $\langle \cdots \rangle_c$ to denote a connected correlation function.
The Eulerian power spectrum of~\S\ref{sec:eulerian-perturbation-theory}
can be recovered from Eq.~\eqref{eq:lagrangian-power-spectrum}
by expanding the exponential and collecting terms at the same
loop-level~\cite{Matsubara:2007wj,Sugiyama:2013gza}.
But we can equally regard~\eqref{eq:lagrangian-power-spectrum}
as a template that controls a subset of terms at all orders in
the Eulerian loop
expansion in terms of the low-order correlation functions
of $\Delta\vect{\Psi}$.

To match the Eulerian power spectrum at one loop requires the
two- and three-point correlation functions of
$\Delta\vect{\Psi}$.
That gives
\begin{equation}
	P(k) = \int \d^3 q \; \e{-\im \vect{k}\cdot\vect{q}}
	\exp\bigg(
		{- \frac{1}{2}} k_i k_j A_{ij}
		+ \frac{\im}{6} k_i k_j k_\ell W_{ij\ell}
		+ \cdots
	\bigg) ,
	\label{eq:lagiangian-power-spectrum-2pf}
\end{equation}
where $A_{ij}$ and $W_{ij\ell}$ are defined by
\begin{subequations}
\begin{equation}
	A_{ij} \equiv
	\langle
		\big[ \Delta\vect{\Psi}(\vect{q}) - \Delta\vect{\Psi}(\vect{0}) \big]_{ij}^2
	\rangle
	=
	X(q) \delta_{ij}
	+ Y(q) \hat{q}_i \hat{q}_j .
	\label{eq:A-def}
\end{equation}
and
\begin{equation}
	W_{ij\ell} \equiv
	\langle
		\big[ \Delta\vect{\Psi}(\vect{q}) - \Delta\vect{\Psi}(\vect{0}) \big]^3_{ij\ell}
	\rangle .
\end{equation}
\end{subequations}
The lowest-order parts
of $X$ and $Y$ are related to the Eulerian power spectrum by
\begin{subequations}
\begin{align}
	\label{eq:X-def}
	X(q) & = D(z)^2 \int_0^\infty \frac{\d k}{\pi^2}
	\Pinit(k)
	\bigg(
		\frac{1}{3} - \frac{j_1(kq)}{kq}
	\bigg) \\
	\label{eq:Y-def}
	Y(q) & = D(z)^2 \int_0^\infty \frac{\d k}{\pi^2}
	\Pinit(k)
	j_2(kq) ,
\end{align}
\end{subequations}
where $j_n(x)$ is the spherical Bessel function of order $n$.
Therefore we can regard $X$ and $Y$ as expansion parameters
similar to $\epsilon_{s<}$
$\epsilon_{s>}$, and $\epsilon_{\delta<}$,
but with suppression in the region $k \lesssim q$
where the factor multiplying $\Pinit$ in each integrand
is of order $(kq)^2$.
For $k \gg q$
the Bessel functions decay,
removing any large contributions near the cutoff that may be
present in $\epsilon_{\delta<}$.
These effects reduce the typical magnitude of high-order loop terms compared
to a na\"{\i}ve estimate using~\eqref{eq:epsilon-sless}--\eqref{eq:epsilon-deltaless}.

\para{`Wiggle' and `no-wiggle' power spectra}
Eqs.~\eqref{eq:lagiangian-power-spectrum-2pf},
\eqref{eq:A-def}
and~\eqref{eq:X-def}--\eqref{eq:Y-def}
have been used as the basis of a resummation scheme by a number of
authors~\cite{Matsubara:2007wj,Okamura:2011nu,Senatore:2014via,
McQuinn:2015tva,Baldauf:2015xfa,Vlah:2015sea,Vlah:2015zda}.
The scheme originally proposed by Matsubara
deduced a template from~\eqref{eq:lagiangian-power-spectrum-2pf}
by taking the $\vect{q}$-independent part of $X$
outside the $\vect{q}$-integral.
This suggests that $P(k)$ should contain a multiplicative
damping factor
$\exp[-(\epsilon_{s<} + \epsilon_{s>})/3]$.
As explained above, this is
a `template' in the sense that the exponential
contains terms at all orders in the
Eulerian loop expansion but is determined entirely
by the Eulerian two-point function.
Unfortunately this scheme is quantitatively acceptable only for low $k$,
and causes unphysical overdamping for $k$ in the quasilinear regime
of interest~\cite{Matsubara:2007wj,Taruya:2010mx,McQuinn:2015tva}.

Vlah, Seljak, Chu \& Feng proposed an alternative
scheme that evades these
difficulties~\cite{Vlah:2015zda}, based on a division
of the power spectrum into `wiggle' and `no-wiggle' components.
(See also Ref.~\cite{Blas:2016sfa}.)
These separate the effect of baryonic oscillations
from the smooth power spectrum predicted from dark matter alone.
We define a `no-wiggle' form of the initial power spectrum
by filtering~\cite{Vlah:2015zda,Baldauf:2015xfa},
\begin{equation}
	\Pinitnw(k) =
	\frac{\Pref(k)}{(2\pi \lambda^2)^{1/2}}
	\int \d \ln q \;
	\frac{\Pinit(q)}{\Pref(q)}
	\exp\bigg(
		{- \frac{(\ln k/q)^2}{2\lambda^2}}
	\bigg) ,
	\label{eq:Pinit-wiggle}
\end{equation}
where $\Pref(k)$ is any suitable smooth reference
power spectrum whose broadband power roughly matches
$\Pinit$. This fixes the normalization of $\Pinitnw$.
In our numerical work we use the Eisentein \& Hu fitting function for the power
spectrum with no baryons~\cite{Eisenstein:1997jh}.
The dimensionless scale $\lambda$ sets the size of the filter window. We use
$\lambda = 0.25 (k/\kpiv)^{0.04}$, where $\kpiv = 0.05 h / \Mpc$ is a fixed
reference scale.
This choice is intended to match the overall amplitude
and scale-dependence suggested in Ref.~\cite{Vlah:2015zda}.

\begin{figure}
    \begin{center}
        \includegraphics{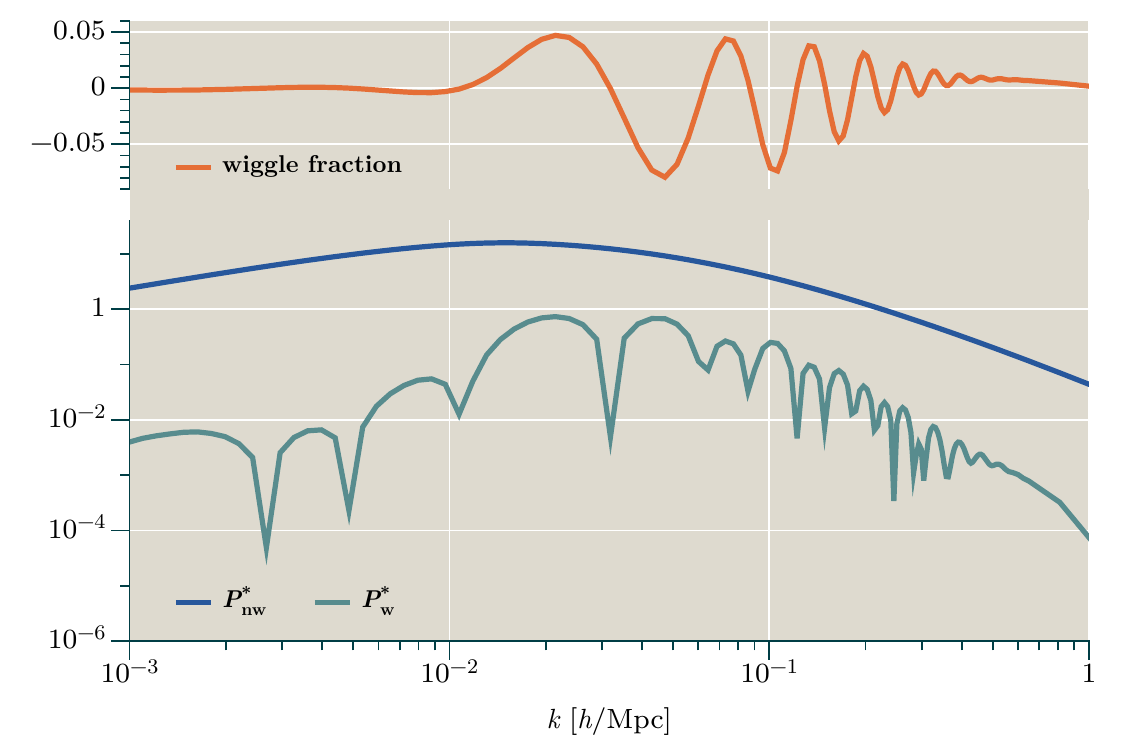}    
    \end{center}
    \caption{\label{fig:wiggle-filter}\semibold{Top panel}: wiggle fraction
    $\Pinitw/\Pinit$.
    \semibold{Bottom panel}:
    Representative
    filtered `wiggle' and `no-wiggle' power
    spectra. 
    The initial power spectrum $\Pinit$ is a Planck2015 cosmology at
    redshift $z^\ast = 50$.}
\end{figure}

Given $\Pinitnw$, the `wiggle' component $\Pinitw$ is defined by
\begin{equation}
	\Pinitw \equiv \Pinit - \Pinitnw .
	\label{eq:Pinit-nowiggle}
\end{equation}
We plot the filtered `wiggle' and `no-wiggle'
components in Fig.~\ref{fig:wiggle-filter}.

\para{Damping the `wiggle' component}
Once $\Pinitnw$ has been computed, it can be used to define `no-wiggle'
versions of $P_{11}$, $P_{22}$ and $P_{13}$,
and corresponding `wiggle' components by analogy with~\eqref{eq:Pinit-nowiggle}.
The same can be done for $A_{ij}$ and $W_{ij\ell}$,
producing `wiggle' and `no-wiggle'
components
$A_{ij}^{\wiggle}$, $A_{ij}^{\nowiggle}$,
$W_{ij\ell}^{\wiggle}$, $W_{ij\ell}^{\nowiggle}$.

To extract a template we expand perturbatively, except that we
keep the interaction between
$A_{ij}^{\nowiggle}$ and the `wiggle'
terms to all orders in the Eulerian loop expansion
for $A_{ij}^{\nowiggle}$.
That yields
\begin{equation}
    \PloopVSCF(k) = \Ploopnw
	+
	\int \d^3 q \; \e{-\im \vect{k} \cdot \vect{q}}
	\exp
	\bigg(
		{-\frac{1}{2} k_i k_j A_{ij}^{\nowiggle,\attreelevel}}
	\bigg)
	\bigg(
		{-\frac{1}{2} k_m k_n A_{mn}^{\wiggle,\uptooneloop}}
		+
		\frac{\im}{6} k_m k_n k_r W_{mnr}^{\wiggle,\uptooneloop}
	\bigg) ,
	\label{eq:P-resum-intermediate}
\end{equation}
where the label `${\leq}\looplevel{n}$'
means that the quantity to which it is attached
includes terms up to and including level $n$ in the Eulerian loop
expansion.
If terms at exactly level $n$ are required we write instead
`${=}\looplevel{n}$'.
Eq.~\eqref{eq:P-resum-intermediate}
will act as a template if we can rewrite the integral
as a combination of the exponential
and the `wiggle'
power spectra $\Ptreew$ and $\Patloopw$.

In general there is no simple way to perform this rewriting.
But since the `wiggle' components have support only over scales near the
baryon bump, and
$A_{ij}^{\text{nw}}$ is relatively slowly varying on these scales,
we can \emph{approximately} factorize~\eqref{eq:P-resum-intermediate}
to obtain~\cite{Vlah:2015sea,Vlah:2015zda}
\begin{equation}
	\PloopVSCF(k) \equiv \Ploopnw
	+ \exp\bigg(
		{- \frac{1}{2}} \llangle k_i k_j A_{ij}^{\nowiggle,\attreelevel} \rrangle
	\bigg)
	\bigg(
	   \Ploopw
	   +
	   \frac{1}{2} \llangle k_i k_j A_{ij}^{\nowiggle,\attreelevel} \rrangle \Ptreew
	\bigg)
	,
	\label{eq:P-resum}
\end{equation}
where $\llangle k_i k_j A_{ij}^{\nowiggle,\attreelevel} \rrangle$
is an average of $k_i k_j A_{ij}^{\nowiggle,\attreelevel}$ over the range of $\vect{q}$
where the `wiggle' components have support.
(We write `$\equiv$' rather than `$=$' to emphasize that this should be regarded as
a definition rather than an equality.)
The second term in the final bracket has appeared because
Eq.~\eqref{eq:Pinit-nowiggle}
makes
$\Ploopw$ contain cross-products between `wiggle' and `no-wiggle' components,
of which the relevant combination at one-loop is
the Zel'dovich-like term
$A_{ij}^{\wiggle,\attreelevel} A_{mn}^{\nowiggle,\attreelevel}$~\cite{Sugiyama:2013gza}.
This component does not appear in~\eqref{eq:P-resum-intermediate}
and should be subtracted.
Its effect makes the expansion of~\eqref{eq:P-resum} up to one-loop
agree with the one-loop Eulerian result.

Eq.~\eqref{eq:P-resum} is our template for the resummed power spectrum,
with the one-loop terms
$\Ploopnw$ and $\Ploopw$ understood to include counterterms when applied to the
effective field theory
of~\S\S\ref{sec:ultraviolet-sensitivity}--\ref{sec:renormalized-operators}.
The precise definition of
$\llangle k_i k_j A_{ij}^{\nowiggle,\attreelevel} \rrangle$
should be regarded as part of the approximate
integration procedure~\eqref{eq:P-resum},
but if $A_{ij}^{\nowiggle}$ is nearly constant
over the relevant $\vect{q}$ then any sensible choice will yield nearly
the same result. We choose
\begin{equation}
    \llangle k_i k_j A_{ij}^{\nowiggle,\attreelevel} \rrangle
    \equiv
    \frac{k_i k_j}{V(\qmin, \qmax)}
    \int_{q=\qmin}^{q=\qmax}
    \d^3 q \;
    A_{ij}^{\nowiggle}(\vect{q}) ,
    \label{eq:A-average-defn}  
\end{equation}
where $V(a, b)$
is the volume of the three-dimensional spherical shell
between radii $r=a$ and $r=b$.
We have verified that our results do not strongly depend on the way this
integral is weighted.
When applied to Eq.~\eqref{eq:A-def}
and Eqs.~\eqref{eq:X-def}--\eqref{eq:Y-def}
this yields
\begin{equation}
    \llangle k_i k_j A_{ij}^{\nowiggle,\attreelevel} \rrangle
    = k^2 \llangle A^{\nowiggle,\attreelevel} \rrangle ,   
\end{equation}
where we have defined
\begin{equation}
    \llangle A^{\nowiggle,\attreelevel} \rrangle
    \equiv
    \frac{D(z)^2}{\pi^2}
    \frac{1}{\qmax^3 - \qmin^3}
    \int_{\qmin}^{\qmax} \d q \, q^2
    \int_0^\infty \d k \,
    \Pinitnw(k)
    \big[
        1 - j_0(kq)
    \big] .
    \label{eq:expectation-A}
\end{equation}
The amplitude of $\llangle A^{\nowiggle,\attreelevel} \rrangle$
is inherited from $X$ and $Y$, which measure
the typical amplitude of the displacement $\vect{\Psi}$ on the scale $q$.
Therefore the degree of damping at momentum $k$ is determined by the ratio
$k/\kdamp$, where
$\kdamp \sim \llangle A^{\nowiggle,\attreelevel} \rrangle^{-1/2}$
is a wavenumber measuring the typical displacement averaged
between the scales $\qmin$ and $\qmax$.
For concrete calculations we choose $\qmin = 10 h^{-1} \, \Mpc$
and $\qmax = 300 h^{-1} \, \Mpc$, which roughly bracket the range
over which the `wiggle' component has support in Fig.~\ref{fig:wiggle-filter}.
The $k$-integral is carried up to the same ultraviolet cutoff we use when
computing the SPT loops.

The exponential provides efficient damping for $k \gtrsim \kdamp$.
For a Planck2015-like cosmology we find $\kdamp \approx 0.18 h/\Mpc$,
and by referring to Fig.~\ref{fig:wiggle-filter}
it can be seen that this is comparable to the scales on which baryon
acoustic oscillations
are visible.
Therefore we expect the outcome of this resummation prescription to
be modest suppression of these oscillations, while leaving the broadband
power unchanged.
The underlying physical reason is that
random motions associated with these
displacements
wash out coherence of the baryon acoustic
oscillation~\cite{Eisenstein:2006nk,Eisenstein:2006nj,Crocce:2005xz,Crocce:2007dt}.

\para{Relation to Senatore--Zaldarriaga resummation}
An alternative resummation prescription was proposed by Senatore \&
Zaldarriaga~\cite{Senatore:2014via}, which is superficially quite different
to the one described here.
The relation between these prescriptions was discussed briefly
by Vlah et al.~\cite{Vlah:2015sea}.
In Appendix~\ref{appendix:senatore-zaldarriaga} we give a slightly different
discussion that emphasizes its relation to the `wiggle' and `no-wiggle'
filtering procedure described above.

\subsection{Comparison of results}
\label{sec:real-space-results}
It was explained in~\S\ref{sec:resummation-methods} that the
resummed expression Eq.~\eqref{eq:P-resum} is a model, not a theorem
about the behaviour of high-order diagrams in SPT.
Its utility should be judged on its ability to reproduce observed features
of the measured or simulated $\delta$ correlation function.
In this section we compare Eq.~\eqref{eq:P-resum} with the
unresummed effective field theory prediction~\eqref{eq:P11}--\eqref{eq:P13}
and~\eqref{eq:P13-UV-parametrization}
and with traditional SPT.

\para{Fitting counterterms}
Like any
effective field theory,
ours is not predictive until we fix the counterterms. For the $\delta$ power
spectrum this means that we must assign a value to $\ctrterm{2}{\delta}/\kNL^2$.

As explained in~\S\ref{sec:renormalized-operators},
we can define a renormalized operator $\renormalized{\delta}$
by imposing whatever condition we wish,
such as fixing $\langle \renormalized{\delta}
\renormalized{\delta} \rangle$ to a prescribed value at some wavenumber
$\kren$.
The physical overdensity would then be related to $\renormalized{\delta}$
by a further finite renormalization, in the same way that the {\MSbar}
running mass is related to the physical pole mass by a finite shift.
For the level of complexity at which we are working
there is nothing to be gained from this freedom,
and we may as well choose $\renormalized{\delta}$ to match the
observed power spectrum as closely as possible. This was the approach adopted
by Carrasco et al.~\cite{Carrasco:2012cv,Carrasco:2013mua}.
Therefore we will determine the counterterm by adjusting
$\langle \renormalized{\delta} \renormalized{\delta} \rangle$
to match a numerical, non-linear power spectrum
over a suitable range of $k$.

There are several ways this can be done.
Carrasco et al.~\cite{Carrasco:2012cv,Carrasco:2013mua}
used an ensemble of $N$-body
simulations to estimate the fully non-linear power spectrum.
We will adopt this approach in~\S\ref{sec:one-loop-rsd}
when we renormalize the redshift-space power spectrum, for which there
is no other way to accurately capture its non-linear effects.
For the real-space overdensity there are alternatives,
such as use of
semianalytic models
that are calibrated to match simulations~\cite{Smith:2002dz,Takahashi:2012em}.
In this section we illustrate the performance of
our models by adjusting the counterterms to match the {\CAMB} {\Halofit}
power spectrum at $z=0$ as closely as possible.

\para{Numerical value of $\ctrterm{2}{\delta}$}
Since we are working at a single redshift
there is no need to divide
the counterterm into $\ctrconst{2}{\delta}$ and $\ctrfree{2}{\delta}$ components, and
we report the single value
$\ctrterm{2}{\delta}(z=0)/\kNL^2$.

\begin{figure}
    \begin{center}
        \includegraphics{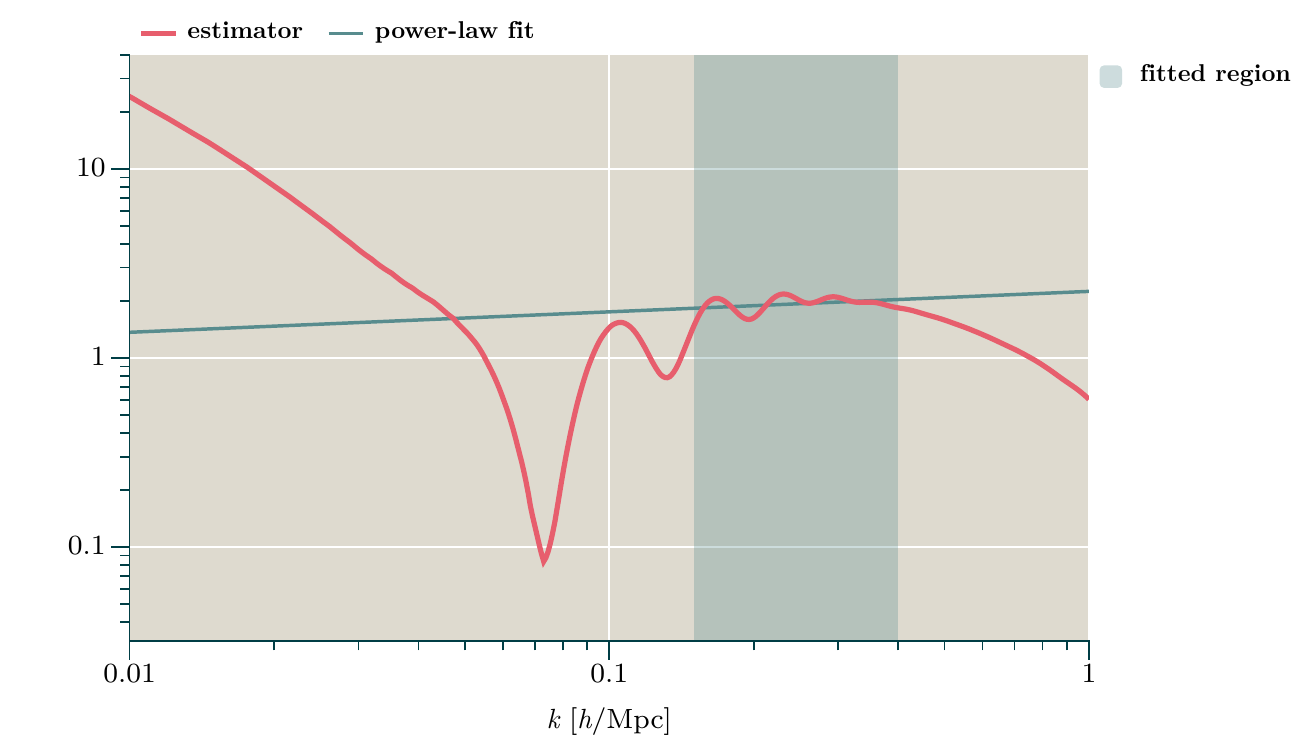}    
    \end{center}
    \caption{\label{fig:counterterm-fit}Fitting for the counterterm.
    The red line shows the estimator,
    $-(\PNL - \PSPT^{\uptooneloop}) / (2k^2 D^2 \Pinit)$,
    where $\PNL$ is the `measured'
    non-linear power spectrum we wish to match.
    It is approximately constant in a region where the difference between
    the 1-loop SPT prediction and the measured power spectrum is
    adequately described by the leading counterterm.
    To obtain an estimate we optimize the fit in the quasilinear region
    $\kminreal \leq k \leq \kmaxreal$,
    shaded light green,
    where we expect the EFT counterterm to improve the prediction.
    The green line shows a least-squares power-law
    approximation to the estimator in this region, which is
    $2.246 \times (k / h \times \Mpc)^{0.1082}$.
    As expected, it is nearly $k$-independent.}
\end{figure}

We estimate the counterterm by performing a least-squares fit
over the range $k=\kminreal$ to $k=\kmaxreal$ where the
SPT result begins to deviate from measurement
and we expect the EFT counterterm
to improve the prediction.
In Fig.~\ref{fig:counterterm-fit} we show
the discrepancy between the one-loop SPT power spectrum $\PSPT$
and the {\CAMB} {\Halofit} power spectrum $\PNL$ and fit it to
a term with the functional
form predicted by the EFT.
The red line shows the estimator
\begin{equation}
    \frac{\ctrterm{2}{\delta}}{\kNL^2}
    \approx
    - \frac{\PNL - \PSPT}{2 k^2 D^2 \Pinit} ,
\end{equation}
which should be approximately $k$-independent
in the fitted region if the
predicted functional form is correct.
The shaded light-green area shows the region included in the fit,
and it can be seen that this region exhibits roughly the expected behaviour.
(The oscillations within the shaded
region arise from misprediction of the amplitude and phase of the
baryon acoustic oscillations, which the EFT counterterm is not expected to improve.)
For guidance, the green line shows a power-law fit to the shaded region.

\begin{figure}
    \begin{center}
        \includegraphics{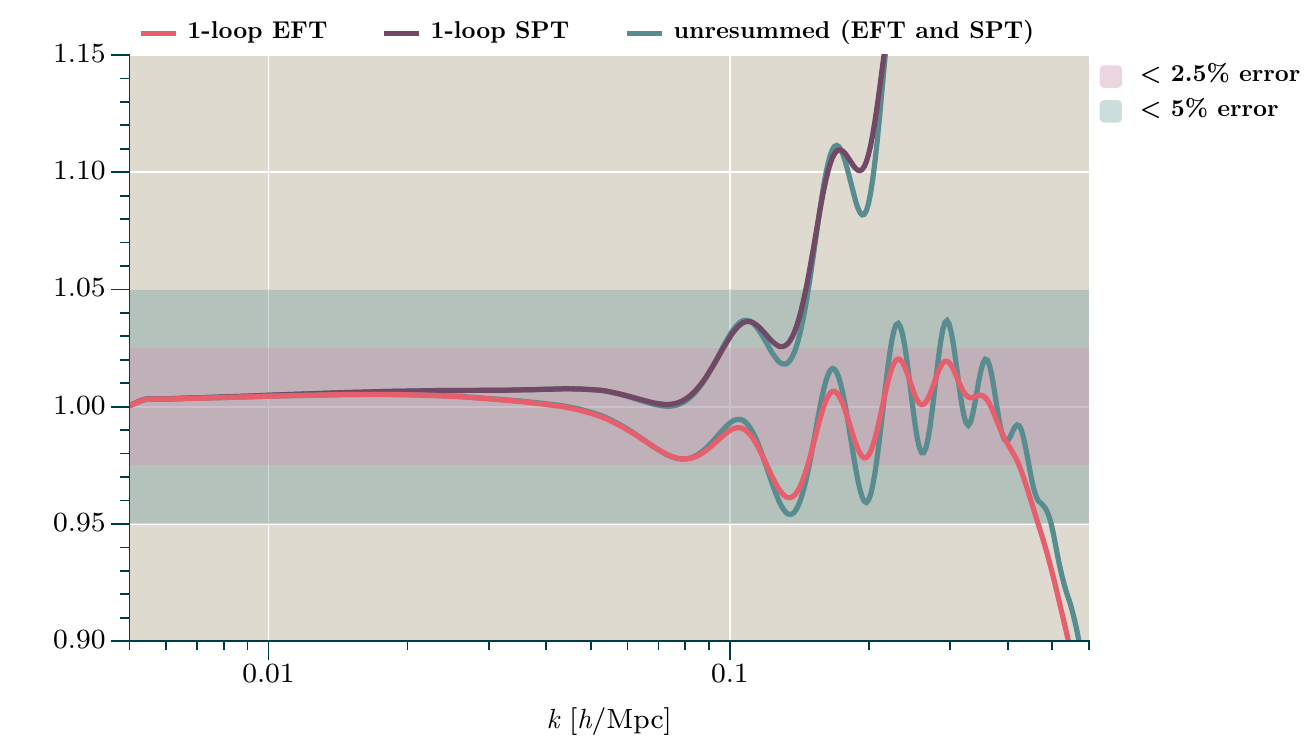}    
    \end{center}
    \caption{\label{fig:fit-comparison}Comparison of fit to {\CAMB}
    non-linear ({\Halofit}) matter power spectrum for 1-loop SPT
    and
    1-loop EFT in their resummed and unresummed variants.
    The quantity plotted is $P/\PNL$,
    and the cosmology matches the Planck2015
    TT+TE+EE+lowP+lensing+ext best-fit
    parameters~\cite{Ade:2015xua}.
    The light-pink region marks where the prediction is within
    $2.5\%$ of the {\CAMB} power spectrum, and the light-green
    region marks where it is within $5\%$.}
\end{figure}

Using a cutoff on the loop momenta
of $1.4 h/\Mpc$, we find
\begin{equation}
    \frac{\ctrterm{2}{\delta}}{\kNL^2} = 1.94 h^{-2} \, \Mpc^2
    \qquad
    \text{at $z=0$}
    .    
\end{equation}
This compares with the value $(1.62 \pm 0.03) h^{-2} \, \Mpc^2$ reported
by Carrasco et al.~\cite{Carrasco:2013mua}
(although for a different cosmology).
In Fig.~\ref{fig:fit-comparison} we compare the predictions of SPT
with the resummed and unresummed EFT.
Our results are consistent with previous analyses,
which all found that including the EFT counterterm led to an improved
fit~\cite{Carrasco:2012cv,Carrasco:2013mua,McQuinn:2015tva,Vlah:2015sea,Vlah:2015zda}.

The suppression of baryon oscillations due to resummation is visible, but the
improvement in fit is modest with residual oscillatory structure remaining
even after resummation.
This is most likely an artefact of the {\Halofit} implementation used by {\CAMB}.
When we measure power spectra directly from simulation in
{\S}\ref{sec:results} we will find that the baryon oscillations are smaller
and resummation successfully smooths the EFT prediction;
cf. the top panel of Fig.~\ref{fig:multipole-comparison}.

\section{One-loop renormalization of the matter power spectrum in redshift space}
\label{sec:one-loop-rsd}

Our aim is to use the machinery reviewed in~\S\ref{sec:one-loop-real}
to renormalize the two-point function of the redshift-space density contrast,
and hence its Legendre multipoles $P_\ell$.
These are potentially sensitive tests of modified gravity; see, eg.,
Refs.~\cite{Jennings:2012pt,Burrage:2015lla}.

\subsection{The redshift-space density contrast}
\label{sec:redshift-space}
Inclusion of redshift-space effects for the two-point function is now
well-understood~\cite{1987MNRAS.227....1K,1989MNRAS.236..851L,
1990MNRAS.242..428M,Cole:1993kh}.
If the Hubble flow accounted for the entire recession velocity
$\vrec$ of an
object at distance $\vect{r}$ then
it would follow that $\vrec = H \vect{r}$.
In practice each object is also embedded in the flow $\vect{v}$
described in~\S\ref{sec:matter-equations},
and therefore its recession velocity is modified
so that
$\vrec = H \vect{r} + (\vect{v} \cdot \hat{\vect{r}}) \hat{\vect{r}}$.
A galaxy survey that measures the redshift corresponding to $\vrec$
and uses it to infer a distance based on the Hubble flow will assign this galaxy
a displaced radial position,
\begin{equation}
    \vect{s} = \vect{r} + \frac{\vect{v} \cdot \hat{\vect{r}}}{H} \hat{\vect{r}} .
    \label{eq:redshift-space-s}
\end{equation}
These displacements systematically distort the measured overdensity field.

\para{Redshift-space overdensity}
The mapping between $\vect{r}$ and $\vect{s}$ conserves mass. Using this
property and Eq.~\eqref{eq:redshift-space-s},
Scoccimarro showed that~\cite{Scoccimarro:2004tg}
\begin{equation}
    \deltarsd(\vect{k})
    =
    \delta(\vect{k})
    +
    \int \d^3 r \;
    \e{-\im \vect{r}\cdot\vect{k}}
    \bigg[
        \exp
        \Big(
            {-\frac{\im}{H}}
            (\vect{k}\cdot\hat{\vect{r}})
            \big[
                \vect{v}(\vect{r})\cdot\hat{\vect{r}}
            \big]
        \Big)
    \bigg]
    \big[
        1 + \delta(\vect{\vect{r}})
    \big] .
    \label{eq:delta-s}
\end{equation}
As for any operator,
it is necessary to
exchange $\deltarsd$ for a renormalized
operator $\renormalizedc{\deltarsd}$
containing counterterms that describe
the unknown ultraviolet part of its loop integrals.
Because~\eqref{eq:delta-s}
is a composite operator in the
language of~\S\ref{sec:renormalized-operators}
these counterterms are not fixed by our definition of $\renormalized{\delta}$.
By analogy with~\eqref{eq:delta-renormalized},
\eqref{eq:velocity-renormalized}
and~\eqref{eq:vdelta-renormalized}
we expect that $\renormalizedc{\deltarsd}$ could involve
both multiplicative renormalization
and mixing with $\partial^2\delta$.
As in~\S\ref{sec:one-loop-real} we will parametrize ultraviolet effects
only up to $\BigO(k^2)$, and therefore we neglect mixing with
$\partial^4 \delta$ or higher-derivative operators.
In~\S\ref{sec:EFT-accuracy}
we determine an upper limit on the region where this approximation
is valid.

We will verify below that there is no multiplicative
renormalization.
In principle there are decaying contributions
from $\vect{v}$ and
$\vect{v} \delta$, but these are projected out by
our boundary conditions for the $D_i$ as explained
in~\S\ref{sec:eulerian-perturbation-theory}.
Therefore, up to one loop, we have
\begin{equation}
    \renormalizedc{\deltarsd}
    =
    \deltarsd
    +
    \frac{\ctrterm{2}{\deltarsd}}{\kNL^2} \partial^2 \delta
    .
\end{equation}
Inspection of~\eqref{eq:delta-s}
shows that the Eulerian expansion of $\deltarsd$
can be written in the form
\begin{subequations}
\begin{equation}
    \label{eq:deltas-expansion}    
    \deltarsd(\vect{k}) =
    \sum_n \hat{r}_{i_1} \cdots \hat{r}_{i_{2n}} (\deltarsdmu{2n})_{i_1 \cdots i_{2n}} ,
\end{equation}
where $(\deltarsdmu{2n})_{i_1 \cdots i_{2n}}$ is an operator
that transforms as a rank-$2n$ tensor under
spatial rotations.
When inserted in a correlation function, isotropy of the background will
convert contraction over $i_1 \cdots i_{2n}$ into a sum of powers of
$\mu^2 = (\hat{\vect{k}}\cdot\hat{\vect{r}})^2$
with highest power $\mu^{2n}$.
Therefore, although $\deltarsd$ does not itself admit a series expansion in
$\mu^2$, the counterterms needed to make its correlation functions
finite will do so.
It follows that
$\ctrterm{2}{\deltarsd}$
can be written in the form
\begin{equation}
    \label{eq:deltas-ctrterm-expansion}
    \ctrterm{2}{\deltarsd} = \sum_n \muctrterm{2}{2n} \mu^{2n} .
\end{equation}
\end{subequations}
There is one counterterm for each available power of $\mu$, although these
need not be independent at every order in the loop expansion.
As usual we expect that
averages over
ultraviolet modes should respect the symmetries
of the low-energy theory and therefore renormalization will not
generate odd powers of $\mu$.

\para{One-loop formulae}
To compute $\langle \deltarsd \deltarsd \rangle$ to one-loop
we calculate the expansion~\eqref{eq:deltas-expansion},
dropping operators that contribute only
at two loops or higher. This
yields~\cite{Heavens:1998es,Scoccimarro:1999ed,Matsubara:2007wj,
Senatore:2014vja,Lewandowski:2015ziq}
\begin{equation}
\begin{split}
    [\deltarsd]_{\vect{k}}
    =
    [\delta]_{\vect{k}}
    &
    \mbox{}
    - \frac{\im}{H} (\vect{k}\cdot\hat{\vect{r}})
        [ \hat{\vect{r}}\cdot\vect{v} ]_{\vect{k}}
    - \frac{\im}{H} (\vect{k}\cdot\hat{\vect{r}})
        [ \hat{\vect{r}}\cdot\vect{v} \delta ]_{\vect{k}}
    - \frac{1}{2!H^2} (\vect{k}\cdot\hat{\vect{r}})^2
        [ (\hat{\vect{r}}\cdot\vect{v})^2 ]_{\vect{k}}
    \\
    &
    \mbox{}
    - \frac{1}{2!H^2} (\vect{k}\cdot\hat{\vect{r}})^2
        [ (\hat{\vect{r}}\cdot\vect{v})^2\delta ]_{\vect{k}}
    + \frac{\im}{3!H^3} (\vect{k}\cdot\hat{\vect{r}})^3
        [ (\hat{\vect{r}}\cdot\vect{v})^3 ]_{\vect{k}}
    +
    \cdots
    .
\end{split}
\label{eq:delta-s-oneloop}
\end{equation}
We have adopted the notation of Ref.~\cite{Senatore:2014vja} in which
$[f]_{\vect{k}}$ denotes the Fourier transform of $f$.
In principle, based on simple power-counting of $\hat{\vect{r}}$,
Eq.~\eqref{eq:delta-s-oneloop}
may produce powers of $\mu^2$ up to $\mu^6$,
but in practice we will see that for
the two-point function at one-loop
the highest-order term is absent.

Eq.~\eqref{eq:delta-s-oneloop} was used by Matsubara~\cite{Matsubara:2007wj},
and rederived by Senatore \& Zaldarriaga~\cite{Senatore:2014via}.
In Refs.~\cite{Senatore:2014via,Lewandowski:2015ziq}
the continuity equation was used to exchange
the $\vect{v}$ and $\vect{v}\delta$
terms
for $\dot{\delta}$, but this is only possible under the assumption
that it is $\vect{\pi}$ rather than $\vect{v}$
that can be written as potential flow.
The two are not equivalent,
and
in SPT the potential flow approximation is normally applied only to $\vect{v}$.
Therefore we should retain $\vect{v}$ and $\vect{v}\delta$ separately
in~Eq.~\eqref{eq:delta-s-oneloop}.
With this choice our final result will match that derived by Matsubara
after replacing all growth functions by their Einstein--de Sitter
counterparts~\cite{Matsubara:2007wj}.
It also agrees with Perko et al.~\cite{Perko:2016puo},
in which the terms $\vect{v}$ and $\vect{v}\delta$
were retained.

\para{Role of composite operators}
Eq.~\eqref{eq:delta-s-oneloop} can be considered as a
single composite operator
renormalized by the counterterm $\ctrterm{2}{\deltarsd}$.
Alternatively, as emphasized in Refs.~\cite{Senatore:2014vja,Lewandowski:2015ziq,Perko:2016puo},
it may be regarded
as a sum of
$\delta$ and $\vect{v}$
with composite operators
$\vect{v}\delta$,
$\vect{v}\vect{v}$,
$\vect{v}\vect{v}\delta$
and $\vect{v}\vect{v}\vect{v}$.
In this second point of view we require
new renormalization conditions
to define
$\renormalizedc{\vect{v}\delta}$,
$\renormalizedc{\vect{v}\vect{v}}$,
$\renormalizedc{\vect{v}\vect{v}\delta}$
and
$\renormalizedc{\vect{v}\vect{v}\vect{v}}$,
in addition to those already
used to define
$\renormalized{\delta}$ and $\renormalized{\vect{v}}$.

As usual, we are free to choose these new
renormalization conditions in any convenient fashion.
In an `off shell' scheme we impose arbitrary
conditions unrelated to any measured correlation function.
Further finite renormalizations would be required at each power of
$\mu^2$ to match this off-shell $\renormalizedc{\deltarsd}$
to an observable quantity.%
    \footnote{In Ref.~\cite{Senatore:2014vja}, some of the
    counterterms appearing in the definition
    of $\renormalizedc{\vect{v}\vect{v}}$,
    $\renormalizedc{\vect{v}\vect{v}\delta}$
    and
    $\renormalizedc{\vect{v}\vect{v}\vect{v}}$ were
    equated.
    This choice is too restrictive, as recognized
    in Refs.~\cite{Lewandowski:2015ziq,Perko:2016puo}.}
Alternatively, we might choose to adjust the definition
of one or more composite operators in such a way that
$\renormalizedc{\deltarsd}$
is matched to some measured correlation function.
This is the choice made in Refs.~\cite{Lewandowski:2015ziq,Perko:2016puo}.
If $\deltarsd$ is broken into a sum of many composite operators
then our renormalization conditions
need not fix the definition of each operator uniquely.
Therefore we should expect degeneracies.
These merely reflect the division of $\deltarsd$ into a collection
of independent operators, when only the sum has physical
significance. By writing the counterterms as the coefficients of
a series expansion in $\mu^2$ we avoid explicit degeneracies of this kind.

The price paid for this convenience is a possibility of overcounting.
The requirement that~\eqref{eq:delta-s-oneloop}
is renormalized by mixing with a set of local operators obeying the
symmetries of the theory---%
principally,
rotational invariance and Galilean invariance---%
places restrictions on the $\muctrterm{2}{2n}$.
By explicit calculation using Eqs.~\eqref{eq:ctrterms-Zbasis-0}--\eqref{eq:ctrterms-Zbasis-8}
below, or
by using the operator
product expansion
(as in Refs.~\cite{Senatore:2014vja,Lewandowski:2015ziq})
to determine how
the composite operators in~\eqref{eq:delta-s-oneloop} mix with
$\partial^2 \delta$ at one loop,
we find that
at one-loop level
the counterterms satisfy the constraints
\begin{subequations}
    \begin{align}
        \label{eq:ctrterm-condition-a}
        \muctrterm{2}{6} & = f^3 \muctrterm{2}{0} - f^2 \muctrterm{2}{2} + f \muctrterm{2}{4} , \\
        \label{eq:ctrterm-condition-b}
        \muctrterm{2}{8} & = 0 .
    \end{align}
\end{subequations}
Therefore 
(neglecting stochastic counterterms)
there is no renormalization of $\mu^8$ at one-loop.

\para{Counterterms for the one-loop power spectrum}
We define the renormalized
redshift-space power spectrum $\Prsdren$ by
\begin{equation}
    \langle
        \renormalizedft{\deltarsd}_{\vect{k}_1}
        \renormalizedft{\deltarsd}_{\vect{k}_2}
    \rangle
    =
    (2\pi)^3 \delta(\vect{k}_1 + \vect{k}_2)
    \Prsdren(k) ,
\end{equation}
where $k = |\vect{k}_1| = |\vect{k}_2|$.
Bearing the foregoing discussion
in mind, it follows that the renormalized
$\deltarsd$ two-point function at one-loop can be written
\begin{equation}
    \Prsdren
    =
    \Prsd^{\SPT,\Lambda}
    -
    2 \sum_{n=0}^3
    \muctrterm{2}{2n} \mu^{2n} \frac{k^2}{\kNL^2} P  ,
    \label{eq:Prsd-counterterms}
\end{equation}
where $P_{s}^{\text{SPT},\Lambda}$
is the one-loop SPT power spectrum following from
Eq.~\eqref{eq:delta-s-oneloop}
with the loop integrals cut off at $q\sim\Lambda$.
The counterterms
$\muctrterm{2}{0}$,
$\muctrterm{2}{2}$,
$\muctrterm{2}{4}$
and
$\muctrterm{2}{6}$
can be chosen independently
subject to the
condition \eqref{eq:ctrterm-condition-a}.
(We have dropped the counterterm for $\mu^8$, which is necessarily absent.)
However, because $\deltarsd$
at $\mu = 0$ is equal to $\delta$
we will find $\muctrterm{2}{0} = \ctrterm{2}{\delta}$.

\para{Comparison with Lewandowski et al}
Eq.~\eqref{eq:Prsd-counterterms}
should be compared with Eq. (2.15) of Lewandowski et al.~\cite{Lewandowski:2015ziq}.
In this reference, the renormalized
$\deltarsd$ power spectrum was expressed in the form
\begin{equation}
\begin{split}
    \Prsdren =
    \Prsd^{\SPT,\Lambda}
    -
    2 (2\pi) D^2
    \Big[
    &
        \cs^2
        + \mu^2 \Big(
            2 f \cs^2
            + \frac{1}{H} \frac{\d \cs^2}{\d t}
            + \frac{1}{2} \bar{c}_1^2
        \Big)
    \\
    &
        + \mu^4 \Big(
            f^2\cs^2
            + \frac{f}{H} \frac{\d \cs^2}{\d t}
            + \frac{f}{2} \bar{c}_1^2
            + \frac{1}{2} \bar{c}_2^2
        \Big)
    + \frac{f}{2} \bar{c}_2^2 \mu^6
    \Big]
    \frac{k^2}{\kNL^2} P ,
\end{split}
\label{eq:Lewandowski-counterterms}
\end{equation}
with $\Prsd^{\SPT,\Lambda}$ now understood to be evaluated
in the
Einstein--de Sitter approximation
where all growth functions are replaced by their counterparts
from Table~\ref{table:EdS-growth}.
The counterterms are $\cs$, $\bar{c}_1$ and $\bar{c}_2$,
with $\bar{c}_1$ and $\bar{c}_2$ constructed from degenerate
combinations of the counterterms for the composite operators
appearing in~\eqref{eq:delta-s-oneloop}
as explained above.
Notice that, despite its appearance,
the $\cs$ used here does not equal the effective speed of sound
appearing in the renormalized Euler equation~\eqref{eq:renormalized-Euler}.
Finally, as usual, $D$ is the linear growth factor.

Eq.~\eqref{eq:Lewandowski-counterterms}
can be used to map the counterterms $\muctrterm{2}{2n}$ used in this paper
to their counterparts in Ref.~\cite{Lewandowski:2015ziq}.
At order-$\mu^2$ it contains
contributions involving the $\mu^0$ counterterm $\cs$ and its
time derivative $\csdot$.
These appear because Ref.~\cite{Lewandowski:2015ziq}
used the continuity equation
to eliminate $\vect{v}$ and $\vect{v}\delta$
in favour of the time derivative $\dot{\delta}$,
and included the counterterms for
$\renormalizedc{\dot{\delta}}$ among the contributions
at $\mu^2$.
As explained above, we believe this exchange
is not compatible with
the assumptions used to obtain Eqs.~\eqref{eq:alpha-eq}--\eqref{eq:beta-eq};
instead, $\vect{v}$ and $\vect{v}\delta$ should be retained separately.

To compute the time derivative $\csdot$, it was assumed
in Ref.~\cite{Lewandowski:2015ziq} that
$\cs^2 \propto D^{8/3}$.
With this choice, and
neglecting further
differences in time-dependent factors, the relations are
\begin{subequations}
    \begin{align}
        \muctrterm{2}{0} & = 2\pi \cs^2 , \\
        \muctrterm{2}{2} & =
            2\pi \bigg( \frac{14 f}{3} \cs^2 + \frac{1}{2} \bar{c}_1^2 \bigg) , \\
        \muctrterm{2}{4} & =
            2\pi \bigg( \frac{11f^2}{3} \cs^2 + \frac{f}{2} \bar{c}_1^2 + \frac{1}{2} \bar{c}_2^2 \bigg) , \\
        \muctrterm{2}{6} & = \pi f \bar{c}_2^2 .
    \end{align}    
\end{subequations}
Note that these quantities satisfy the linear constraint~\eqref{eq:ctrterm-condition-a}.

\subsection{Evaluation of the one-loop two-point function}
\label{sec:evaluate-rsd-twopf}

The principal challenge is to compute the one-loop two-point function
$P_{s}^{\text{SPT},\Lambda}$.
The calculation is technically straightforward, but very lengthy.
Its complexity arises partly from the number of terms
that appear in~\eqref{eq:delta-s-oneloop},
but also from the fact that the loop integrals
for the composite operators
$\vect{v}\delta$, $\vect{v}\vect{v}$, $\vect{v}\vect{v}\delta$ and
$\vect{v}\vect{v}\vect{v}$ are tensorial.
In this section we collect the necessary expressions.
The computation was first performed by Matsubara using the
Einstein--de Sitter approximation described on p.\pageref{eq:growth-factor-def}.
Here we give the result with its exact time dependence for the first time.

To simplify the computation we introduce a new method to evaluate the tensor
integrals. Matsubara's computation used the traditional approach of rotational
covariance to reduce these integrals
to scalar form-factors multiplying
fixed tensors with the correct transformation properties under rotations.
To solve for these form-factors one applies suitable contractions to yield
a system of scalar simultaneous equations.
This is a standard method, widely used
to reduce tensor integrals in field theory~\cite{Weinzierl:2006qs}.
The disadvantage is that
the final step of solving for the scalar form-factors can be algebraically expensive.
As we now describe,
our new method simplifies the calculation by
extracting the form factors directly.

\para{Application to 22 integrals}
To illustrate the method, consider the 22-type integration arising
from the $\langle [\vect{v}\delta]_{\vect{k}_1} [\vect{v}\delta]_{\vect{k}_2} \rangle$
contribution to $\langle \deltarsd \deltarsd \rangle$.
Then
\begin{equation}
    P_s(k) \supseteq
    - f^2 D^4 k^2 \mu^2
    \int
    \frac{\d^3 q}{(2\pi)^3}
    \frac{\d^3 s}{(2\pi)^3}
    (2\pi)^3 \delta(\vect{q} + \vect{s} - \vect{k}_1)
    \Pinit(q) \Pinit(s)
    \hat{r}_i \hat{r}_j
    \left(
        \frac{q_i q_j}{q^2 s^2}
        -
        \frac{q_i q_j}{q^4}
        -
        \frac{q_i k_{1j}}{q^2 s^2}
    \right)
    \label{eq:22-example}
\end{equation}
In principle the term $q_i k_{1j}$ should be symmetrized
over $i$ and $j$, but since it is
contracted with the symmetric combination $\hat{r}_i \hat{r}_j$
there is no need to do so explicitly.

Now replace the $\delta$-function by its Fourier representation, and
expand the resulting exponential using the Rayleigh plane wave formula,
\begin{equation}
    \e{\im \vect{k}\cdot\vect{x}}
    =
    \sum_{\ell=0}^\infty
    (2\ell+1)
    \im^\ell
    j_\ell(kx)
    \Legendre{\ell}{\hat{\vect{k}}\cdot\hat{\vect{x}}}
    .
    \label{eq:rayleigh-formula}
\end{equation}
Here, $j_\ell$ is the spherical Bessel function of order $\ell$
and $\Legendre{\ell}{x}$
is the $\ell^{\text{th}}$ Legendre polynomial.
That yields
\begin{equation}
\begin{split}
    P_s&(k) \supseteq
    - f^2 D^4 k^2 \mu^2
    \int
    \frac{\d^3 q}{(2\pi)^3}
    \frac{\d^3 s}{(2\pi)^3}
    \; \d^3 x \; \Pinit(q) \Pinit(s)
    \\
    & \mbox{}
    \times
    \bigg[
        \bigg(
            \frac{2}{3} \Legendre{2}{\hat{\vect{q}}\cdot\hat{\vect{r}}}
            +
            \frac{1}{2} \Legendre{0}{\hat{\vect{q}}\cdot\hat{\vect{r}}}
        \bigg)
        \bigg(
            \frac{1}{s^2} 
            -
            \frac{1}{q^2}
        \bigg)
        -
        \frac{k \mu}{qs^2} \Legendre{1}{\hat{\vect{q}}\cdot\hat{\vect{r}}}
    \bigg]
    \\
    & \mbox{}
    \times
    \sum_{\ell,\ell',\ell''}
    (2\ell+1)(2\ell'+1)(2\ell''+1)
    \im^{\ell+\ell'+\ell''}
    j_{\ell}(sx)
    j_{\ell'}(kx)
    j_{\ell''}(qx)
    \Legendre{\ell}{\hat{\vect{s}}\cdot\hat{\vect{x}}}
    \Legendre{\ell'}{-\hat{\vect{k}}_1\cdot\hat{\vect{x}}}
    \Legendre{\ell''}{\hat{\vect{q}}\cdot\hat{\vect{x}}}
    .
\end{split}
\end{equation}
The angular part of the $\vect{q}$, $\vect{s}$ and $\vect{x}$
integrations can be done using
the generalized orthogonality relation
\begin{equation}
    \int \d^2 \hat{\vect{x}} \;
    \Legendre{\ell}{\hat{\vect{a}}\cdot\hat{\vect{x}}}
    \Legendre{\ell'}{\hat{\vect{b}}\cdot\hat{\vect{x}}}
    =
    \frac{4\pi}{2\ell+1}
    \delta_{\ell\ell'}
    \Legendre{\ell}{\hat{\vect{a}}\cdot\hat{\vect{b}}} .
    \label{eq:generalized-Legendre-orthogonality}
\end{equation}
The result is
\begin{equation}
    P_s(k)
    \supseteq
    - f^2 D^4 k^2 \mu^2 8
    \Legendre{2}{-\mu}
    \int
    \frac{q^2 s^2 \; \d q \, \d s}{(2\pi)^3}
    \Pinit(q) \Pinit(s)
    \bigg[
        \frac{2}{3}
        \bigg( \frac{1}{s^2} - \frac{1}{q^2} \bigg)
        \FabrikantThree{2}{2}{0}
        +
        \frac{k \mu}{q s^2}
        \FabrikantThree{2}{1}{1}
    \bigg] ,
    \label{eq:fabrikant-22-example}
\end{equation}
where we have defined the 3-Bessel integral
$\FabrikantThree{\mu}{\nu}{\sigma}$ by
\begin{equation}
    \FabrikantThree{\mu}{\nu}{\sigma}
    \equiv
    \int_0^\infty x^2 j_\mu(k x) j_\nu(q x) j_\sigma(s x) \, \d x .
    \label{eq:fabrikant-3J}
\end{equation}
To reduce clutter we have suppressed explicit dependence on
the wavenumbers $k$, $q$ and $s$,
but this should be understood via the associations
$\mu \mapsto k$, $\nu \mapsto q$ and $\sigma \mapsto s$.
The problem of computing these integrals
analytically
for general $k$, $q$ and $s$
and arbitrary orders
$\mu$, $\nu$ and $\sigma$
was solved by
Gervois \& Navelet~\cite{doi:10.1137/0520067}
and Fabrikant~\cite{fabrikant2013elementary}.
We summarize Fabrikant's method in Appendix~\ref{appendix:fabrikant},
and as part of the bundle of software products accompanying this paper
we include a Mathematica notebook that implements the computation.

In general, the $\FabrikantThree{\mu}{\nu}{\sigma}$
vanish except where $k$, $q$ and $s$ satisfy the triangle condition
$|k-q| < s < |k+q|$.
Accordingly we may write $s = (q^2 + k^2 - 2 k q \cos\theta)^{1/2}$
and change variable from $s$ to $\theta$.
Therefore the $\FabrikantThree{\mu}{\nu}{\sigma}$ can be regarded as enforcing
the $\delta$-function $\delta(\vect{q} + \vect{s} - \vect{k}_1)$
with which we began.
The result is a scalar integral over $q$ and $\theta$.
The complexities of all tensor form factors have been absorbed by the
Legendre polynomial
$\Legendre{2}{-\mu}$.
In more general cases we may encounter a sum of Legendre polynomials
if the integrals over $\hat{\vect{q}}$, $\hat{\vect{s}}$
and $\hat{\vect{x}}$
generate nonzero contributions for more than one assignment of
$\ell$, $\ell'$ and $\ell''$.

\para{Comparison with method of covariance}
Had we used rotational covariance, the first step would have been to
introduce form-factors $A$ and $B$
and
express the integral~\eqref{eq:22-example}
in the form $A \delta_{ij} + B \hat{k}_{1i} \hat{k}_{1j}$.
Next, this should be converted to a system of
scalar equations by taking suitable contractions
with $i$ and $j$.
Finally,
after solving this system for $A$ and $B$
we contract
with $\hat{r}_i \hat{r}_j$ to yield the final result
$\mu^2 B + A$.
The solution will have $A = -B/3$, allowing it to be
expressed in the form
$-2 \Legendre{2}{\mu} A$
and reproducing the conclusion of Eq.~\eqref{eq:fabrikant-22-example}.
This approach becomes cumbersome
because of the manipulations needed
to extract the scalar integral $A$.
In our new method these manipulations are replaced by
the requirement to compute the integrals
$\FabrikantThree{\mu}{\nu}{\sigma}$,
but these are easy to tabulate in advance.
The substitution can be automated using a symbolic algebra
tool such as Mathematica.

In more complex cases the saving is greater.
As the tensor structures become more elaborate, the method
of rotational covariance would require us to introduce an increasing
number of form factors and decouple the resulting equations.
In contrast,
the method described here does not suffer from a
corresponding increase in algebraic
complexity; these more elaborate structures merely manifest themselves in
the appearance of
higher-order Legendre polynomials
generated by the
$\hat{\vect{q}}$, $\hat{\vect{s}}$ and $\hat{\vect{x}}$ integrals.
Using Fabrikant's method, the corresponding
$\FabrikantThree{\mu}{\nu}{\sigma}$
are no harder to obtain than those of lower order.

A similar procedure can be used to compute any 22-type integral.
In some cases we encounter
products of Legendre polynomials of the same argument.
In order to use the orthogonality relation such products must be
rewritten as a sum of individual Legendre polynomials, which can be accomplished
using the Neumann--Adams formula or an equivalent~\cite{neumann,adams,bailey1933}.

\para{Application to 13 integrals}
A very similar procedure can be used to perform 13-type integrals.
These are typically simpler because they involve integration only
over $\Pinit(q)$,
not $\Pinit(q)\Pinit(|\vect{k}-\vect{q}|)$ as for a 22-type integral,
and therefore the analogue of the $\vect{s}$-integral in Eq.~\eqref{eq:22-example}
can be performed analytically using
the Fourier transform $\int \d^3 s \, s^{-2} \e{\im \vect{s}\cdot\vect{x}} = 2\pi^2/x$.
Consequently, 13-type integrals require only 2-Bessel integrals of the form
\begin{equation}
   \FabrikantOne{\mu} 
    \equiv
    \int_0^\infty x j_\mu(k x) j_\mu(q x) \, \d x ,
    \label{eq:fabrikant-2J}
\end{equation}
and not the 3-Bessel form~\eqref{eq:fabrikant-3J}.
Tabulated analytic results for such integrals are relatively easy to obtain;
for example, integrals of this type can be performed by Mathematica.
(It is also possible to evaluate them by the method described in
Appendix~\ref{appendix:fabrikant}.)

\para{Alternative evaluation techniques}
We remark that the procedure described in this section
can be regarded as an alternative to the FAST-PT algorithm
recently proposed by McEwen et al.~\cite{McEwen:2016fjn,Fang:2016wcf}.
A similar algorithm was suggested by
Schmittful \& Vlah~\cite{Schmittfull:2016jsw,Schmittfull:2016yqx}.
These methods also utilise the Rayleigh expansion~\eqref{eq:rayleigh-formula},
and agree with our computation of the 13-type integrals.
For the 22 case, however, the FAST-PT approach involves
re-ordering the integrals to obtain [cf.~\eqref{eq:fabrikant-22-example}]
\begin{equation}
   \int \d q \, \d s \; q^{2+\alpha} s^{2+\beta}
    \Pinit(q) \Pinit(s)
        \FabrikantThree{\mu}{\nu}{\sigma} =
   \int \d x \; x^2 j_\mu(k x) I_{\alpha \nu}(x) I_{\beta \sigma}(x) ,
   \label{eq:fast-pt}
\end{equation}
where $I_{\alpha \nu}(x) \equiv \int \d q \, q^{2+\alpha} j_\nu(q x) \Pinit(q)$
is a Hankel transform of the initial power spectrum $\Pinit$.
This should be contrasted with the direct evaluation of
$\FabrikantThree{\mu}{\nu}{\sigma}$ described in Appendix~\ref{appendix:fabrikant}.

In FAST-PT the computation is reduced to numerical evaluation of
the one-dimensional transforms $I_{\alpha\nu}(x)$
and the final one-dimensional $x$-integral in~\eqref{eq:fast-pt}.
This algorithm therefore has complexity
$\BigO(N_1 \log N_1)$.
In comparison, our strategy of direct evaluation leaves
a two-dimensional integral over
$q$ and $s$ (or $q$ and $\theta$ after imposing the
triangle condition),
and therefore has approximate complexity $\BigO(N_2^2)$.
Notice that the constants $N_1$ and $N_2$ measuring the
size of the integrals
can be different;
in practice, we find that $N_2 \sim 10^2$
whereas $N_1$ is at least an order of magnitude larger.
This typically renders the methods equally fast.
An advantage of direct evaluation is that
(as much as possible)
it preserves
the algebraic structure of the integrals.
In addition, because the Bessel integrals are performed
analytically, there are no complications related
to convergence of the Hankel transforms $I_{\alpha \nu}$.
\vspace{3mm}
\hrule
\vspace{2mm}
{\small
\para{Tree-level}
We now summarize the outcome of the complete computation.
To all orders, the tree-level contribution is the Kaiser formula,
\begin{equation}
	P_s \supseteq D^2 ( 1 + f \mu^2 )^2 \Pinit .	
\end{equation}

\para{22-type terms}
At loop level we organize the calculation by defining
coefficients of a series expansion in $\mu$,
\begin{equation}
    P_s \equiv \sum_{n=0}^\infty P_{s,2n} \mu^{2n} .
    \label{eq:Prsd-series-expansion}
\end{equation}
As explained below Eq.~\eqref{eq:delta-s-oneloop},
in principle the one-loop expression for $\deltarsd$ includes
even powers of $\mu$ up to
$\mu^6$, and therefore $P_s$ may contain terms
in principle up to $\mu^8$.
However, in practice, the $\mu^6$ term is missing
and therefore at one-loop the only contribution
at $\mu^8$ comes from the 22-type term formed
from $\langle [\hat{\vect{r}}\cdot\vect{v}]^2 [\hat{\vect{r}}\cdot\vect{v}]^2 \rangle$.

The 22-type contributions can be split into scalar
and tensor terms, the latter arising from the composite
operators in~\eqref{eq:delta-s-oneloop}.
The scalar terms are
\begin{subequations}
\begin{align}
	P_{s,0} & \supseteq
	\DA^2 P_{AA} + \DA \DB P_{AB} + \DB^2 P_{BB} , \\
	P_{s,2} & \supseteq
	2 \DA \DK P_{AA} + (\DB \DK + \DA \DL) P_{AB} + 2 \DB \DL P_{BB}	 , \\
	P_{s,4} & \supseteq
	(\fA^2 \DA^2 - f D^2)^2 P_{AA} + \fB \DB (\fA \DA - f D^2) P_{AB} + \fB^2 \DB^2 P_{BB} .
\end{align}
\end{subequations}
The tensor contributions of 22-type can be written in the form
\begin{equation}
    P_{s,n} \supseteq \frac{f D^2 k^4}{8\pi^2} \int_0^\Lambda \d q \int_{-1}^{+1} \frac{\d x}{k^2 + q^2 - 2kqx}
    \Pinit(q) \Pinit[ (k^2 + q^2 - 2kqx)^{1/2} ]
    S_n .
\end{equation}
The integrand should be set to zero if the
quantity $(k^2 + q^2 - 2 k q x)^{1/2}$ exceeds the
cutoff $\Lambda$.
The quantities $S_n$ are
\begin{subequations}
\begin{align}
    \nonumber
    S_2 & =
    \big[ 2 \DB x(k-qx) + \DA (q + kx - 2q x^2) \big]
    \big[ 2kx + q(2 - f + x^2[f-4]) \big]
    \\ & \quad\; \mbox{}
    - \frac{f D^2}{2} (k - 2 q x)^2(x^2-1) , \\
    \nonumber
    S_4 & =
    f(q + 2kx - 3qx^2)
    \Big[
        \DA(q + kx - 2qx^2)
        + 2\DB x(k-qx)
    \big]
    \\ \nonumber & \quad\; \mbox{}
    +
    \big[ f q(x^2-1) + 2(q + kx - 2qx^2) \big]\big[ \DK q + (\DK + 2\DL) kx - 2(\DK + \DL) q x^2 \big]
    \\ \nonumber & \quad\; \mbox{}
    +
    \frac{f D^2}{8} \big[
        8 q^2 + 8 kq x(4 - 5f + x^2 [5f-6])
        + q(x^2-1)([8-3f]f + 3[16 - 16f + f^2]x^2)
    \\ & \qquad\qquad\quad \mbox{}
    +
        k^2(12x^2 - 4 - 8f[x^2-1])
    \Big] , \\
    \nonumber
    S_6 & =
    f(q + 2kx - 3qx^2)
    \big[
        \DK q + (\DK + 2\DL) kx - 2(\DK + \DL) qx^2
    \big]
    \\ \nonumber & \quad\; \mbox{}
    + \frac{f^2 D^2}{4}
    \Big[
        2k^2 (f-2 -[f-6]x^2)
        + 4kqx(7-3f+[3f-11]x^2)
    \\ & \qquad\qquad\quad \mbox{}
    +
        q^2(4-3f+18[f-2]x^2+5[8-3f]x^4)
    \Big] , \\
    S_8 & =
    \frac{f^3 D^2}{8} \Big[
        8 k q x(3-5x^2)
        + 4 k^2(3x^2 -1)
        + q^2 (3 - 30 x^2 + 35x^4)
    \Big] .
\end{align}
\end{subequations}
\para{13-type terms}
The same division can be made for 13-type terms.
The scalar components are
\begin{subequations}
\begin{equation}
	P_{s,0} \supseteq
	D \Pinit
	\Big[
		(\DD - \DJ) P_D + \DE P_E + (\DF + \DJ) P_F + \DG P_G
		+ \frac{\DJ}{2} \big[ P_{J2} - 2 P_{J1} \big]
	\Big] ,
\end{equation}
\begin{equation}
\begin{split}
	P_{s,2} \supseteq
	D \Pinit
	\Big[
		&
		(\DM + f [\DD - \DJ]) P_D
		+ (\DN + f \DE) P_E
		+ (\DP + f [\DF + \DJ]) P_F
	\\ & \mbox{}
		+ (\DQ + f \DG) P_G
		+ \frac{\DR - 2 f \DJ}{2} P_{J1}
		+ \frac{\DS + f \DJ}{2} P_{J2}
	\Big] ,
\end{split}
\end{equation}
\begin{equation}
\begin{split}
	P_{s,4} \supseteq
	f D \Pinit
	\Big[
		&
		(\fD \DD - \fJ \DJ) P_D + \fE \DE P_E + (\fF \DF + \fJ \DJ - f D \DA) P_F +
		(\fG \DG  - f D \DB) P_G
	\\ & \mbox{}
		+ ( [f - \fA] D \DA + f D^3 - 2 \fJ \DJ ) \frac{P_{J1}}{2}
		+ ( [f - \fB] D \DB + \fJ \DJ ) \frac{P_{J2}}{2}
	\Big] .
\end{split}
\end{equation}
\end{subequations}
The tensor contributions can be written as a single $\d q$ integral in the form
\begin{equation}
    P_{s,n} = \frac{D^2 k^2}{48 \pi^2} \Pinit(k) \int_0^\Lambda \d q \; \Pinit(q)
    T_n .
\end{equation}
The integrands $T_n$ are
\begin{subequations}
\begin{align}
    \nonumber
    T_2 & = 3 f (\DK + \DL) \frac{k^2}{q^2}
    + \Big[
        12 (\DK + \DL) - f (16 \DA + 32 \DB + 8 f D^2 + 11 [\DK + \DL])
    \Big]
    \\ \nonumber & \quad\; \mbox{}
    + \Big[
        16(3\DK + 4\DL) - f (16\DA + 32\DB + 11[\DK + \DL])
    \Big]
    \frac{q^2}{k^2}
    \\ & \quad\; \mbox{}
    + 3(f-4)(\DK + \DL) \frac{q^4}{k^4}
    - \frac{3}{2} (\DK + \DL) \frac{(k^2 - q^2)^3}{k^5 q}
    \Big(
        f \frac{k^2}{q^2} - f + 4
    \Big)
    \ln \left| \frac{1+r}{1-r} \right| , \\
    \nonumber
    \frac{T_4}{f} & = 3 (1+f)(\DK + \DL) \frac{k^2}{q^2}
    - \Big[
        3 \DK + 19 \DL + f(16 \DA + 32 \DB + 16 f D^2 + 11[\DK + \DL])
    \Big]
    \\ \nonumber & \quad\; \mbox{}
    + \Big[
        69 \DK + 85 \DL - f(16 \DA + 32\DB + 11[\DK + \DL])
    \Big] \frac{q^2}{k^2}
    \\ & \quad\; \mbox{}
    + 3(f-7)(\DK + \DL) \frac{q^4}{k^4}
    - \frac{3}{2} (\DK + \DL) \frac{(k^2 - q^2)^3}{k^5 q}
    \Big[8
        (1+f) \frac{k^2}{q^2} - f + 7
    \Big]
    \ln \left| \frac{1+r}{1-r} \right| , \\
    \nonumber
    \frac{T_6}{f^2} & =
    - \Big(
        8 f^2 D^2 + 15 \DK + 31 \DL
    \Big)
    + (\DK + \DL) \Big(
        3 \frac{k^2}{q^2}
        + 21 \frac{q^2}{k^2}
        - 9 \frac{q^4}{k^4}
    \Big)
    \\ & \quad\; \mbox{}
    - \frac{3}{2} (\DK + \DL)
    \frac{(k^2 - q^2)^3}{k^5 q}
    \Big( \frac{k^2}{q^2} + 3 \Big)
    \ln \left| \frac{1+r}{1-r} \right| .
\end{align}    
\end{subequations}

\para{Counterterms}
The appearance
of the counterterms depends on the basis of local operators
in which we choose to express $\deltarsd$.
If we choose to renormalize the basis
$\renormalizedc{\vect{v}\delta}$,
$\renormalizedc{\vect{v}\vect{v}}$,
$\renormalizedc{\vect{v}\vect{v}\delta}$
and
$\renormalizedc{\vect{v}\vect{v}\vect{v}}$,
then the counterterm
for each $P_{s,n}$
will be a linear combination of the loop-level
time dependence for each of these operators,
with coefficients
$\ctrconst{2}{\vect{v}\delta}$,
\ldots,
$\ctrconst{2}{\vect{v}\vect{v}\vect{v}}$,
together with a linear combination of the arbitrary
functions
$\ctrfree{2}{\vect{v}\delta}$,
\dots,
$\ctrfree{2}{\vect{v}\vect{v}\vect{v}}$.%
    \footnote{If we are using an `off-shell' scheme
    in which some or all of the composite operators are defined
    by arbitrary conditions,
    then additional finite renormalizations may
    be needed to allow the $P_{s,n}$ to be matched to measurements.}
In this basis we find,
suppressing the unknown time-dependent terms
$\ctrfree{2}{\delta}$, \ldots, $\ctrfree{2}{\vect{v}\vect{v}\vect{v}}$
associated with each operator,
\begin{subequations}
\begin{align}
    \label{eq:ctrterms-Zbasis-0}
    \frac{\muctrterm{2}{0}}{\kNL^2} & =
        - \frac{1}{D}
        \Big(
            18 \DD + 28 \DE - 7 \DF - 2 \DG - 13 \DJ
        \Big)
        \ctrconst{2}{\delta} , \\
    \nonumber
    \frac{\muctrterm{2}{2}}{\kNL^2} & =
        -\frac{f}{D}
        \Big(
            18 \DD + 28 \DE - 7 \DF - 2 \DG - 13 \DJ
        \Big) \ctrconst{2}{\delta}
    \\ \nonumber & \quad\;
        -\frac{1}{D}
        \Big(
            18 \fD \DD + 28 \fE \DE - 7 \fF \DF - 2 \fG \DG - 13 \fJ \DJ
            + (12 \fA - 5 f) D \DA
    \\ \nonumber & \quad\quad\quad\quad
            + (12 \fB - 10 f) D \DB
            - 12 f D^3
        \Big) \ctrconst{2}{\vect{v}}
    \\ \nonumber & \quad\;
        + \Big(
            (12 \fA - 5 f) \DA
            + (12 \fB - 10 f) \DB
            - 12 f D^2
        \Big) \ctrconst{2}{\vect{v}\delta}
    \\ & \quad\;
        - \frac{4f}{3} \Big(
            \fA \DA + \fB \DB - f D^2
        \Big) (5 \ctrconst{2}{\vect{v}\vect{v},A} + \ctrconst{2}{\vect{v}\vect{v},B} )
        - \frac{5}{2} f^2 D^2 \ctrconst{2}{\vect{v}\vect{v}\delta}
    \\
    \nonumber
    \frac{\muctrterm{2}{4}}{\kNL^2} & =
        - \frac{f}{D} \Big(
            18 \fD \DD + 28 \fE \DE - 7 \fF \DF - 2 \fG \DG - 13 \fJ \DJ
            + 12 (\fA - 5 f) D \DA
    \\ \nonumber & \quad\quad\quad\quad
            + 12 (\fB - 10 f) D \DB
            - 12 f D^3
        \Big) \ctrconst{2}{\vect{v}}
    \\ \nonumber & \quad\;
        + f \Big(
            (12 \fA - 5 f) \DA
            + (12 \fB - 10 f) \DB
            - 12 f D^2
        \Big) \ctrconst{2}{\vect{v}\delta}
    \\ \nonumber & \quad\;
        - \frac{f}{3} \Big(
            (3 + 4 f) \fA \DA
            + 2 (9 + 2f) \fB \DB
            - (3 + 4f) f D^2
        \Big) \ctrconst{2}{\vect{v}\vect{v},B}
    \\ & \quad\;
        - \frac{20 f^2}{3} \Big(
            \fA \DA + \fB \DB - f D^2
        \Big) \ctrconst{2}{\vect{v}\vect{v},A}
        - \frac{5}{2} f^3 D^2
        (\ctrconst{2}{\vect{v}\vect{v}\delta} + \ctrconst{2}{\vect{v}\vect{v}\vect{v}} )
    \\
    \frac{\muctrterm{2}{6}}{\kNL^2} & =
        - f^2 ( \fA \DA + 6 \fB \DB - f D^2) \ctrconst{2}{\vect{v}\vect{v},B}
        - \frac{5}{2} f^4 D^2 \ctrconst{2}{\vect{v}\vect{v}\vect{v}}
    \\
    \label{eq:ctrterms-Zbasis-8}
    \frac{\muctrterm{2}{8}}{\kNL^2} & = 0 .
\end{align}    
\end{subequations}
Notice that there are two renormalization constants associated with the
operator
$\renormalizedc{\vect{v}\vect{v}}$, because
this can mix independently with the tensor factors
$\delta_{ij}$ and
$\hat{k}_i \hat{k}_j$, ie.,
\begin{equation}
    \renormalizedc{\vect{v}\vect{v}}_{ij}
    = (\vect{v}\vect{v})_{ij}
    +
    \Big(
        \ctrterm{2}{\vect{v}\vect{v},A} \delta_{ij}
        + \ctrterm{2}{\vect{v}\vect{v},B} \hat{k}_{i} \hat{k}_j
    \Big) \frac{H^2}{\kNL^2} \frac{1}{k^2} \partial^2 \delta .
\end{equation}
The constants
$\ctrconst{2}{\vect{v}\vect{v},A}$ and
$\ctrconst{2}{\vect{v}\vect{v},B}$ are the corresponding $Z$-parameters.
These correspond to the
Wilson coefficients $c_1$ and $c_2$ defined by Lewandowski et al. in
their Eq. (6.6)~\cite{Lewandowski:2015ziq}.
In principle there could be similar mixing with different tensor factors in
the OPE for $\renormalizedc{\vect{v}\vect{v}\delta}$,
but at one-loop the $\hat{k}_i \hat{k}_j$ tensor does not enter.

Alternatively, if we choose to renormalize the
coefficients of the $\mu$-expansion
$\renormalizedc{\deltarsdmu{n}}$,
as in Eq.~\eqref{eq:Prsd-counterterms},
then we find,
again omitting the possibility of unknown
time-dependent terms,
\begin{subequations}
\begin{align}
    \frac{\muctrterm{2}{0}}{\kNL^2} & = 
    - \frac{1}{D} \Big( 18 \DD + 28 \DE - 8 \DF - 2 \DG - 13 \DJ \Big)
    \ctrconst{2}{\deltarsd,0} , \\
    \nonumber
    \frac{\muctrterm{2}{2}}{\kNL^2} & =
    -\frac{1}{D} \Big(
        18 (\fD + f) \DD + 28 (\fE + f) \DE - 7 (\fF + f) \DF - 2 (\fG + f) \DG
        - 13 (\fJ + J) \DJ
    \Big)
    \\ & \quad\, +
    \Big(
        4(3-2f) (\DK + \DL)
        - 12 \fA \DA - 12 \fB \DB + (12 - \frac{5}{2} f) D^2
    \Big)
    \ctrconst{2}{\deltarsd,2} , \\
    \nonumber
    \frac{\muctrterm{2}{4}}{\kNL^2} & =
    - \frac{f}{D} \Big(
        18 \fD \DD + 28 \fE \DE - 7 \fF \DF - 2 \fG \DG - 13 \fJ \DJ
    \Big)
    \\ \nonumber & \quad\, -f \Big(
        12 \fA \DA + 12 \fB \DB - 11 \DK - 6 \DL
        +f [ (5f - 12) D^2 + 8 \DK + 8 \DL ]
    \Big) , \\
    \frac{\muctrterm{2}{6}}{\kNL^2} & =
    - \frac{f^2}{2} \Big(
        5 f^2 D^2 + 2\DK + 12 \DL)
    \Big)
    \ctrconst{2}{\deltarsd,6} , \\
    \frac{\ctrterm{2}{\deltarsd,8}}{\kNL^2} & = 0 .
\end{align}    
\end{subequations}
As explained above, these cannot all be varied independently
but only subject to the constraint~\eqref{eq:ctrterm-condition-a}.
Notice there is no divergence at $\mu^8$ in agreement with~\eqref{eq:ctrterm-condition-b}.

There are no multiplicative renormalizations of the $P_{s,n}$.
This can be regarded as a nontrivial check of the computation.
Since the mapping between real and redshift space conserves mass,
the same conservation-of-mass argument that
prohibits multiplicative renormalization of $\delta$
will apply to $\deltarsd$;
see footnote~\ref{footnote:no-kzero-in-delta}
on p.\pageref{footnote:no-kzero-in-delta}.
}

\vspace{2mm}
\hrule
\vspace{3mm}

\subsection{Resummation}
\label{sec:rsd-resummation}

If there are large-scale random motions then
the redshift-space power spectrum will require resummation for the same
reasons described in~\S\ref{sec:resummation-methods}.
This can be accomplished by a modification of the procedure used in real space.

The key tool
is still the use of Lagrangian perturbation theory to provide
a template.
The redshift distortion~\eqref{eq:redshift-space-s}
now applies to the Lagrangian picture displacement field $\vect{\Psi}$
with $\vect{v} = \d \vect{\Psi} / \d t$,
so at linear level we have
\begin{equation}
    \vect{\Psi}_{s,1} = \vect{\Psi}_{1} + f (\hat{\vect{r}} \cdot \dot{\vect{\Psi}}_1) \hat{\vect{r}}
    = \vect{R} \cdot \vect{\Psi}_1
\end{equation}
where $f = \d \ln D / \d \ln a$ is defined by Eq.~\eqref{eq:growth-factor} as above
and the `redshift-space distortion tensor' $R_{ij}$ satisfies
\begin{equation}
    R_{ij} = \delta_{ij} + f \hat{r}_i \hat{r}_j .    
\end{equation}
It follows that correlation functions of $\vect{\Psi}_1$ can be converted to redshift space
by projecting all indices with $R_{ij}$.
Therefore, at lowest order, the two-point function $A_{ij}$ becomes
\begin{equation}
\begin{split}
    A_{s,ij}^{\attreelevel} =
    \langle
        [ \Delta \vect{\Psi}_s(\vect{q}) - \Delta \vect{\Psi}_s(\vect{0}) ]^2_{ij}
    \rangle
    =
    \mbox{}
    &
    (\delta_{ij} + 2 f \hat{r}_i \hat{r}_j + f^2 \hat{r}_i \hat{r}_j)
    X(q)
    \\
    & \mbox{}
    +
    \Big[
        \hat{q}_i \hat{q}_j + f (\hat{\vect{q}} \cdot \hat{\vect{r}}) ( \hat{q}_i \hat{r}_j + \hat{q}_j \hat{r}_i )
        + f^2 (\hat{\vect{q}}\cdot \hat{\vect{r}})^2 \hat{r}_i \hat{r}_j
    \Big]
    Y(q) ,
\end{split}
\label{eq:tree-Aij-rsd}
\end{equation}
where $X(q)$ and $Y(q)$ continue to be defined by Eqs.~\eqref{eq:X-def}--\eqref{eq:Y-def}.

\para{VSCF formula}
We may now apply the prescription of Vlah, Seljak, Chu \& Feng to arrive at
an expression for the redshift-space power spectrum analogous to
Eq.~\eqref{eq:P-resum-intermediate},
\begin{equation}
    \PrsdloopVSCF(k) \equiv \Prsdloopnw
	+
	\int \d^3 q \; \e{-\im \vect{k} \cdot \vect{q}}
	\exp
	\bigg(
		{-\frac{1}{2} k_i k_j A_{s,ij}^{\nowiggle,\attreelevel}}
	\bigg)
	\bigg(
		{-\frac{1}{2} k_m k_n A_{s,mn}^{\wiggle,\uptooneloop}}
		+
		\frac{\im}{6} k_m k_n k_r W_{s,mnr}^{\wiggle,\uptooneloop}
	\bigg) .
\end{equation}
The `wiggle' and `no-wiggle' combinations
have the same meaning used in~\S\ref{sec:resummation-methods},
with `no-wiggle' components at one-loop and higher being
built exclusively from the `no-wiggle' initial power spectrum
and the `wiggle' terms absorbing the remainder.
The combination $k_i k_j A_{s,ij}^{\nowiggle,\attreelevel}$ can be evaluated
using~\eqref{eq:tree-Aij-rsd}, which yields
\begin{equation}
    k_i k_j A_{s,ij}^{\nowiggle,\attreelevel}
    =
    k^2 \Big(
        [1 + f(f+2)\mu^2 ] X^{\nowiggle}(q)
        +
        \Big[
            (\hat{\vect{k}}\cdot\hat{\vect{q}})^2
            + 2 f \mu (\hat{\vect{q}}\cdot\hat{\vect{r}})(\hat{\vect{q}}\cdot\hat{\vect{k}})
            + f^2 \mu^2 (\hat{\vect{q}}\cdot\hat{\vect{r}})^2
        \Big] Y^{\nowiggle}(q)
    \Big)
    \label{eq:rsd-damping-factor}
\end{equation}
Eq.~\eqref{eq:rsd-damping-factor} should be compared with
Eq.~(4.14) of Lewandowski et al.~\cite{Lewandowski:2015ziq}.
Since $X^{\nowiggle}$ and $Y^{\nowiggle}$ are still
slowly varying on scales where the `wiggle' components have support,
we are entitled to perform an approximate integration
as in Eq.~\eqref{eq:P-resum},
with the result
\begin{equation}
    \PrsdloopVSCF(k) =
    \Prsdloopnw
    +
    \exp\bigg( {-\frac{1}{2}} \llangle k_i k_j A_{s,ij}^{\nowiggle,\attreelevel} \rrangle \bigg)
	\bigg(
	   \Prsdloopw
	   +
	   \frac{1}{2} \llangle k_i k_j A_{s,ij}^{\nowiggle,\attreelevel} \rrangle \Prsdtreew
	\bigg)
	\label{eq:Prsd-resum}
\end{equation}
The average can be performed as in Eq.~\eqref{eq:A-average-defn}, which yields
\begin{equation}
    \llangle k_i k_j A_{s,ij}^{\nowiggle,\attreelevel} \rrangle
    = k^2 \Big[ 1 + f(f+2) \mu^2 \Big] \llangle A^{\nowiggle,\attreelevel} \rrangle
    \label{eq:Prsd-damping}
\end{equation}
and $\llangle A^{\nowiggle,\attreelevel} \rrangle$
is the same quantity defined in Eq.~\eqref{eq:expectation-A}
that appears in the real-space resummation template.
We evaluate it using the same choices
$\qmin = 10 h^{-1} \, \Mpc$
and
$\qmax = 300 h^{-1} \, \Mpc$
used for the real-space power spectrum, and
similarly we perform the $k$-integration
up to the cutoff $k = 1.4 h/\Mpc$ used to compute the loops.

\para{Application to renormalized power spectrum}
In practice we wish to apply this resummation prescription to the renormalized
power spectrum predicted by the effective field theory.
We denote the resulting power spectrum by $\PrsdrenVSCF$.
It is defined by Eq.~\eqref{eq:Prsd-resum}
with $\Prsdloopnw$ and $\Prsdloopw$ understood to include
the counterterms~\eqref{eq:Prsd-counterterms},
or explicitly
\begin{subequations}
\begin{align}
    \label{eq:EFT-Prsd-nw}
    \Prsdloopnw & = \Prsdnw^{\SPT,\Lambda,\uptooneloop}
    - 2 \sum_{n=0}^4 \muctrterm{2}{2n} \mu^{2n} \frac{k^2}{\kNL^2} \Ptreenw ,
    \\
    \label{eq:EFT-Prsd-w}
    \Prsdloopw & = \Prsdw^{\SPT,\Lambda,\uptooneloop}
    - 2 \sum_{n=0}^4 \muctrterm{2}{2n} \mu^{2n} \frac{k^2}{\kNL^2} \Ptreew .
\end{align}
\end{subequations}

\para{Fingers-of-God suppression}
It was explained in~\S\ref{sec:resummation-methods}
that Matsubara's resummation prescription in real space
produces a universal damping factor $\sim \exp[ (\epsilon_{s<} + \epsilon_{s>})/3 ]$.
In redshift-space the argument of the exponential
is modified by the factor
$1 + f(f+2)\mu^2$ appearing in Eq.~\eqref{eq:Prsd-damping}.
Matsubara observed~\cite{Matsubara:2007wj}
that the resulting suppression factor
resembled the damping factor
$\exp(-k^2 \mu^2 \sigmav^2)$ sometimes used
as a phenomenological description of
power suppression on small scales due to
the velocity dispersion $\sigmav$ within virialized halos,
the so-called `fingers of God' effect~\cite{Peacock:1993xg}.
In perturbation theory we can estimate $\sigmav$ by computing
the isotropic part of the velocity two-point function,
\begin{equation}
    \sigmav^2 = \langle v_i(\vect{x}) v_j(\vect{x}) \rangle_{\text{isotropic}}
    = f^2 D^2 \int \frac{\d k}{6\pi^2} \Pinit(k) .
\end{equation}
The scale $\sigmav^2$ is the same as
our factor $\llangle A^{\nowiggle,\attreelevel} \rrangle$
if the Bessel function in the integrand of~\eqref{eq:expectation-A} is dropped.
Matsubara's observation suggests that one could regard the damping
produced by resummation as a description of the
power suppression from the `fingers of God' effect.
However, this is not physically satisfactory because
the `fingers of God' damping is an ultraviolet effect
that has no clear connexion with the
large-scale random motions
that necessitate resummation.

As explained in~\S\ref{sec:resummation-methods},
Matsubara's prescription leads to excessive damping
on quasilinear scales~\cite{Taruya:2010mx}.
In an effective field theory description
with a Galilean-invariant resummation scheme the conclusion
is different.
(The details of the resummation scheme do not matter for this argument.
The Vlah, Seljak, Chu \& Feng scheme described above is one candidate,
but this discussion would apply equally to the scheme
proposed by Senatore \& Zaldarriaga~\cite{Senatore:2014via}
or the schemes discussed in Refs.~\cite{Vlah:2015sea,McQuinn:2015tva}.
See also Taruya, Nishimichi \& Saito, who used a different procedure to produce
damping of the acoustic oscillations~\cite{Taruya:2010mx}.)
The damping factor is now applied only to the `wiggle' component of the
power spectrum,
and
subtraction of power for $\mu \neq 0$ is provided instead by the counterterms
$\ctrterm{2}{\deltarsd,2n}$ for $n \geq 1$.
Therefore the effective field theory description can \emph{separately} accommodate
suppression of the baryon acoustic oscillations
due to large-scale motions
and suppression of the small-scale power
due to random motion within halos.
This is physically reasonable:
the counterterms encode the averaged small-structure of the theory
and therefore provide a natural description
for the subtraction of power due to virialized velocities.

\subsection{Multipole power spectra}
\label{sec:multipoles}

The outcome of~\S\ref{sec:rsd-resummation} is a very simple prescription
for resummation of the redshift-space power spectrum:
the `no-wiggle' terms are unaffected,
whereas the `wiggle' terms are damped by a term of the
form $\exp(-A-B\mu^2)$.
The simplicity of this $\mu$-dependence makes it
straightforward to extract
Legendre modes from
Eqs.~\eqref{eq:Prsd-resum}--\eqref{eq:Prsd-damping}.
Observational data are typically reported as measurements
of these modes.
Specifically,
Cole, Fisher \& Weinberg defined the
\emph{multipole power spectra} $P_\ell$
to satisfy~\cite{Cole:1993kh}
\begin{equation}
    \Prsd(k,\mu) \equiv \sum_\ell P_{\ell}(k) \Legendre{\ell}{\mu} .
    \label{eq:multipole-Pk}
\end{equation}
We wish to compute the multipole power spectra for the resummed,
renormalized power spectrum, which we denote
$\PrsdrenVSCF$.
They can be computed using the Legendre orthogonality relation~\eqref{eq:generalized-Legendre-orthogonality}
with $\vect{a} = \vect{b}$,
which yields
\begin{equation}
    P_\ell(k) = \frac{2\ell + 1}{2} \int_{-1}^{+1} \d \mu \;
    \PrsdrenVSCF(k, \mu) \Legendre{\ell}{\mu} .
    \label{eq:multipole-decompose}
\end{equation}
We have not added distinguishing labels to $P_\ell(k)$,
but there is no ambiguity because the only multipole power
spectra we will discuss are those defined by the resummed,
renormalized power spectrum
$\PrsdrenVSCF$.
The $P_\ell$ are identically zero for odd $\ell$
because $\Prsdren$ is a function of $\mu^2$.
Measurements exist for the lowest
multipoles $\ell = 0$ (the monopole),
$\ell = 2$ (the quadrupole) and
$\ell = 4$ (the hexadecapole)~\cite{Yamamoto:2008gr,Blake:2011rj,Oka:2013cba,Beutler:2013yhm}.

\para{Counterterms}
Eqs.~\eqref{eq:multipole-decompose} and \eqref{eq:EFT-Prsd-nw}--\eqref{eq:EFT-Prsd-w}
show that we can write
\begin{equation}
    P_\ell(k) = P_{\ell}^{\SPT,\Lambda,\uptooneloop} - 2 \ellctrterm{2}{\ell}
    \frac{k^2}{\kNL^2} \Ptree.
    \label{eq:Legendre-counterterms}
\end{equation}
where the labels `$\SPT$', `$\Lambda$',
and `$\uptooneloop$' have their usual meanings,
and $\Ptree$ is the tree-level power spectrum in real space.
As explained above,
there is just one counterterm
$\ellctrterm{2}{\ell}$ for each multipole $\ell$.
It is a linear combination of the  counterterms
$\muctrterm{2}{2n}$
defined in Eq.~\eqref{eq:Prsd-counterterms}
and
associated with the power series
expansion in $\mu^n$.
If we apply the VSCF resummation scheme to the linear power spectrum
$\Ptree$
that
appears in the counterterms then the coefficients of this combination
become weakly dependent on cosmology
via the damping factor
$\llangle A^{\nowiggle,\attreelevel} \rrangle$.
In practice, however, it makes very little difference whether or not
we choose to apply resummation to the counterterms.

\para{Numerical considerations}
The possibility of analytically extracting the $P_\ell$
is an advantage of the VSCF resummation
prescription. For example, using the resummation scheme
proposed by Senatore \& Zaldarriaga~\cite{Senatore:2014via,Lewandowski:2015ziq},
the resummed expression involves multiple
integrations that do not decouple from $\mu$.
The Legendre multipoles must be computed by performing
the $\mu$ integration in~\eqref{eq:multipole-decompose}
numerically, giving the final result
\begin{equation}
    P_\ell^{\uptooneloop}(k) =
    \sum_{j=0}^1
    \sum_{\ell'}
    \int \frac{\d k' \, (k')^2}{2\pi^2} M_{\| {N-j}}(k, k')_{\ell \ell'} P_{\ell'}^{\renormalizedtag}(k')_j
\end{equation}
where $P_{\ell}^{\renormalizedtag}(k)_j$ is the $\ell^{\text{th}}$ multipole of the
renormalized power spectrum at order $j$ in the Eulerian expansion,
and $M_{\| {N-j}}(k, k')_{\ell \ell'}$
is a mode-coupling matrix
whose definition is given in Eq.~(4.18) of Ref.~\cite{Lewandowski:2015ziq}.
It involves the $\mu$ integral
together with a three-dimensional integration over the Lagrangian coordinate
$\vect{q}$.
The final prescription
therefore requires five-dimensional integration
and summation over $\ell'$, and is numerically expensive.
Lewandowski et al. mitigated this
difficulty by
developing approximate analytic
estimates for some of these integrations,
which could be regarded
as a counterpart of the approximate
integration used in Eq.~\eqref{eq:P-resum}.
However,
their final procedure is still more complex than
the VSCF method employed here.

The `no-wiggle' part of the power spectrum
is unchanged by the VSCF procedure.
Extracting Legendre multipoles is therefore
no more complex than
a trivial change-of-basis in the series representation
from $\mu^n$ to $\Legendre{\ell}{\mu}$.
The damped `wiggle' part requires evaluation
of the integrals
\begin{equation}
    I_\ell \equiv \frac{2\ell+1}{2} \int_{-1}^{+1}
    \e{- A - B \mu^2} \Legendre{\ell}{\mu} \mu^n \, \d \mu .
\end{equation}

\vspace{3mm}
\hrule
\vspace{2mm}
{\small
\noindent The $I_\ell$ can be expressed in terms of the incomplete $\Gamma$-function
$\Gamma_x(s)$, defined by
\begin{equation}
    \Gamma_x(a) \equiv \int_x^\infty t^{s-1} \e{-t} \, \d t .    
\end{equation}
The required results are
\begin{align}
    I_0 & = \frac{1 + (-1)^n}{4} B^{-n_1} \e{-A}
    \Big[ \Gamma(n_1) - \Gamma_B(n_1) \Big] , \\
    I_2 & = 5 \frac{1 + (-1)^n}{16} B^{-n_3} \e{-A}
    \Big[ (3-2B+3n) \Gamma(n_1) + 2B \Gamma_B(n_1) - 6 \Gamma_B(n_3) \Big] , \\
    \nonumber
    I_4 & = 9 \frac{1 + (-1)^n}{128} B^{-n_5} \e{-A}
    \Big[ \big( 12B^2 - 120B n_1 + 140 n_1 n_3 \big) \Gamma(n_1) - 12 B^2 \Gamma_B(n_1)
    \\ & \mbox{} \qquad \qquad \qquad \qquad \qquad \quad {- 120} B \Gamma_B(n_3) + 150 \Gamma_B(n_5) \Big] ,
\end{align}
where $n_p = (n+p)/2$. For numerical evaluation it is sometimes helpful
to rewrite the incomplete $\Gamma$-function in terms of $\erf(z)$,%
    \footnote{The incomplete $\Gamma$ function itself is not commonly included
    as a standard function in numerical libraries,
    but the error function is; eg. it is available as
    \href{http://en.cppreference.com/w/cpp/numeric/math/erf}{\texttt{std::erf()} for
    $\geq$ {\CC}11}.}
using the recurrence formula $\Gamma_x(n+1) = n \Gamma_x(n) + x^n \e{-x}$
and
\begin{equation}
    \Gamma_x(\frac{1}{2}) = \sqrt{\pi} [ 1 - \erf(\sqrt{x}) ] .
\end{equation}}
\hrule
\vspace{2mm}

\begin{figure}
    \begin{center}
        \includegraphics{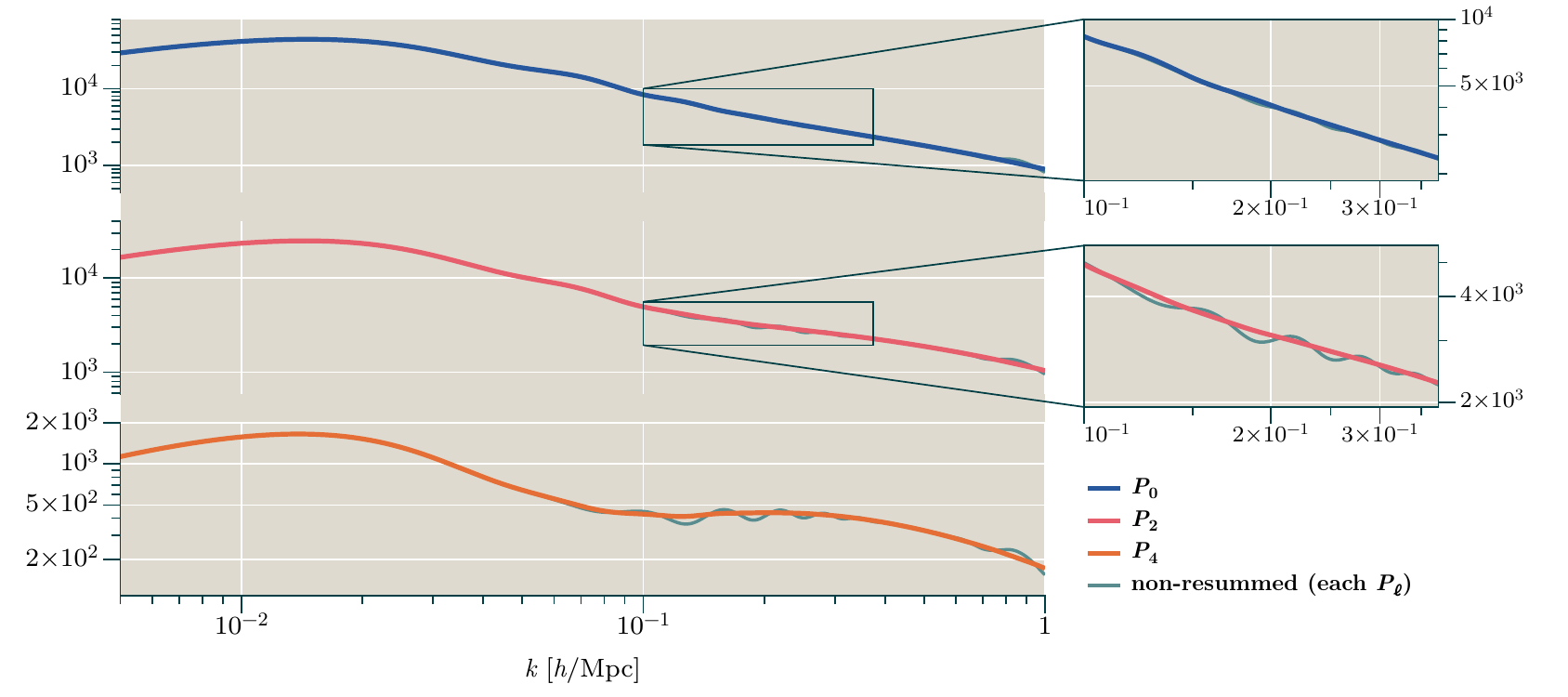}    
    \end{center}
    \caption{\label{fig:multipole-SPT}Resummed and non-resummed versions
    of the multipole power spectra $P_\ell$.
    The blue, red and orange lines
    show the resummed multipoles $P_0$, $P_2$ and $P_4$, respectively.
    For each multipole, a green line shows the same power spectrum without resummation.
    Notice that $P_2$ and $P_4$ exhibit very significant
    damping of the `wiggle' component.
    The background cosmology matches the
    \href{https://www.cosmosim.org/cms/simulations/mdr1/}{MDR1 MultiDark simulation}~\cite{Prada:2011jf}
    and also our own simulations~(\S\ref{sec:simulations}).}
\end{figure}

\para{Effect of resummation}
In Fig.~\ref{fig:multipole-SPT} we plot the $P_\ell$ for $\ell=0$, $\ell=2$ and $\ell=4$.
We use a background cosmology adjusted to match that used in the
\href{https://www.cosmosim.org/cms/simulations/mdr1/}{MDR1 MultiDark simulation}.
This is the same background cosmology we will use in~\S\ref{sec:simulations}
to obtain non-linear estimates
for these multipoles
from our own simulations.

Whereas the effect of resummation on the real-space power spectrum was
small (roughly a percent-level effect), 
Fig.~\ref{fig:multipole-SPT} shows that its influence on the redshift-space
multipoles is more significant.
For the cosmology considered here,
the suppression of oscillations in $P_0$ and $P_2$ is roughly
a $10\%$ effect, and the suppression in $P_4$ is roughly
a $15\%$ effect.

\subsection{Numerical calculation of the non-linear redshift-space power spectrum}
\label{sec:simulations}

Our task is now to renormalize the multipole power spectra in a similar fashion
to~\S\ref{sec:real-space-results}.
The `on-shell' scheme consists of adjusting the
counterterms to optimize the fit to those $P_\ell$ for which we have
measurements.

\subsubsection{Power-spectrum methodology}
There are not yet any well-calibrated fitting formulae
comparable to {\Halofit}
for the
non-linear multipole power spectra,
and therefore we must obtain direct estimates.
To do so we
use the public
\href{https://github.com/gevolution-code/gevolution-1.1}{\gevolution}
code%
    \footnote{The most interesting feature of {\gevolution} is that it can include
    relativistic effects in the weak field limit. We do not make use of this
    feature, instead running {\gevolution} in Newtonian mode, but in principle
    this could be used to test the validity of the non-relativistic limit
    described in~\S\ref{sec:matter-equations}.}
\cite{Adamek:2015eda} to perform a custom simulation with $1024^3$ particles and a
box-size of $(2000 \, \Mpc/h)^3$. This corresponds to an $N$-body particle mass
of $\simeq 6\times 10^{11} M_{\odot}/h$, approximately matching the mass of 
Milky Way-sized galaxies.
The background cosmology matches the MultiDark MDR1 simulation~\cite{Prada:2011jf},
which we use to cross-check the validity of our results at $z=0$.
Our simulations can be reproduced by downloading a
{\gevolution} settings file, as explained in Appendix~\ref{appendix:software}.

We record snapshots at $z\in \{0,0.25,0.5,0.75,1\}$.
To estimate the real-space power spectrum we construct a density field using
cloud-in-cell interpolation of the particle locations. For the redshift-space power
spectra we adjust the location of each particle using Eq.~\eqref{eq:redshift-space-s}
and construct a density field from these adjusted locations.
This can be done in three different ways, by choosing the line-of-sight
to be oriented along each of the three axes of the simulation.
To reduce numerical noise in the power spectra our final comparisons
use an average of these three possibilities.

The amplitude of the real-space power spectrum is estimated by binning Fourier modes
of the density field and applying the anti-aliasing prescription of
Jeong \& Komatsu~\cite{Jeong:2008rj}.
The redshift-space power spectra are handled in the same way.
To extract multipoles we perform a least-squares fit to the expansion~\eqref{eq:multipole-Pk}
in each $\vect{k}$-bin.
Where the noise is small the outcome of this procedure closely matches the direct
projection~\eqref{eq:multipole-decompose}.
Where the noise is more significant, we find that the least-squares fit produces
more stable results.

\subsubsection{Difficulties encountered when simulating redshift-space distortions}
In the remainder of this paper we discuss only the $1024^3$-particle, $(2000\, \Mpc/h)^3$-side
simulation. However, to validate our numerical estimates we have tested their
convergence using a larger suite of simulations.
As part of this procedure we encounter two clear difficulties:

\begin{itemize}
    \item The transformation from $\delta$ to $\deltarsd$ 
    described by Eq.~\eqref{eq:delta-s} couples different scales:
    small-scale velocities can affect the redshift-space density on larger scales.
    While Eq.~\eqref{eq:delta-s} is not explicitly used to construct
    $\deltarsd$ from the simulation, our methodology will reproduce its
    effects.
    Therefore, accurate redshift-space power spectrum
    measurements
    require higher resolution simulations than those needed for the real-space
    power spectrum.

    \item The redshift-space power spectrum is sensitive to large-scale bulk flows,
    for which the sample variance is larger than the sample variance in the
    density field on the same scales. Therefore, if a simulation does not have
    sufficiently large volume, the redshift-space power spectrum will differ
    from the predictions of linear theory even on large scales.
    Similar issues were discussed in Jennings et al.~\cite{2011MNRAS.410.2081J}.
\end{itemize}

To understand these issues we analyse a set of simulations,
of which the most relevant are:
(\emph{a}) the $1024^3$-particle, $(2000 \, \Mpc/h)^3$-box simulation already mentioned;
(\emph{b}) a $1024^3$-particle, $(1000 \, \Mpc/h)^3$-box simulation;
(\emph{c}) a $1024^3$-particle, $(330 \, \Mpc/h)^3$-box simulation; and
(\emph{d}) a set of $512^3$-particle simulations of box size $(1000 \, \Mpc/h)^3$.

\para{Small-scale convergence}
We find that, for scales in the range $k =0.2 h/\Mpc$ to $1 h/\Mpc$,
the $1024^3$-particle, $330 \, \Mpc/h$-side simulation and the $1024^3$-particle,
$1000 \, \Mpc/h$-side simulation match closely.
We interpret this to indicate that velocities on scales smaller than those resolved by the
$1000 \, \Mpc/h$-side simulation do not contaminate the redshift-space density
for this $k$-range.
However, for the $\ell=2$ mode on the same scales, we observe a difference between
these high-resolution simulations and the $512^3$-particle, $1000 \, \Mpc/h$-side simulation.%
    \footnote{It is difficult to quantify the magnitude of this difference, because the
    $\ell=2$ mode undergoes a zero-crossing in the same range.
    However, at $k=0.3 h/\Mpc$, where the difference is largest (and close to the
    zero-crossing), the difference in amplitude between the lower- and higher-resolution
    simulations is $\sim 20\%$ of their shared value at $k=0.2 h/\Mpc$ (far from the
    crossing).}
This suggests that some effects due to the non-linear velocity field are
not captured by the resolution of our reference simulation.
These scales typically enclose a mass smaller than a Milky Way-sized
galaxy.
Therefore it is unclear whether observations resolve masses down to the scale
where these velocities become relevant.
A full investigation of these effects would require an analysis of the redshift-space
density field of \emph{halos}.
This is beyond the scope of the present analysis, where we study only the dark matter field.

\para{Large-scale convergence}
We find that all the simulations with box sizes smaller than our reference
$2000 \, \Mpc/h$-side simulation show increased scatter on the largest scales. For example,
the $1000 \, \Mpc/h$-side simulations exhibit a scatter of $\sim 30\%$ in the $\ell=2$
mode even for $k\lesssim 0.06 h/\Mpc$. We interpret this as a consequence of
slow convergence of the bulk flows in each simulation. In fact, even for our largest
$2000\, \Mpc/h$-side simulation, the scatter in the $\ell=4$ mode
is substantial on the largest scales.

The difficulty entailed by using a box size large enough to
suppress sample variance of the bulk flows, while retaining enough resolution
to capture the effect of non-linear velocities on small scales,
indicates that accurate simulations of redshift-space distortions
is computationally expensive.

\para{Covariance}
Finally, the variance in different $k$-bins of our redshift-space power spectra---%
primarily in the $\ell=2$ mode---%
appears to be correlated, even on large scales.
We suspect this occurs because a bulk flow that boosts the power spectrum at one scale
will provide a correlated uplift over a range of nearby scales.
Where precision fits are made to cosmological models this covariance should
be appropriately modelled and taken into account.

In practice this is unlikely to be straightforward.
To determine covariances accurately from simulations will
require many independent realizations,
even for a single cosmological model and choice of background parameters.
Determining how the power spectrum and its covariance changes over the entire parameter
range of multiple cosmological models will require very many more.
In this paper, the computational expense of performing these simulations means that
we do not address this issue. Instead, we assign uncorrelated error estimates
to each $k$-bin,
in order to assess general properties of the EFT prediction.
For precision work, however, the covariances should be taken into account.

\subsection{Results}
\label{sec:results}

In this section
we report our measurements of the counterterms at redshift $z=0$
where the effect of non-linearities is expected to be most pronounced.
We express the counterterms
in the $\mu^{2n}$ basis defined in~\eqref{eq:Prsd-counterterms}
and therefore quote values for the quantities
$\muctrterm{2}{2n} / \kNL^2$.%
    \footnote{It is a matter of convenience whether we renormalize
    by adjusting the counterterms $\muctrterm{2}{2n}$
    for the power-series expansion in $\mu$ [see~\eqref{eq:Prsd-counterterms}],
    or for the counterterms
    $\ellctrterm{2}{\ell}$
    defined for the Legendre-mode expansion [see~\eqref{eq:Legendre-counterterms}].
    In practice we will use the $\muctrterm{2}{2n}$ because
    $\muctrterm{2}{0}$ should coincide with the
    counterterm $\ctrterm{2}{\delta}$ obtained from renormalizing the
    real-space power spectrum.
    This provides a simple way to assess compatibility of the two procedures.
    In addition, it is
    straightforward to impose the
    constraints~\eqref{eq:ctrterm-condition-a}--\eqref{eq:ctrterm-condition-b}
    in this basis.}
Our parameter choices match those in~\S\ref{sec:rsd-resummation},
with an ultraviolet cutoff at $k = 1.4 h/\Mpc$.
The $X$ and $Y$ parameters used in the infrared resummation are
averaged between
$\qmin = 10 h^{-1} \, \Mpc$ and
$\qmax = 300 h^{-1} \, \Mpc$,
and their wavenumber integral is carried up to the same ultraviolet cutoff.
The non-linear measurements forming our renormalization conditions
are taken from
the $1024^3$-particle, $(2000/h \, \Mpc)^3$
simulation volume described in~\S\ref{sec:simulations}.
As explained in~\S\ref{sec:simulations},
we reduce noise on the redshift-space multipole measurements by averaging
over projections oriented along each of the three coordinate axes.

For most of this section we discuss only the $z=0$ results.
The EFT description for
redshifts $z > 0$ is considered in~\S\ref{sec:redshift-dependence}.

\subsubsection{Fitting for counterterms}
\begin{figure}
    \begin{center}
        \includegraphics{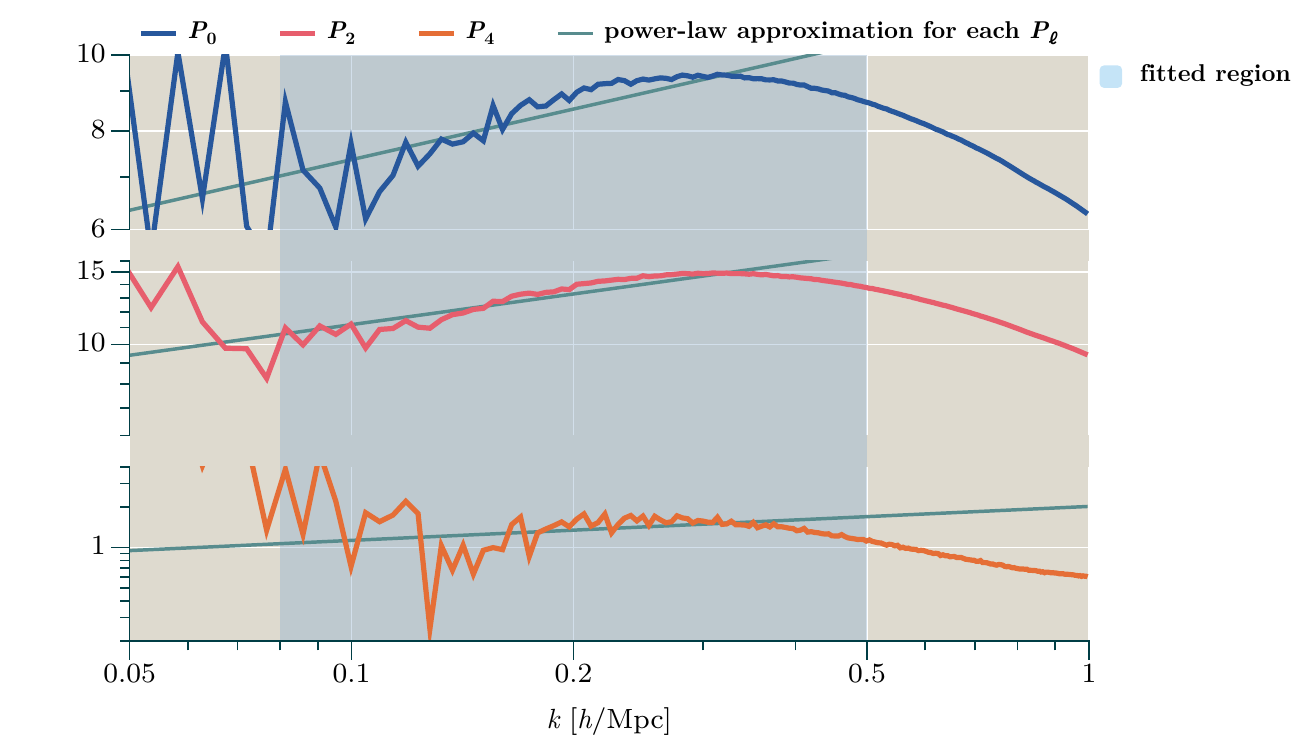}    
    \end{center}
    \caption{\label{fig:multipole-counterterms}Fitting counterterms for the multipole
    power spectra $P_\ell$ at $z=0$.
    Each panel shows the estimator $-(\PNL - P_\ell^{\SPT,\uptooneloop})/2k^2 D^2 \Pinit$,
    where $\PNL$ is the non-linear power spectrum for multipole $\ell$
    obtained from numerical simulations.
    The green lines show a least-squares power-law approximation to these
    estimators in the region $\kmin \leq k \leq \kmax$ where we
    optimize the fit (shaded light blue on the plot);
    the parameters of these fits appear in Table~\ref{table:counterterm-fits}.
    The choice of region to be used in the fit should be regarded as part of
    the renormalization scheme.}
\end{figure}

In Fig.~\ref{fig:multipole-counterterms}
we plot estimators for each $\ell$-counterterm at $z=0$,
together with power-law fits to the region included in the optimization.
As in~\S\ref{sec:real-space-results},
these estimators should be approximately $k$-independent
in a region where the difference
between the SPT and $N$-body power spectra
is described by lowest-order operator mixing.
We take this fitting region to $\kmin \leq k \leq \kmax$,
beginning roughly where the SPT prediction
becomes inaccurate and EFT counterterms are
required; cf.~\S\ref{sec:real-space-results}.
The parameters of the power-laws are shown in Table~\ref{table:counterterm-fits}.

\ctable[
    caption = {Power-law fits to the counterterm estimators
    of Fig.~\ref{fig:multipole-counterterms} over the region
    $\kmin \leq k \leq \kmax$ at $z=0$. The estimators
    are reasonably good fits to a constant, although not as good
    as in real space.},
    label = table:counterterm-fits
]{ll}{}
{
    \toprule
        \semibold{multipole} & \semibold{power-law fit} {\scriptsize [$\kmin \leq k \leq \kmax$]} \NN
        $P_0$ & $12.04 \times (k / h \times \Mpc)^{0.2135}$ \NN
        $P_2$ & $19.81 \times (k / h \times \Mpc)^{0.2484}$ \NN
        $P_4$ & $2.021 \times (k / h \times \Mpc)^{0.2531}$ \NN
    \bottomrule
}

The conclusion is that $P_0$ and $P_2$ can be reasonably well-described by
lowest-order operator mixing
in this region,
but the fit is not as good as in real space.
This may indicate that higher-derivative mixing,
stochastic counterterms,
or higher-order terms in the loop expansion
are already required for $k \gtrsim 0.1 h/\Mpc$.
Although we are not including stochastic counterterms in our analysis we have
verified that their contribution does not improve these fits.
For $k \gtrsim 0.5 h/\Mpc$ the estimators
show significant $k$-dependence, which
we interpret to mean that
higher-derivative mixing
or
higher-order loops
are definitively relevant.
Therefore, if predictions using only lowest-order mixing happen to
match the measured power spectra on these scales, this should be
regarded as accidental.
We caution that the fit for $P_4$ should be treated cautiously because our
estimates are noisy.
Extracting multipole power spectra from the $N$-body data
becomes increasingly difficult at high $\ell$.

\para{Numerical estimates and degeneracies}
To determine numerical values for the counterterms
$\muctrterm{2}{0}$, $\muctrterm{2}{2}$
and $\muctrterm{2}{4}$
we have the option to fit simultaneously to both real- and redshift-space
power spectra, or just to the redshift-space measurements.
In Table~\ref{table:MLE-counterterms}
we list
the maximum likelihood estimates for each case.
\ctable[
    caption = {Maximum-likelihood estimates for the
    counterterms $\muctrterm{2}{0}$, $\muctrterm{2}{2}$, $\muctrterm{2}{4}$
    at $z=0$.
    As explained in~\S\ref{sec:simulations},
    for reasons of computational expense
    we do not include covariances between $k$-bins of
    the different power spectra,
    but instead assign $5\%$ uncorrelated errors to each bin.
    However, the results are not strongly sensitive to the size of the error bar we
    assume.},
    label = table:MLE-counterterms
]{llllll}{
    \tnote[a]{\tiny Ref.~\cite{Lewandowski:2015ziq}. We have converted
    their results using a growth factor $f$ matching the
    \href{https://www.cosmosim.org/cms/simulations/bigmdpl/}{Big MultiDark Planck simulation}~\cite{2016MNRAS.457.4340K},
    which was used to estimate the non-linear multipole power spectra
    used as renormalization conditions in this reference.}
}
{
    \toprule
        & \multicolumn{2}{c}{$P$ + $P_\ell$} & \semibold{$P_\ell$ only} 
        & \semibold{$P$ only} & \setcounter{rownum}{0}\semibold{Lewandowski et al.}\tmark[a] \\ \cmidrule(r{.75em}l){2-3}
        & \semibold{resummed} & \semibold{unresummed} & & & \NN
        $\muctrterm{2}{0}/\kNL^2$ $[h^{-2} \, \Mpc^2]$
            & $2.52$ & $2.50$ & $2.52$ & $2.51$ & $1.26$ \NN
        $\muctrterm{2}{2}/\kNL^2$ $[h^{-2} \, \Mpc^2]$
            & $16.2$ & $15.9$ & $16.2$ & & $4.08$ \NN
        $\muctrterm{2}{4}/\kNL^2$ $[h^{-2} \, \Mpc^2]$
            & $6.91$ & $6.69$ & $6.91$ & & $2.03$ \NN
    \bottomrule
}
Marginalized constraints obtained from a Monte Carlo Markov chain
analysis
give similar values.
(For an MCMC analysis we assume a wide, flat prior on each parameter
that comfortably encloses the posterior parameter range.)
Notice that 
$\muctrterm{2}{0}/\kNL^2 = \ctrterm{2}{\delta}/\kNL^2$ is
not equal to the value derived
for this counterterm in~\S\ref{sec:real-space-results}
by renormalizing against the Planck2015 cosmology.
The measured value is cosmology-dependent~\cite{Cataneo:2016suz}.

Our results depend weakly on the $k$-range used in the fit.
Increasing the lower limit
$k_{\text{min}} = 0.1 h/\Mpc$
to $0.2 h/\Mpc$ adjusts $\muctrterm{2}{0}$ by $\sim 10\%$,
but $\muctrterm{2}{2}$ and $\muctrterm{2}{5}$ by only $\sim 1\%$.
We do not quote error estimates for these
counterterms because that would require an estimate for the covariance
between the measured $P(k)$ and $P_\ell(k)$.
As explained in~\S\ref{sec:simulations},
to obtain reliable
estimates of these covariance matrices would require more simulations than
we were able to perform.
We hope to return to this issue in the future.

\begin{figure}
    \begin{center}
        \includegraphics{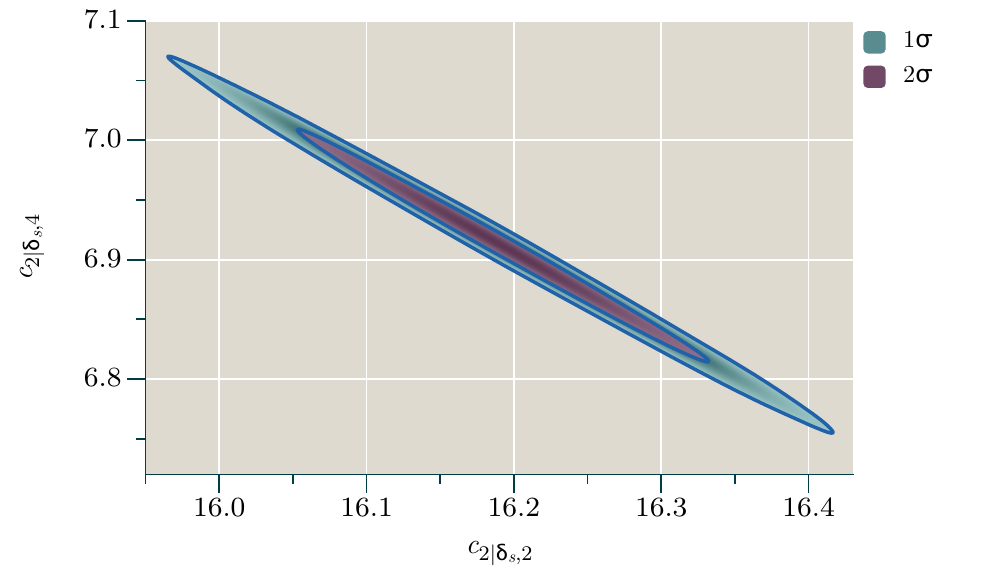}    
    \end{center}
    \caption{\label{fig:multipole-degeneracy}Representative
    degeneracy between $\muctrterm{2}{2}$
    and $\muctrterm{2}{4}$ counterterms for the redshift-space
    power spectrum.
    The size of the error ellipses is merely indicative;
    they are constructed assuming a wide, flat prior on each parameter
    and taking independent errors of $5\%$ and $20\%$ per $k$-bin
    for the real- and
    redshift-space power spectra, respectively. (See the discussion in the main text.)
    With these assumptions, the green ellipse encloses the two-$\sigma$
    region and the purple ellipse encloses the one-$\sigma$ region.}
\end{figure}

Under our assumptions, the $\mu^0$ counterterm $\muctrterm{2}{0}$ is well-determined
no matter which measurements we choose to include in the fit.
One linear combination of $\mu^2$ and $\mu^4$
counterterms $\muctrterm{2}{2}$ and $\muctrterm{2}{4}$
is tightly constrained, whereas the orthogonal combination
exhibits a similar uncertainty to $\muctrterm{2}{0}$; see
Fig.~\ref{fig:multipole-degeneracy}.
In this plot we exhibit representative one- and two-$\sigma$
contours showing the shape of this degeneracy,
computed assuming independent errors of $5\%$ per bin on the real-space power
spectrum, and $20\%$ per bin on each multipole.%
    \footnote{Notice that this error assignment for the $P_\ell$ is larger
    than that used to construct Table~\ref{table:MLE-counterterms}.
    This has been done in order to resolve the contours more clearly.
    Using these larger estimates shift the maximum-likelihood
    estimate for
    $\muctrterm{2}{0} / \kNL^2$ by $-0.02 h^{-2} \Mpc^2$
    but leaves
    $\muctrterm{2}{2} / \kNL^2$ and
    $\muctrterm{2}{4} / \kNL^2$ unchanged,
    and therefore makes
    negligible difference to the predicted power spectra.}
These are merely fiducial values,
so we caution that the \emph{size} of these error ellipses
has little meaning.

The degeneracy represents what would occur in the ideal scenario
that the $\ell = 0, 2, 4$ multipoles can all be measured equally well.
The linear combination $\muctrterm{2}{2} - \muctrterm{2}{4}$ appears
with greatest significance
in the amplitude of the $\ell=4$ counterterm, and therefore inherits the
most stringent
constraint because
$P_4$ has substantially smaller amplitude than $P_0$ or $P_2$.
In a realistic present-day scenario where $P_4$ carries largest measurement uncertainty
we would expect this degeneracy to be relaxed.

\subsubsection{Accuracy of EFT prediction}
\label{sec:EFT-accuracy}
\begin{figure}
    \begin{center}
        \includegraphics{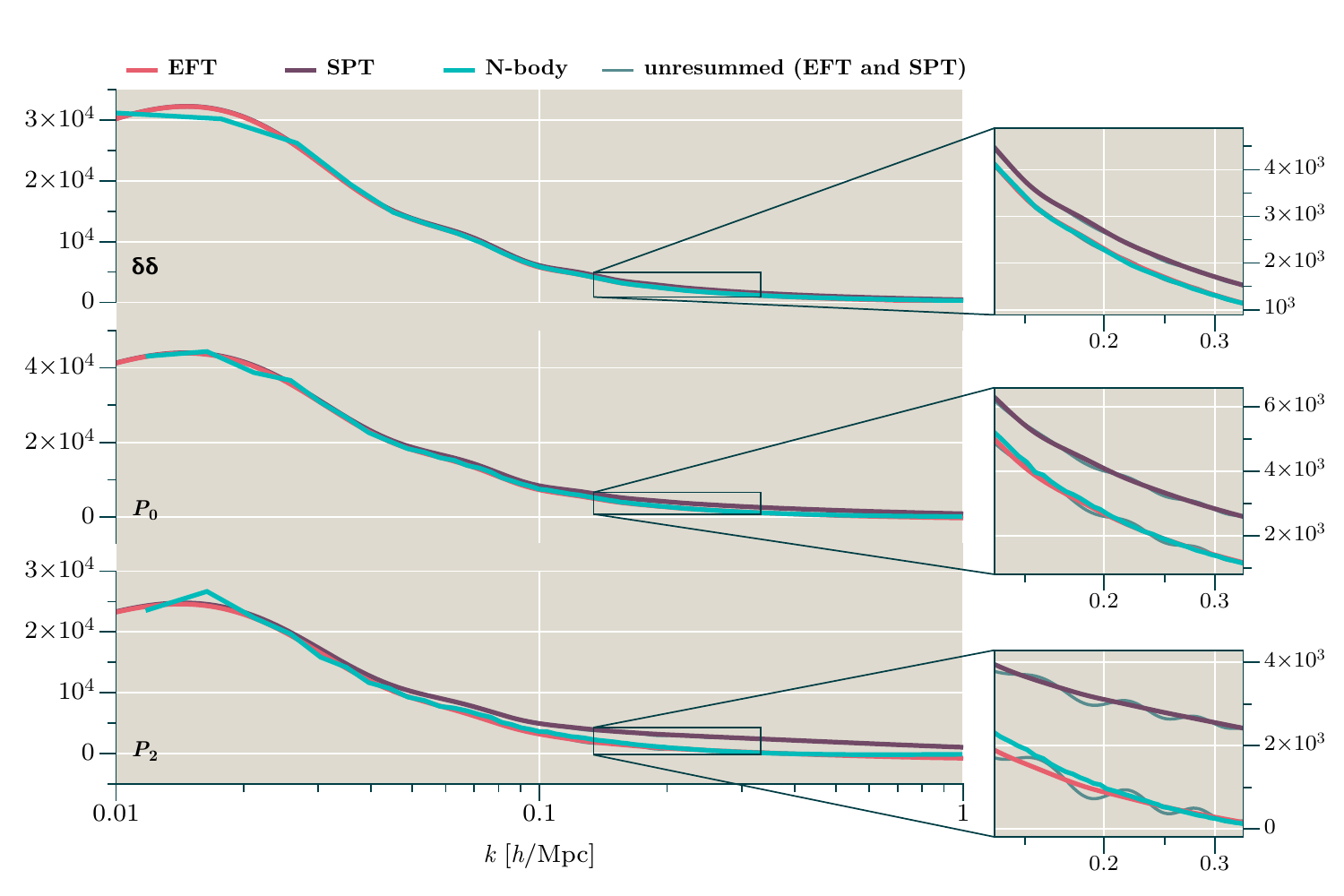}    
    \end{center}
    \caption{\label{fig:multipole-results}Real-space
    and multipole power spectra $P_0$ and $P_2$
    predicted by the effective field theory framework
    (red lines), compared to the
    predictions in SPT (purple lines).
    The evaluation is at $z=0$.
    The turquoise lines represent the non-linear power spectra recovered
    from $N$-body simulations.
    For both the EFT and SPT predictions, the associated green line shows
    the unresummed result.
    The zoomed panels highlight regions where resummation plays a significant
    role in improving the prediction for $P_0$ and $P_2$.
    As in~\S\ref{sec:real-space-results},
    its importance for the real-space power spectrum is modest.}
\end{figure}

\begin{figure}
    \begin{center}
        \includegraphics{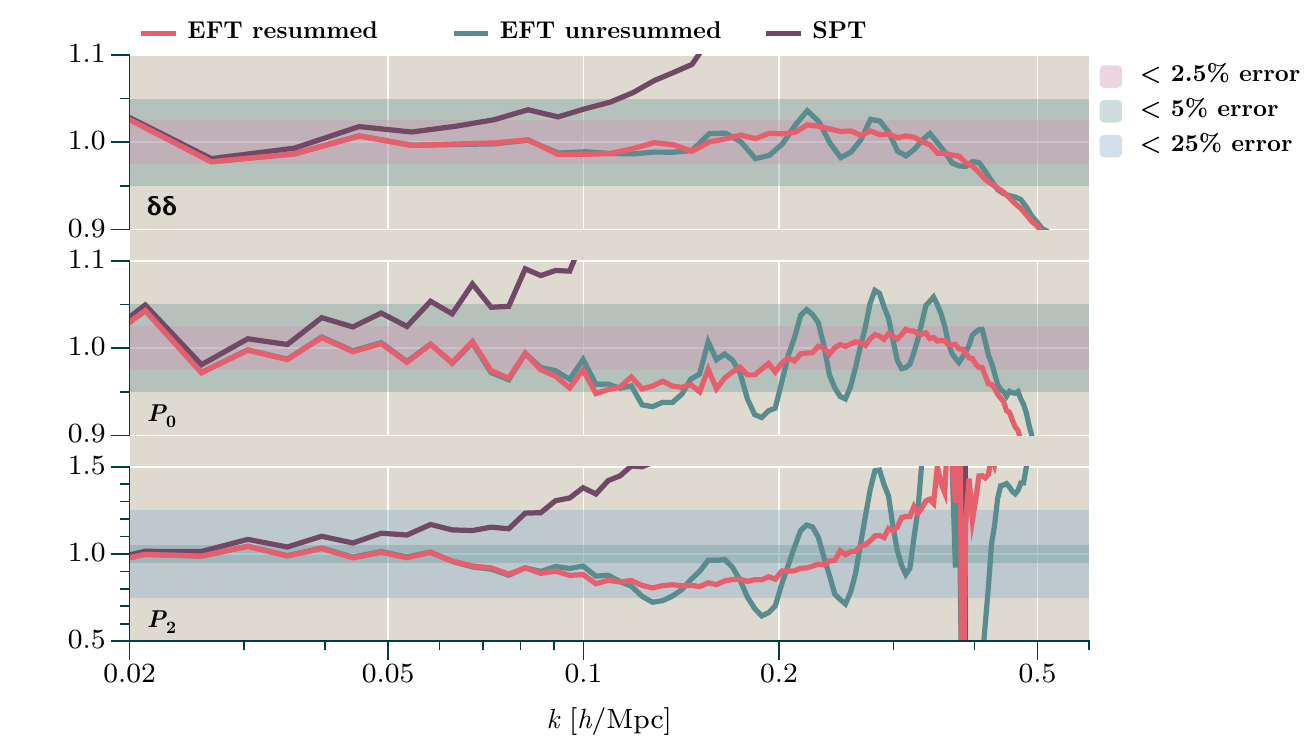}    
    \end{center}
    \caption{\label{fig:multipole-comparison}Comparison of fit
    between the predicted one-loop EFT power spectra
    and the non-linear power spectra at $z=0$.
    The EFT predictions are shown in both resummed (red line) and non-resummed
    (green line) versions.
    Corresponding results from non-resummed SPT are included
    for comparison.
    The quantity plotted is $P/\PNL$ and the background cosmology
    matches the
    \href{https://www.cosmosim.org/cms/simulations/mdr1/}{MDR1 MultiDark simulation}~\cite{Prada:2011jf}.
    The shaded light-pink region marks where the
    prediction is within $2.5\%$ of the measured value;
    the light-green region marks where it is within $5\%$;
    and the light-blue region marks where it is within $25\%$.}
\end{figure}

In Fig.~\ref{fig:multipole-results}
we plot results for the real-space power spectrum and the
$P_0$ and $P_2$ modes using both SPT and renormalized EFT.
In Fig.~\ref{fig:multipole-comparison}
we show the relative accuracy achieved by
each prediction for the same spectra.
In both figures
the EFT power spectra are taken to be constructed using the
$P + P_\ell$ counterterms from Table~\ref{table:MLE-counterterms}.
Red lines indicate the resummed EFT prediction;
for comparison,
their unresummed counterparts are shown in green.
We also plot the unresummed SPT prediction in purple.
The shaded regions indicate where the prediction has $\leq 2.5\%$
accuracy (light pink), $\leq 5\%$ accuracy
(light green) and $\leq 25\%$ accuracy (light blue).

\para{Real-space $P(k)$ and redshift-space $\ell=0$ mode}
As for the real-space power spectrum, the resummed prediction gives better
accuracy for $k \gtrsim 0.1 h/\Mpc$ where SPT tends to overpredict the
amplitude of baryon oscillations. Although the effect is visible in both
real space and redshift space, it is more visible in the
higher redshift-space multipoles.
The general performance of the one-loop renormalized result is good.
The resummed, renormalized real-space power spectrum is typically within $2.5\%$
of the measured value up to $k \sim 0.4 h/\Mpc$.
The performance of the $P_0$ mode is similar but marginally less good, with some
excursions into the $5\%$ accuracy band for
$0.1 h/\Mpc \lesssim k \lesssim 0.2 h/\Mpc$.

Both the real-space power spectrum and $P_0$ exhibit a downturn near $k \sim 0.5 h/\Mpc$,
dipping significantly below the measured non-linear result.
In the case of $P_0$ the power spectrum becomes negative near
$k \approx 0.74 h/\Mpc$. This is unphysical because the monopole power spectrum should
be positive, and therefore its zero-crossing must be removed
by higher-order loop corrections
or higher-derivative mixing that we have not included.
Fig.~\ref{fig:multipole-counterterms} already suggests that
such contributions
become important for $k \gtrsim 0.5 h/\Mpc$,
but requiring positivity of $P_0$ implies that we may deduce
a firm upper limit for the validity of one-loop EFT predictions using only
the leading-order counterterm.
For the MDR1 cosmology considered here,
higher-order effects must become
significant before $k \approx 0.74 h/\Mpc$.

\begin{figure}
    \begin{center}
        \includegraphics{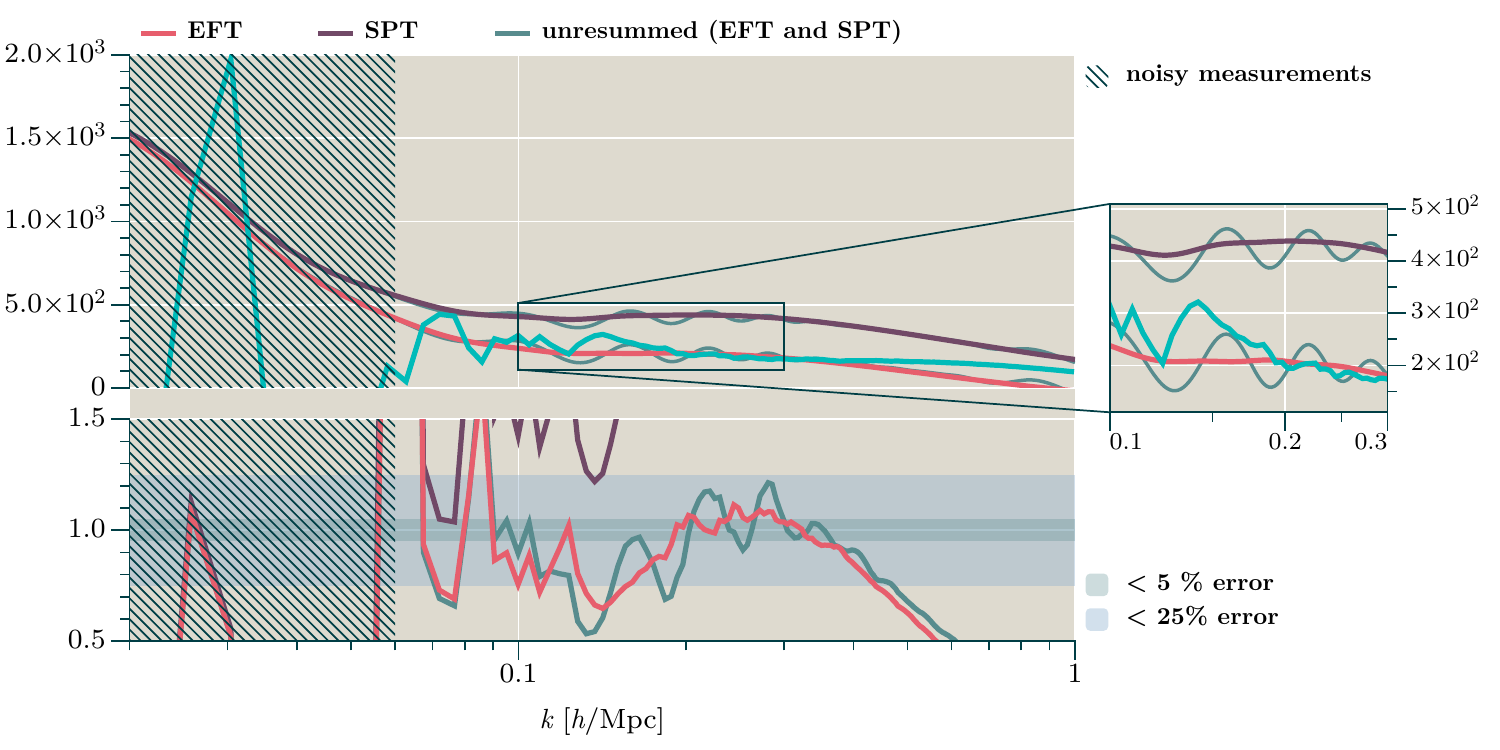}    
    \end{center}
    \caption{\label{fig:P4-fit}\semibold{Top panel}: smoothed
    $P_4$ power spectrum measured from simulations
    (turquoise line) compared to EFT predictions (red line) and
    SPT (purple line). Green lines show the corresponding
    unresummed predictions. This panel should be compared to
    Fig.~\ref{fig:multipole-results}.
    Note that although we plot the comparison
    to a smoothed power spectrum, our
    fit for the counterterms uses raw measurements.
    \semibold{Bottom panel}: relative accuracy of the $P_4$ mode compared
    to the smoothed $N$-body power spectrum.
    This panel should be compared to Fig.~\ref{fig:multipole-comparison}.
    The hatched region $k < 0.06 h/\Mpc$
    marks where we believe the measurements are
    too poorly-determined for a comparison to be meaningful.}
\end{figure}

\para{Redshift-space $\ell=2$ mode}
For $P_2$, which is more strongly sensitive to velocity information,
the EFT prediction is still typically within $25\%$ of the measured value up to
$k \sim 0.4 h/\Mpc$.
(This number should be interpreted in light of the discussion in the following
paragraph.)
The feature near $k=0.4 h/\Mpc$ in $P_2$ arises from
a sign change where the $\ell=2$ mode becomes negative.
Unlike the $\ell=0$ mode, this sign change is physical~\cite{Cole:1993kh}.
It occurs at slightly different locations for the predicted and measured
power spectra, causing the relative error to diverge.
This divergence is therefore an artefact of the plot
and does not have real physical significance.
We collect the $\ell=4$ results separately in Fig.~\ref{fig:P4-fit}.
They show similar accuracy to the $\ell=2$ mode for
$0.1 h/\Mpc \lesssim k \lesssim 0.4 h/\Mpc$, but at smaller
$k$ they
are too noisy to allow a meaningful comparison.

Based on inspection of
Figs.~\ref{fig:multipole-comparison} and~\ref{fig:P4-fit},
it may appear that we achieve only modest accuracy for $P_2$ and $P_4$.
While this is true for the \emph{relative} accuracy of the prediction,
it should be noted that the improvement compared to SPT is dramatic.
However, Fig.~\ref{fig:multipole-results}
clearly shows that the amplitude of the one-loop SPT estimate must be
adjusted \emph{significantly} downward
in the quasilinear region
in order to achieve
an acceptable prediction.
A similar effect was observed by Taruya, Nishimichi \& Saito~\cite{Taruya:2010mx},
who compared $N$-body
simulations with the predictions of an `improved' perturbation theory
intended to damp acoustic oscillations in a similar way to the
resummation schemes discussed in \S\S\ref{sec:resummation-methods}
and~\ref{sec:rsd-resummation}.
Bearing this in mind,
the renormalized, resummed EFT prediction is strikingly
successful in matching the amplitude of $P_2$ and $P_4$
for quasilinear $k$.
This is especially true at $\ell=4$, for which small variations in the
counterterms leave a good match to $P_0$ and $P_2$ but
push the $P_4$ prediction well away from its measured value.
Therefore
obtaining a reasonable match to $P_0$, $P_2$ and $P_4$ simultaneously
is nontrivial.
Despite this optimistic result,
it seems clear that
matching the $\ell = 2, 4$ modes to $\lesssim 5\%$
almost certainly requires inclusion
of higher-order loop contributions and counterterms.
A similar conclusion was reached by
Lewandowski et al.~\cite{Lewandowski:2015ziq}.

Notice also that amplitudes of the $P_\ell$ become
quite small, which inflates the significance of the relative error.
Indeed, as stated above,
the measured $P_2$ changes sign: this is a consequence of
suppression due to the fingers-of-God effect~\cite{Cole:1993kh}.
In our framework this sign change is not present before renormalization.%
    \footnote{The sign change occurs near $k = 0.38 h/\Mpc$, and is therefore
    well before the scale $k = 0.74 h/\Mpc$ where we have reason to believe
    the linear counterterm leads to oversubtraction.}
Its appearance in the
final result is entirely attributable to parametrization of small-scale
physics by counterterms,
as anticipated in the discussion of~\S\ref{sec:rsd-resummation}.

\subsubsection{Accuracy of Einstein--de Sitter approximation}
\begin{figure}
	\begin{center}
		\includegraphics{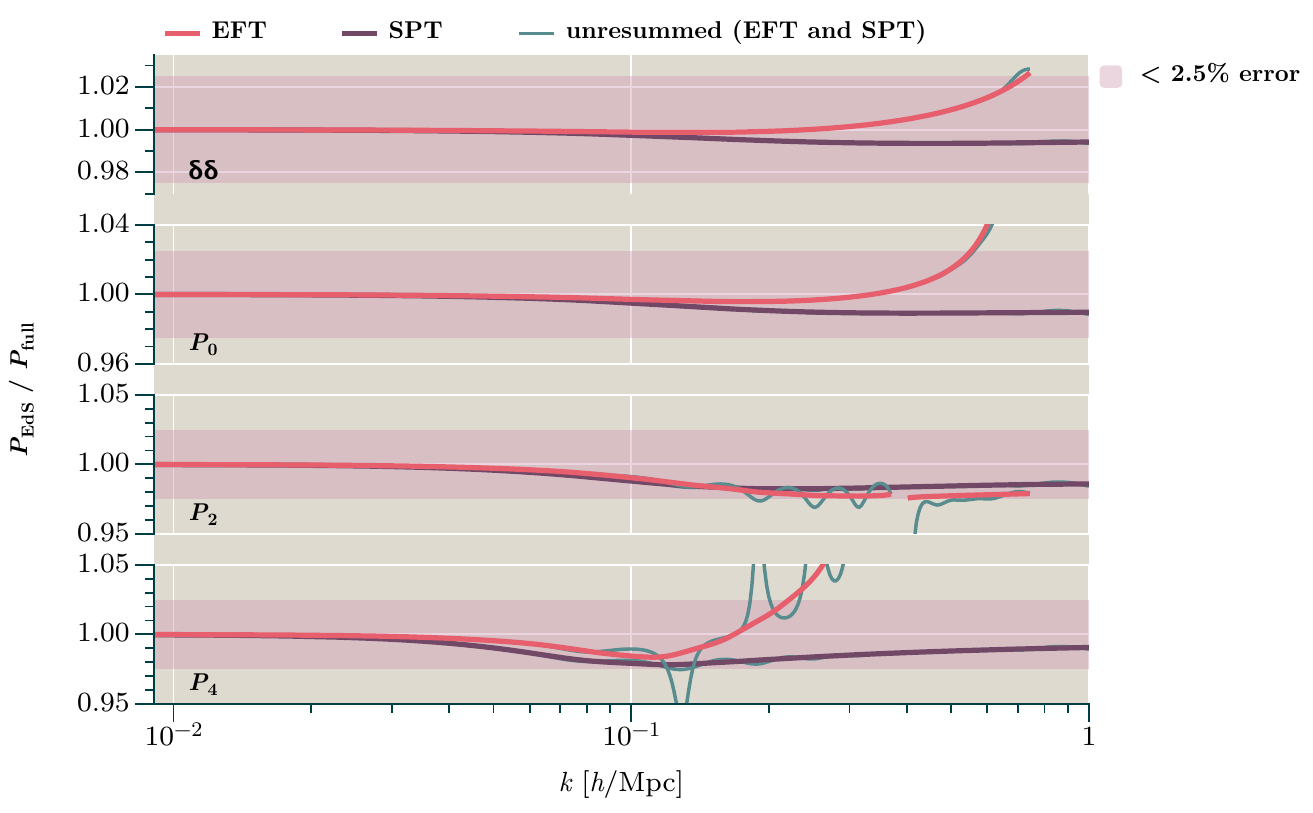}	
	\end{center}
	\caption{\label{fig:EdS-comparison}Comparison of Einstein--de Sitter
	approximation with full time-dependence (at $z=0$)
	for the effective field theory prediction (red lines)
	and SPT (purple lines).
	In each case, the associated green line shows the unresummed result.
	The plotted quantity is $P_{\text{EdS}} / P_{\text{full}}$,
	where $P_{\text{EdS}}$ is computed in the Einstein--de Sitter approximation
	and $P_{\text{full}}$ is the result including the full one-loop
	time dependence.
	EFT lines are cut off for $k > 0.74 h/\Mpc$ where the leading-order
	counterterm must be supplemented by higher-order contributions.
	For the EdS--EFT prediction we make a separate fit for the counterterms;
	they are $\muctrterm{2}{0} = 2.44 h^{-2} \, \Mpc^2$,
	$\muctrterm{2}{2} = 16.0 h^{-2} \, \Mpc^2$,
	$\muctrterm{2}{4} = 6.76 h^{-2} \, \Mpc^2$ in the resummed case, and
	$\muctrterm{2}{0} = 2.43 h^{-2} \, \Mpc^2$,
	$\muctrterm{2}{2} = 15.7 h^{-2} \, \Mpc^2$
	and $\muctrterm{2}{4} = 6.54 h^{-2} \, \Mpc^2$ in the unresummed case.
	The light-pink shaded region marks where the Einstein--de Sitter
	approximation is within $2.5\%$ of the prediction using the
	full time-dependence.}
\end{figure}

Since we retain the full time-dependence of the one-loop
redshift-space power spectrum
it is possible to assess the accuracy of the
Einstein--de Sitter approximation.
As discussed in~\S\ref{sec:eulerian-perturbation-theory},
this consists in replacing the growth functions $D_i$
and growth factors $f_i$ with their
counterparts from Table~\ref{table:EdS-growth}.
In Fig.~\ref{fig:EdS-comparison}
we show the relative accuracy of the Einstein--de Sitter approximation
for the real-space power spectrum, and the $\ell=0, 2, 4$
modes of the redshift-space power spectrum.

For the real-space power spectrum, and the $\ell = 0, 2$ multipoles,
the Einstein--de Sitter approximation is excellent up to
$k \sim 0.1 h/\Mpc$.
For $P_4$ it is excellent up to
$k \sim 0.05 h/\Mpc$.
For larger $k$ the EdS approximation
marginally underpredicts the
amplitude of the 1-loop SPT power spectrum.
The sign of the effect
can be understood by comparison with Fig.~\ref{fig:EdS-growth},
which shows that the largest effect of retaining the full
time dependence is a $\sim 2\%$ enhancement for $\DF$ and $\DG$.

This underprediction does not automatically translate into an
underprediction for the EFT power spectrum, because in principle
the
counterterm subtractions
can mask the effect.
Some compensation is visible for both the real-space
power spectrum and $P_0$.
In each case the EFT prediction
using the Einstein--de Sitter approximation
is very close to EFT prediction using the
full time dependence, up to values of $k$ where
oversubtraction from the leading-order counterterm
becomes problematic.
Near these scales the Einstein--de Sitter approximation
begins to \emph{relatively}
overpredict $P_0$, because the zero-crossing
point occurs at smaller $k$ if the full time dependence
is used.
This causes an unphysical divergence in the relative
error, for the same reasons outline above.
This dramatic feature
would not survive if higher-order
counterterms were introduced, and so its presence should
be treated with caution.

For $P_2$ and $P_4$ the EFT
subtractions do not completely absorb the error in the
Einstein--de Sitter approximation.
For each of these multipoles the EFT
power spectrum has a net $\sim 2\%$
underprediction
in the region
$k \gtrsim 0.1 h/\Mpc$.
The size of this error is
somewhat smaller than the relative error in the EFT prediction
itself; see Fig.~\ref{fig:multipole-comparison}.
However, if predictions at the $\lesssim 5\%$ level are required
for $P_2$ and $P_4$,
we conclude that the Einstein--de Sitter approximation would
no longer be acceptable.
A very similar conclusion was reached by
Fasiello \& Vlah~\cite{Fasiello:2016qpn}.

\subsubsection{Redshift dependence}
\label{sec:redshift-dependence}

Finally, we consider the EFT prediction for $z > 0$. At high redshift we expect
non-linearities to be less significant, and therefore the net contribution of
the counterterms to be smaller.

\begin{figure}
    \begin{center}
        \raisebox{-0.5\height}{\includegraphics{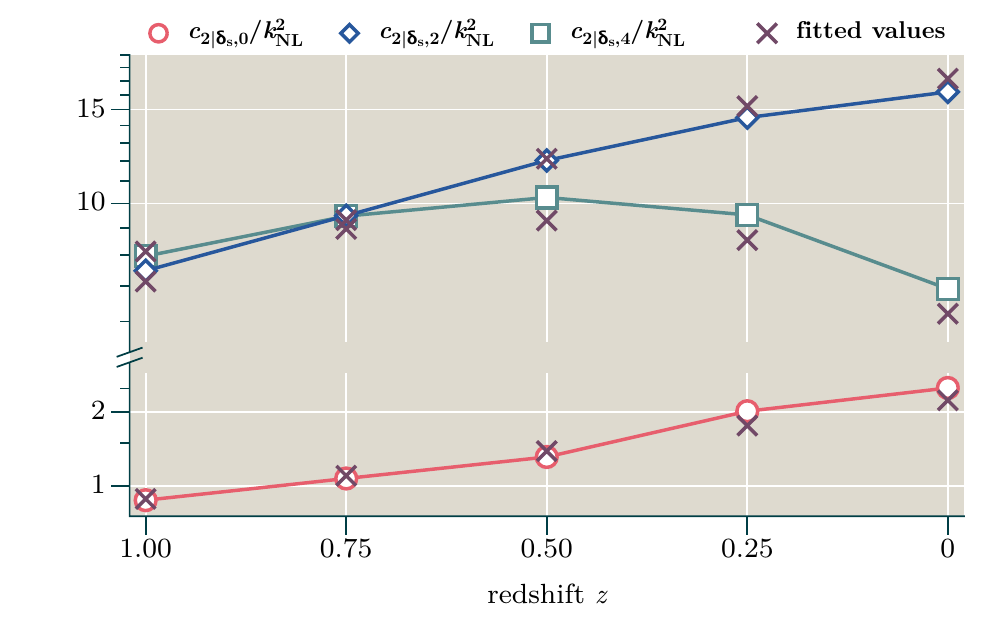}}
    \end{center}
    \caption{\label{fig:z-dependence}
    Time dependence of EFT counterterms.
    The plotted values are taken from Table~\ref{table:z-dependence}.
    The points marked `fitted values' match the time
    dependence predicted from
    Eqs.~\eqref{eq:ctrterms-Zbasis-0}--\eqref{eq:ctrterms-Zbasis-8}
    assuming no unknown ultraviolet contributions,
    with values for the constants $Z_{2|i}$ taken from
    Table~\ref{table:z-values}.}
\end{figure}

To determine how the counterterms vary with redshift, we extract
power spectra at $z \in \{ 0.25, 0.5, 0.75, 1 \}$ to accompany the results at
$z=0$ described above. Fitting for the counterterms independently at each redshift
yields the results of Table~\ref{table:z-dependence}, which we plot in
Fig.~\ref{fig:z-dependence}.
In this redshift interval, both the $\mu^0$ and $\mu^2$ counterterm are
increasing. In comparison the $\mu^4$ counterterm is very roughly stable,
becoming marginally more important at intermediate redshifts $z \sim 0.5$.

It was explained in~\S\ref{sec:ultraviolet-sensitivity} that the time dependence
of the counterterms is not predicted by the effective theory,
because by construction their values depend on the evolution of modes that are not
adequately described by the low-energy theory.
Nevertheless, one can ask whether the redshift dependence of Table~\ref{table:z-dependence}
\emph{requires} new types of time dependence beyond what is visible in the
perturbative description, or whether the perturbative description could already be
sufficient.
For example, virialized modes are believed to decouple completely from the
evolution of perturbations at low wavenumber except for a small renormalization
of the background~\cite{Baumann:2010tm}.
If this decoupling persists to large enough scales the net effect might be
equivalent to a cutoff on the loop integrals at a fairly modest wavenumber,
low enough that the time dependence predicted by perturbation theory is not yet
inadequate
(excepting the possibility of non-local
memory effects~\cite{Carroll:2013oxa,Fuhrer:2015cia}).
A discussion of the time dependence of the counterterms in the context
of the Einstein--de Sitter approximation was previously
given by
Hertzberg~\cite{Hertzberg:2012qn}.

\begin{figure}
    \begin{center}
        \raisebox{-0.5\height}{\includegraphics{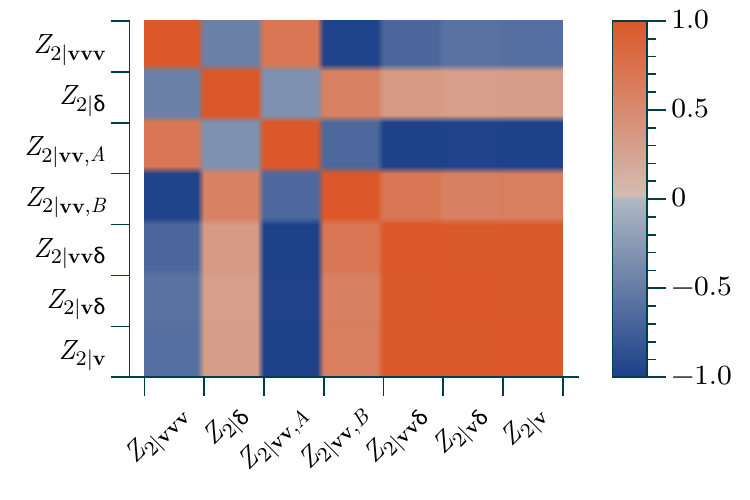}}
    \end{center}
    \caption{\label{fig:z-correlation}Correlation matrix for the $Z_{2|i}$.}
\end{figure}

To check whether Table~\ref{table:z-dependence} is compatible with the
perturbative prediction for ultraviolet contributions to the loop integrals,
we perform a global fit for the parameters
$\ctrconst{2}{\delta}$,
$\ctrconst{2}{\vect{v}}$,
$\ctrconst{2}{\vect{v}\delta}$,
$\ctrconst{2}{\vect{v}\vect{v},A}$,
$\ctrconst{2}{\vect{v}\vect{v},B}$,
$\ctrconst{2}{\vect{v}\vect{v}\delta}$
and
$\ctrconst{2}{\vect{v}\vect{v}\vect{v}}$,
assuming all unpredicted ultraviolet time dependence to be absent.
We use the same error estimate of $5\%$ in each $k$-bin
used to measure the $\muctrterm{2}{2n}$
and impose a flat prior over the interval $[-1, 1]$ on each parameter.
We give the marginalized posterior
parameter values in Table~\ref{table:z-values}
and
plot the $\muctrterm{2}{2n}$ predicted by these values as the points
marked `fitted values' in Fig.~\ref{fig:z-dependence}.
The fit matches the measured values closely. Notice that
under the conditions used to perform the fit,
$\muctrterm{2}{0}$ is determined entirely by
$\ctrconst{2}{\delta}$
and therefore
Fig.~\ref{fig:z-dependence} shows that---in conjunction with the
perturbatively-predicted time-dependent factors---this single
parameter is enough to fit all five data points accurately.
The lines for
$\muctrterm{2}{2}$
and
$\muctrterm{2}{4}$ depend on all seven $Z$-parameters, but it is still
nontrivial that an accurately-fitting combination can be found to match
the ten sample points.
We find that there are degeneracies between groups of the $Z$ parameters.
Their correlation matrix is plotted in Fig.~\ref{fig:z-correlation}.
(We do not report error bars for the $Z_{2|i}$
for the same reason discussed above, that
we do not have reliable estimates of the covariance between our measured
power spectra.)
The values we have reported include the full time-dependence at one-loop,
but the performance of the Einstein--de Sitter approximation is comparable.

It is not possible to draw strong conclusions from this analysis.
To the degree that the simulations provide a description of dark-matter clustering
in the real universe, there seems no evidence that the time-dependence of
deeply ultraviolet modes strongly influences the evolution of modes within the EFT.
To some extent, however, this outcome was already embedded in the simulations
because
these assume that feedback from gas dynamics and other unmodelled baryonic processes
does not significantly influence the clustering of modes on much larger scales.

The normalization of the $Z$-parameters is chosen so that, in the perturbative
description, they equal the common value
\begin{equation}
    Z = \frac{1}{15\pi^2} \int \d q \; \Pinit(q) .    
\end{equation}
Although this is a firm prediction of perturbation theory, we would normally
disregard it. The values assumed by the $Z$s make a statement about the
ultraviolet completion, and any such statements
derived from the low-energy
theory alone cannot be trustworthy.
Nevertheless, if the time-dependence predicted by perturbation theory is accurate
one might wonder whether the ultraviolet modes decouple to the extent that
approximate equality of the $Z$s is restored.
However, inspection of the values in Table~\ref{table:z-dependence} shows that
this is not the case.

\ctable[
    caption = {Variation of counterterms with redshift. We fit the
    resummed prediction to the
    real-space power spectrum and the $\ell = 0, 2, 4$
    multipoles over the
    region $\kmin \leq k \leq \kmax$ at each redshift.},
    label = table:z-dependence
]{llllll}{}
{
    \toprule
        \semibold{counterterm} & $z=0$ & $z=0.25$ & $z=0.5$ & $z=0.75$ & $z=1$ \NN
        $\muctrterm{2}{0}$ $[h^{-2} \, \Mpc^2]$
            & $2.52$ & $2.02$ & $1.31$ & $1.07$ & $0.873$ \NN
        $\muctrterm{2}{2}$ $[h^{-2} \, \Mpc^2]$
            & $16.2$ & $14.5$ & $12.0$ & $9.48$ & $7.47$ \NN
        $\muctrterm{2}{4}$ $[h^{-2} \, \Mpc^2]$
            & $6.91$ & $9.51$ & $10.3$ & $9.46$ & $7.96$ \NN
    \bottomrule
}
\ctable[
    caption = {Global fit for the $Z$-parameters. We fit simultaneously to
    measurements of the real-space power spectrum and the $\ell = 0, 2, 4$
    multipoles measured from our simulations
    at redshifts $z \in \{ 0, 0.25, 0.5, 0.75, 1 \}$.
    All values are reported in units of $h^{-2} \, \Mpc^2$.},
    label = table:z-values
]{ccccccc}{}
{
    \toprule
        $\ctrconst{2}{\delta}$ &
        $\ctrconst{2}{\vect{v}}$ &
        $\ctrconst{2}{\vect{v}\delta}$ &
        $\ctrconst{2}{\vect{v}\vect{v},A}$ &
        $\ctrconst{2}{\vect{v}\vect{v},B}$ &
        $\ctrconst{2}{\vect{v}\vect{v}\delta}$ &
        $\ctrconst{2}{\vect{v}\vect{v}\vect{v}}$ \NN
        $1.1 \times 10^{-3}$ &
        $-1.5 \times 10^{-2}$ &
        $-7.6 \times 10^{-2}$ &
        $-1.1 \times 10^{-2}$ &
        $5.9 \times 10^{-3}$ &
        $3.1 \times 10^{-2}$ &
        $-1.2 \times 10^{-2}$ \NN
    \bottomrule
}

\section{Conclusions}
\label{sec:conclusions}

In this paper we have presented the complete one-loop renormalization of the
redshift-space power spectrum and its Legendre multipoles
$\ell = 0$, $\ell = 2$ and $\ell=4$.
The same principles apply to modes with $\ell \geq 6$,
but our numerical results for the hexadecapole 
are already noisy
and present-day observational constraints on this multipole are
not yet competitive with the monopole or quadrupole.

The outcome of a similar renormalization has already been reported by
Lewandowski et al.~\cite{Lewandowski:2015ziq}
in an approximation where all growth functions are replaced by their
Einstein--de Sitter counterparts.
In this paper we include the exact time dependence for the first time,
showing that---at least within the EFT framework, although not for SPT---it
is an excellent approximation for the
real-space power spectrum and $\ell=0$ mode, but leads to $\sim 2\%$
errors in the $\ell=2, 4$ modes for $k \gtrsim 0.1 h/\Mpc$.
Results including exact time dependence were given in
Refs.~\cite{Carrasco:2012cv,Lewandowski:2017kes},
and applied to redshift space in models more general than {\LCDM} by
Fasiello \& Vlah~\cite{Fasiello:2016qpn}.
However, because they did not commit to a specific scenario they quoted their
results as unevaluated quadratures.
The explicit time dependence
of the SPT redshift-space power spectrum
is a new result.

\para{Comparison with previous results}
Our formalism is broadly in agreement with the methods used in
Refs.~\cite{Senatore:2014vja,Lewandowski:2015ziq,Perko:2016puo}.
In our presentation we have emphasized the role of the counterterms
in parametrizing ultraviolet contributions to loop integrals,
rather than
arising from smoothed equations of motion.
The resulting language is closer to familiar applications of
effective field theory in particle physics.
In addition,
our formalism differs from
that presented by Lewandowski et al.~\cite{Lewandowski:2015ziq}
in certain technical details, and in our procedure to fit for
the counterterms of the redshift-space power spectrum.

For a cosmology matching the MultiDark MDR1 simulation,
we find that the real-space power spectrum and the
$\ell=0$ mode of the redshift-space power spectrum can be matched
within $\sim 5\%$
using the leading EFT counterterm
up to roughly $k \lesssim 0.4 h/\Mpc$ at $z=0$,
and with a firm upper limit of $k \lesssim 0.74 h/\Mpc$
that follows from imposing positivity of the $\ell=0$ mode.
In practice the higher-order counterterms that
restore positivity are presumably
already important at substantially smaller
wavenumbers.

These maximum $k$-values
are somewhat larger than those found by Lewandowski et al.,
who reported a fit with $< 2\%$ error
to a redshift-space power spectrum extracted from
the BigMDPL simulation up to $k \lesssim 0.13 h/\Mpc$ at $z=0.56$.
At this redshift, they estimated that higher-order counterterms might already be
important at $k=0.2 h/\Mpc$,
and suggested that
the non-linear scale that controls breakdown of the EFT expansion
might sit near $k \approx 0.8 h/\Mpc$.
Our results are more comparable to those reported by Perko et al.,
who worked with the halo power spectrum
and found a good match up to $k = 0.43 h/\Mpc$ at
$z=0.67$.
Their fit included more counterterms
(roughly, four bias parameters and three stochastic counterterms)
and therefore the degree to which our predictions can be compared is
not entirely clear. Nevertheless, the qualitative features are very
similar.

Our results could also be compared with the `improved' perturbation theory
of Taruya, Nishimichi \& Saito~\cite{Taruya:2010mx}.
Their prediction can be written as a suppressed Kaiser power spectrum
with corrections,
\begin{equation}
    \Prsd = D_{\text{FoG}}(k \mu f \sigmav)
    \Big(
        P_{\delta\delta}
        + 2 f \mu^2 P_{\delta\theta}
        + f^2 \mu^4 P_{\theta\theta}
    \Big)
    + A(k, \mu) + B(k, \mu) ,
\end{equation}
where $D_{\text{FoG}}$ is a fingers-of-God suppression factor to be chosen
by hand,
and $A(k, \mu)$ and $B(k, \mu)$ represent a subset of the terms
generated by the composite operators
$\vect{v}\delta$,
$\vect{v}\vect{v}$,
$\vect{v}\vect{v}\delta$
and
$\vect{v}\vect{v}\vect{v}$ in~\eqref{eq:delta-s-oneloop}.
If the power spectra $P_{\delta\delta}$, $P_{\delta\theta}$ and
$P_{\theta\theta}$ are evaluated at one-loop
and
we take $D_{\text{FoG}} \approx \exp(-k^2 \mu^2 f^2 \sigmav^2)
\approx  1 - k^2 \mu^2 f^2 \sigmav^2$,
then this very nearly
reproduces Matsubara's SPT result for $\Prsd$~\cite{Taruya:2010mx}.
Instead, Taruya et al. obtained their improvement by
evaluating the power spectra
using an alternative prescription~\cite{Taruya:2007xy,Taruya:2009ir}.
By comparing this model to $N$-body simulations they were able to demonstrate
$\sim 1\%$ accuracy up to $k \lesssim 0.2 h/\Mpc$
for the monopole and quadrupole
at $z=0.5$.
This model is intended to capture physical effects similar to those
used in the EFT model, but these effects appear in different ways:
subtraction of power for quasilinear $k$ from $D_{\text{FoG}}$
rather than counterterms, and
damping of the acoustic oscillations from a combination
of $D_{\text{FoG}}$ and the modified computation of
$P_{\delta\delta}$, $P_{\delta\theta}$, $P_{\theta\theta}$
rather than resummation.
The final predictions are broadly comparable, and
it would be interesting to understand more clearly
how these descriptions are related.

\para{Outlook}
Although we cannot rely on perturbation theory to determine the time-dependence
of the counterterms, we show that independent fits for their values
at $z \in \{ 0, 0.25, 0.5, 0.75, 1 \}$ are compatible with
the time-dependence predicted by the perturbative expansion.
In addition, 
as part of our calculation we have introduced a number of technical innovations:
\begin{itemize}
    \item We use a new method to decompose the tensor integrals
    that appear in $\Prsd$ at one-loop level (\S\ref{sec:evaluate-rsd-twopf}),
    and use it to extract their $\mu$-dependence.
    This method is
    based on the Rayleigh plane-wave expansion and analytic integrals over
    two or three spherical Bessel functions.

	\item We have extended the resummation scheme proposed by
	Vlah, Seljak, Chu \& Feng~\cite{Vlah:2015zda}
	to redshift space (\S\ref{sec:rsd-resummation}).
	This simplifies calculation of the resummed $P_\ell$
	by comparison with the resummation scheme suggested by
	Senatore \& Zaldarriaga~\cite{Senatore:2014via}.
	
	In redshift-space this
	and similar schemes appear similar to the suppression
	factors used phenomenologically to model
	the fingers-of-God effect. However we argue that it is more
	appropriate to interpret the redshift-space counterterms
	as the source of this suppression,
	which arises (at least in part)
	from virialized motions on small scales~\cite{Scoccimarro:2004tg}.
	Specifically, we show that the redshift-space
	EFT counterterms successfully reproduce
	the zero-crossing of the $\ell=2$ mode, which is
	associated with this suppression.
\end{itemize}
We find that the effective field-theory framework successfully
produces fits that extend the reach of perturbation theory
by a factor of a few in $k$.
While this is a considerable achievement,
the practical value of these fitting functions has not yet been demonstrated.
First,
without a prediction for the time-dependence of the counterterms we are obliged
to fit independently at each redshift.
This reduces the predictivity of the formalism.
Second,
the values of the counterterms vary
even between relatively nearby cosmologies such as the Planck2015 and MDR1
models studied in this paper.
A proposal to evade the requirement to renormalize
on a model-by-model basis has been given
by Cantaneo, Foreman \& Senatore~\cite{Cataneo:2016suz}.
In cases where this or a similar method can be used,
the EFT method may be advantageous for
parameter fitting
or Fisher forecasts.
Specifically, we can reduce the computational requirements
if it is possible to produce
high-precision predictions over a region of parameter
space using sparser coverage with $N$-body
simulations than if we were to achieve the same
precision by interpolating the power spectra from
these simulations directly.

An alternative use case is to compute covariance matrices that extend to small scales,
as suggested by Bertolini et al.~\cite{Bertolini:2015fya}
and Mohammed, Seljak \& Vlah~\cite{Mohammed:2016sre}.
Our results suggest that
accurately modelling
redshift-space measurements gives enhanced value to both these use cases.
As explained in~\S\ref{sec:simulations}, we find that bulk flows converge very slowly
and exhibit large sample variance,
while small-scale velocity effects
give important contributions to the redshift-space power spectrum
on larger scales.
This requirement for high-resolution simulations
in large volumes
implies that numerical estimation of redshift-space measurements
is substantially more expensive than simulation of real-space measurements
at the same accuracy.
If EFT methods can be used to mitigate this expense then their
deployment becomes even more attractive.

\begin{acknowledgments}
The work reported in this paper
has been supported by the
European Research Council under the European Union's
Seventh Framework Programme (FP/2007--2013) and ERC Grant Agreement No. 308082
(DR, DS).
LFdlB acknowledges support from the UK Science and Technology
Facilities Council via Research Training Grant
ST/M503836/1.

\para{Data availability statement}
Computation of the 1-loop matter power spectrum and its
multipole decomposition was performed by computer codes,
as described in Appendix~\ref{appendix:software}.
These codes
are available for download under open-source licenses.
The specific datasets used to construct
the EFT power spectra reported in \S\ref{sec:real-space-results}
and~\S\ref{sec:results}
have been made available at \href{http://zenodo.org}{zenodo.org}.
This deposit also includes the
\href{http://camb.info}{{\CAMB}} parameter files used to construct
the linear power spectra,
and settings files needed for
\href{https://github.com/gevolution-code/gevolution-1.1}{\gevolution}
to perform the $N$-body simulations described in~\S\ref{sec:simulations}.

Digital identifiers, attribution information, and licensing conditions
are listed in Appendix~\ref{appendix:software} for each of these products.

\end{acknowledgments}

\appendix

\section{Resummation using the Senatore--Zaldarriaga procedure}
\label{appendix:senatore-zaldarriaga}

In~\S\ref{sec:resummation-methods} we described the resummation method
of Vlah, Seljak, Chu \& Feng (the `VSCF scheme'),
which makes explicit use of a decomposition into `wiggle' and `no-wiggle' components.
This decomposition was critical in allowing the
formal Lagrangian-theory expression~\eqref{eq:lagiangian-power-spectrum-2pf}
to be rewritten in terms of
$\Ploopw$ and $\Ploopnw$
even when the exponential is not completely expanded.
Without this step it would not have been possible
to extract a simple, analytic resummation `template'.

Senatore \& Zaldarriaga suggested a different resummation prescription
that does not make explicit use of this
decomposition~\cite{Senatore:2014via,Senatore:2014vja}.
Therefore the relation between these schemes is not
completely clear. In this Appendix we briefly sketch the
Senatore--Zaldarriaga procedure
and explain how it is related to the method
of Vlah, Seljak, Chu \& Feng.

\para{Isolate infrared contributions}
We define $K(\vect{q}, \vect{k})$ to be the exponential
kernel in the Lagrangian formula for the power spectrum,
\begin{equation}
    K(\vect{q},\vect{k}) \equiv
    \exp
    \bigg(
        {- \frac{1}{2}} k_i k_j A_{ij}
        + \frac{\im}{6} k_i k_j k_\ell W_{ij\ell}
        + \cdots
    \bigg)
    .
\end{equation}
The main strategy in the VSCF scheme is to
separate the `no-wiggle' component of $K$ from the remainder;
cf.~\eqref{eq:P-resum-intermediate}.
Senatore \& Zaldarriaga instead chose to isolate the infrared contribution
from the two-point function $A_{ij}$.
This yields a new kernel $\KIR$ that satisfies
\begin{equation}
    \KIR(\vect{q},\vect{k}) \approx
    \exp    
    \bigg(
        {-\frac{1}{2}} k_i k_j \AIR_{ij}
    \bigg) ,
    \label{eq:KIR-def}
\end{equation}
where $\AIR_{ij}$ continues to be defined 
by~\eqref{eq:A-def},
but with the form-factors $X$ and $Y$
in Eqs.~\eqref{eq:X-def}--\eqref{eq:Y-def}
evaluated at tree-level and
restricted to wavenumbers in the infrared.
Then the power spectrum~\eqref{eq:lagiangian-power-spectrum-2pf}
can be written
\begin{equation}
    \label{eq:resummation_SZ}
    P(k) = \int \d^3 q \; \e{-\im \vect{q}\cdot\vect{k}}
    \Big[
        \KIR(\vect{q},\vect{k})
    \Big]
    \Big[
        \KIR^{-1}(\vect{q},\vect{k})
        K(\vect{q},\vect{k})
    \Big] .
\end{equation}
Notice that both factors in square brackets $[ \cdots ]$ contain
`wiggle` and `no-wiggle' contributions,
although the `wiggle' terms in $\KIR$ will be very small
and could be dropped.

If all quantities were expanded to one-loop
then this expression must reproduce the one-loop Eulerian
power spectrum.
Therefore the infrared-subtracted kernel
$\KIR^{-1}K$ by itself can differ from the
one-loop Eulerian result
only by terms that involve $\AIR_{ij}$,
\begin{equation}
    \int \d^3 q \; \e{-\im \vect{q}\cdot\vect{k}}
    \Big[
        \KIR^{-1}(\vect{q},\vect{k})
        K(\vect{q},\vect{k})
    \Big]
    \approx
    P^{\uptooneloop}(k)
    -
    \PiIR(k) ,
\end{equation}
where $\PiIR$ is defined by
\begin{equation}
    \PiIR(k)
    =
    P^{\attreelevel}_{\IRtag}(k)
    +
    \frac{1}{4}
    \int \d^3 q \; \e{-\im \vect{q}\cdot\vect{k}} \,
    k_i k_j
    k_m k_n
    \AIR_{ij}(\vect{q})
    \Big(
         A_{mn}^{\attreelevel}(\vect{q})
         -
         \AIR_{mn}(\vect{q})
    \Big) .
\end{equation}
This correction subtracts some of the power in
$P^{\uptooneloop}$ arising from infrared modes.
To rewrite the factor $\KIR^{-1}K$ in~\eqref{eq:resummation_SZ}
using this result,
insert a decomposition of unity
in the form
\begin{equation}
    1
    \equiv \int \d^3 q' \; \DiracD(\vect{q}-\vect{q}')
    \equiv \int \frac{\d^3 q' \, \d^3 k'}{(2\pi)^3} \e{\im\vect{k}'\cdot(\vect{q}-\vect{q}')} ,
\end{equation}
where $\DiracD$ is the Dirac $\delta$-function.
This yields
\begin{equation}
    P(k) =
    \int \d^3 q 
    \int \d^3 q'
    \int \frac{\d^3 k'}{(2\pi)^3}
    \e{-\im \vect{q}\cdot(\vect{k}-\vect{k}')}
    \e{-\im \vect{k}'\cdot\vect{q}'}
    \KIR(\vect{q},\vect{k})
    \Big[
        \KIR^{-1}(\vect{q}',\vect{k}) K(\vect{q}',\vect{k})
    \Big] .
\end{equation}

\para{Integrate $\KIR$ to a smoothing kernel}
If $\KIR$ depends only weakly on $\vect{q}$
then the integral over $\d^3 q$
produces a kernel that has support only in a narrow region where
$\vect{k} \approx \vect{k}'$.
This relation becomes exact in the limit that
$\KIR$ has no dependence on $\vect{q}$.
Although this is not the case it practice, it gives
a simple scenario in which to visualize the outcome 
of the integration.
The kernel is proportional to
$\DiracD(\vect{k}-\vect{k}')$
if $\KIR$ also has no dependence
on $\vect{k}$, and convolution with it has no effect.
Otherwise, the kernel can be expanded as a series
in derivatives of $\DiracD(\vect{k}-\vect{k}')$,
and convolution with it represents a local smoothing.
For~\eqref{eq:KIR-def},
the shape of the smoothing kernel is determined by the Gaussian
$k$-dependence of $\KIR$,
and the width of its smoothing window
is determined by the amplitude of $X$ and $Y$.

Returning finally to the case where $X$, $Y$ and
$\KIR$ have weak $\vect{q}$-dependence, we can make a Taylor expansion
in $\vect{q}$
around some fiducial value
and exchange explicit powers of $\vect{q}$ for
further derivatives with
respect to $\vect{k}$ or $\vect{k}'$.
Therefore each term in the series expansion integrates to an
increasingly high-derivative smoothing kernel.
The result can be regarded as a superposition of
smoothing kernels with
varying widths determined by the variation of
$X$ and $Y$ with $\vect{q}$.
Therefore the smoothing is modulated on the scale
of the infrared modes retained in these
form-factors.
This modulation partially restores
the infrared power
subtracted by $\PiIR$.

These arguments
are strictly valid only when it is safe to
commute limits and summations with the integration
over $\vect{q}$.
Assuming such exchanges to be acceptable, however,
we can collect all these together to obtain a net
smoothing kernel $M(\vect{k}, \vect{k}')$ defined by
\begin{equation}
    M(\vect{k}, \vect{k}') =
    \int \d^3 q \, \e{-\im \vect{q} \cdot (\vect{k}-\vect{k}')}
    \KIR(\vect{q}, \vect{k}) .    
\end{equation}

\para{Resummed template is smoothed power spectrum}
The final step is to
use the approximation that $M(\vect{k}, \vect{k}')$ has support
only near $\vect{k} \approx \vect{k}'$
to exchange the $\vect{k}$-dependence of
$\KIR^{-1}K$ for $\vect{k}'$-dependence.
The effect is to average
$\e{\im \vect{k}' \cdot \vect{q}'} \KIR^{-1}(\vect{q}',\vect{k}')
K(\vect{q}',\vect{k}')$ over a range of $\vect{k}'$ near $\vect{k}$.
The error in this approximation comes from the inclusion of
$\e{\im \vect{k}' \cdot \vect{q}'}$ in the average,
and can be expressed in terms of gradients of the smoothing
kernel in $\vect{k}$.
If the smoothing does not vary rapidly with wavenumber we may
hope it is not large.
Proceeding in this way,
Senatore \& Zaldarriaga
obtained the template~\cite{Senatore:2014via,Lewandowski:2015ziq}
\begin{equation}
    \label{eq:resummation_SZ_2}
    P_{\text{SZ}}(k) =
    \int \frac{\d^3 k'}{(2\pi)^3}
    M(\vect{k},\vect{k}')
    \left[
        P^{\uptooneloop}(k')
        -
        \PiIR(k')
    \right] .
\end{equation}
If the power spectrum is nearly constant in $k$ then it is unaffected
by the smoothing kernel, and therefore the `no-wiggle' component
will be practically unchanged.
But the `wiggle' component is averaged, causing it to be suppressed.
Therefore, in the Senatore \& Zaldarriaga scheme, the separation into
`wiggle' and `no-wiggle' components becomes important only in the
final average. However, the net effect is still to suppress the
acoustic oscillations while leaving the broadband power unchanged.

Note that in the VSCF scheme
the amount of suppression applied to the `wiggle'
component at wavenumber $k$ is determined by $k/\kdamp$, where
as explained in~\S\ref{sec:resummation-methods}
the wavenumber $\kdamp$ measures the typical
\emph{total} displacement of particles
averaged between $\qmin$ and $\qmax$;
the infrared modes are not treated separately.
In the Senatore--Zaldarriaga scheme we smooth the power spectrum over a window
set by the typical displacement induced by infrared modes only, averaged
over all scales.
The final resummed power spectra are qualitatively similar, but there is
no simple relation between the two procedures.

Although the Senatore--Zaldarriaga scheme is elegant, it has computational drawbacks.
When applied to redshift-space distortions it is necessary
to treat the angular dependence of the integrals numerically.
This significantly increases the computational burden.
By comparison, in the VSCF scheme the angular dependence can be
extracted analytically
[cf. Eqs.~\eqref{eq:Prsd-resum} and~\eqref{eq:Prsd-damping}],
which simplifies the resummation procedure.

\section{Fabrikant's procedure to evaluate the three-Bessel integrals}
\label{appendix:fabrikant}

In~\S\ref{sec:evaluate-rsd-twopf} we described
a new procedure for computing the redshift-space one-loop SPT power
spectrum $P_s^{\text{SPT}}$, based on the Rayleigh plane-wave expansion.
To reduce the resulting expressions to closed form we must
integrate
over the Bessel functions appearing in the Rayleigh formula.
For 13-type integrals this requires weighted integrals over two Bessel functions,
which are relatively well-understood.
For 22-type integrals it requires weighted integrals over \emph{three} Bessel functions.
These are substantially more difficult to evaluate.

\para{Context}
In 1936, Bailey gave the formula (for positive $a$, $b$ and $c$)
\begin{equation}
\begin{split}
    \int_0^\infty t^{\lambda-1} J_\mu(at) J_\nu(bt) J_\rho(ct) \, \d t
    = \mbox{}
    &
    \frac{2^{\lambda-1} a^\mu b^\nu \Gamma( \frac{\lambda+\mu+\nu+\rho}{2} )}
    {c^{\lambda+\mu+\nu} \Gamma(\mu+1) \Gamma(\nu+1) \Gamma(1-\frac{\lambda+\mu+\nu+\rho}{2})}
    \\
    & \mbox{} \times
    F_4 \bigg(
        \begin{array}{l@{\hspace{2mm}}l@{\hspace{2mm}}l}
            {[}\lambda+\mu+\nu-\rho{]}/2, & \mu + 1, & a^2/c^2 \\
            {[}\lambda+\mu+\nu+\rho{]}/2, & \nu + 1, & b^2/c^2
        \end{array}
    \bigg)
    ,
\end{split}
\label{eq:bailey-formula}
\end{equation}
where $\Re(\lambda+\mu+\nu+\rho) > 0$, $\Re(\lambda) < 5/2$, and $c > a + b$;
the function $F_4$ is the fourth type of Appell hypergeometric function,
\begin{equation}
    F_4 \bigg(
        \begin{array}{l@{\hspace{2mm}}l@{\hspace{2mm}}l}
            a, & c_1, & x \\
            b, & c_2, & y
        \end{array}
    \bigg)
    \equiv
    \sum_{m,n=0}^\infty
    \frac{\pochhammer{a}{m+n} \pochhammer{b}{m+n}}{\pochhammer{c_1}{m} \pochhammer{c_2}{n} m! n!}
    x^m y^n .
\end{equation}
Here, $\pochhammer{x}{a}$ is the Pochhammer symbol (or `rising factorial') defined by
$\pochhammer{x}{a} = \Gamma(x+a)/\Gamma(a)$.
The condition $c > a + b$ implies that the lengths $a$, $b$, $c$ do not form the sides of a
triangle.
Bailey's methods did not determine the integral when this condition is not satisfied.
In particular, there is no reason for the result to be analytic in $a$, $b$ and $c$
and therefore we cannot extend~\eqref{eq:bailey-formula} by analytic continuation.
(Some results based on analytic continuation are known in special cases;
see the discussion in Ref.~\cite{doi:10.1137/0520067}.)

Various extensions of Bailey's results are known.
Fabrikant and D\^{o}me used a different computational technique
to find integral representations that could be evaluated
explicitly \cite{fabrikant2001elementary,fabrikant2003computation}, but still in
the non-triangular case.
Mehrem used the Rayleigh plane-wave expansion~\eqref{eq:rayleigh-formula}
to determine various integrals over two and three spherical Bessel functions,
again in the non-triangular case and only when a Clebsch--Gordon coefficient involving the orders
$\mu$, $\nu$, $\sigma$ is not zero~\cite{mehrem2011plane,Mehrem:2010qk}.
Earlier, Gervois \& Navelet~\cite{Gervois:1984ck,doi:10.1137/0520067}
had studied the (spherical)
three-Bessel integral even in the case where $(a,b,c)$ do form a triangle.
Their result appears in its most developed form in Table 2 of Ref.~\cite{doi:10.1137/0520067},
which can be applied whenever $\lambda + \mu + \nu + \rho$ is an integer.

This result of Gervois \& Navelet
is already sufficient for the purposes of this paper
(where $\mu$, $\nu$, $\sigma$ and $\lambda$ are individually integers),
but in our practical calculations we make use of a more recent
formalism due to Fabrikant~\cite{fabrikant2013elementary}.
The Fabrikant method is equivalent to that of Gervois \& Navelet when
$\lambda + \mu + \nu + \rho$ is an integer, but is marginally more general because it
allows for non-integer $\lambda$.

\para{Fabrikant's method}
We briefly summarize the procedure. The aim is to compute a generalization
of the integral~\eqref{eq:fabrikant-3J},
\begin{equation}
  I = \int_0^\infty x^\lambda j_\mu(kx) j_\nu(qx) j_\sigma(sx) \, \d x,
\end{equation}
where $\mu$, $\nu$ and $\sigma$ are integers,
and $\lambda$, $k$, $q$, and $s$ are real.
Fabrikant rewrote the spherical Bessel functions as derivatives of
trigonometric functions,
\begin{equation}
  j_n(z) = (-1)^{n} z^{n} \left( \frac{\d}{z \, \d z} \right) ^{n} \frac{\sin z}{z} ,
\end{equation}
which allows $I$ to be expressed in the form
\begin{equation}
    I = (-1)^{\mu + \nu + \sigma} a^\mu b^\nu c^\sigma
    \frac{\partial^\mu}{(a \, \partial a)^\mu}
    \frac{\partial^\nu}{(b \, \partial b)^\nu}
    \frac{\partial^\sigma}{(c \, \partial c)^\sigma}
    \int_0^\infty
    \frac{\sin at \sin bt \sin ct}{abc \, t^{\mu + \nu + \sigma + 3 - \lambda}}
    \, \d t .   
\end{equation}
The integral in this expression may be formally divergent, but where $I$ exists the
differentiation will yield a finite result.
It can be performed using standard trigonometric identities, yielding
\begin{equation}
\begin{split}
    I = \mbox{} & (-1)^{\mu + \nu + \sigma} a^\mu b^\nu c^\sigma \cos \frac{\pi}{2} (\mu + \nu + \sigma + 3 - \lambda)
    \frac{\partial^\mu}{(a \, \partial a)^\mu}
    \frac{\partial^\nu}{(b \, \partial b)^\nu}
    \frac{\partial^\sigma}{(c \, \partial c)^\sigma}
    \frac{\Gamma(\lambda - \mu - \nu - \sigma - 2)}{4abc}
    \\
    & \mbox{} \times
    \bigg(
        |c + a - b|^{\mu + \nu + \sigma + 2 - \lambda} \sgn (c + a - b)
        + \text{2 cyclic perms}
        -
        (a + b + c)^{\mu + \nu + \sigma + 2 - \lambda}
    \bigg)
\end{split}
\label{eq:fabrikant-basic-expr}
\end{equation}
where $\sgn(x)$ is the sign function, defined by $\sgn(x) = x/|x|$ for $x \neq 0$ and $\sgn(0) = 0$.
Compare Eq.~\eqref{eq:fabrikant-basic-expr} with Eqs.~(3.8), (3.9) and~(3.10) of Ref.~\cite{doi:10.1137/0520067}.

If the argument of the $\cos$ or $\Gamma$ functions is zero, then the result should be computed via a limit.
The differentiations can be performed using Mathematica, or by using the formula
\begin{equation}
    z^{n+1} \frac{\partial^{n+1} f(z)}{(z \, \partial z)^{n+1}}
    =
    \sum_{k=0}^{\lfloor n/2 \rfloor} \frac{\Gamma(n+2k+1) f^{(n - 2k + 1)}(z)}{\Gamma(2k+1) \Gamma(n-2k+1) (2z)^{2k}}
    -
    \sum_{k=0}^{\lfloor (n-1)/2 \rfloor} \frac{\Gamma(n+2k+2) f^{(n - 2k)}(z)}{\Gamma(2k+2) \Gamma(n-2k) (2z)^{2k+1}} .
    \label{eq:fabrikant-basic-derivs}    
\end{equation}
The quantity $\lfloor x \rfloor$ is the floor of $x$, ie. the largest integer that does not exceed $x$,
and $f^{(n)}(x)$ is the $n^{\text{th}}$ derivative of $f$.
As part of the software bundle accompanying this paper
we provide a Mathematica notebook to
evaluate Eqs.~\eqref{eq:fabrikant-basic-expr} and~\eqref{eq:fabrikant-basic-derivs}.
It will automatically test the resulting expression against results obtained
using Mathematica's built-in integration strategies for highly oscillatory
integrals.

\vspace{3mm}
\hrule
\vspace{2mm}

{\small
\para{Specific results}
We collect the results needed for the computation of $P_s^{\text{SPT}}$.
\begin{subequations}
\begin{align}
    \FabrikantThree{0}{0}{0} & = \frac{\pi}{4 k q s} \\
    \FabrikantThree{1}{1}{0} & = \frac{\pi}{8} \frac{k^2 + q^2 - s^2}{k^2 q^2 s} \\
    \FabrikantThree{2}{2}{0} & = \frac{\pi}{32} \frac{3k^4 + 2k^2(q^2 - 3s^2) + 3(q^2 - s^2)^2}{k^3 q^3 s} \\
    \FabrikantThree{2}{2}{2} & = \frac{\pi}{64} \frac{(3k^4 + 2k^2 q^2 + 3 q^4) s^2 + 3(k^2 + q^2)s^4 -
        3 (k^2 - q^2)^2(k^2 + q^2) - 3 s^6}{k^3 q^3 s^3} \\
    \label{eq:fabrikant231}
    \FabrikantThree{2}{3}{1} & = \frac{\pi}{64} \frac{3k^4(q^2 + 5s^2) + (q^2 -s^2)^2(q^2 + 5s^2) + k^2(q^4 + 6 q^2 s^2 - 15 s^4) - 5k^6}
        {k^3 q^4 s^2} \\
    \nonumber
    \FabrikantThree{2}{4}{2} & = \frac{\pi}{512} \frac{1}{k^3 q^5 s^3} \Big(
        35 k^8 - 20 k^6(3q^2 + 7s^2) + 6k^4(3q^4 + 10 q^2 s^2 + 35s^2)
    \\ & \qquad\qquad \mbox{} + (q^2 - s^2)^2(3q^4 + 10 q^2 s^2 + 35 s^4)
        + 4 k^2 (q^6 + 3q^4 s^2 + 15 q^2 s^4 - 35 s^6)
    \Big) \\
    \label{eq:fabrikant330}
    \FabrikantThree{3}{3}{0} & = \frac{\pi}{64} \frac{(k^2 + q^2 - s^2)\big[5k^4 + 5(q^2 - s^2)^2 - 2k^2(q^2 + 5s^2)\big]}
        {k^4 q^4 s} \\
    \nonumber
    \FabrikantThree{4}{4}{0} & = \frac{\pi}{512} \frac{1}{k^4 q^5 s} \Big(
        35 k^8 + 20 k^2(q^2 - 7s^2)\big[ k^4 + (q^2 - s^2)^2\big] + 35(q^2 - s^2)^4
    \\ & \qquad\qquad \mbox{} + 6k^4(3q^4 - 30q^2 s^2 + 35 s^4)
    \Big)
\end{align}
\end{subequations}
Index permutations can be obtained by making suitable exchanges of $k$, $q$ and $s$;
for example, $\FabrikantThree{2}{1}{3}$ can be obtained from~\eqref{eq:fabrikant231}
by exchanging $q$ and $s$,
and $\FabrikantThree{0}{3}{3}$ can be obtained from~\eqref{eq:fabrikant330}
by exchanging $k$ and $s$.
}

\section{Accompanying software bundle}
\label{appendix:software}

The calculations needed to obtain the redshift-space power spectrum are complex.
To assist those wishing to replicate our results
we have made available a large collection of resources,
including Mathematica notebooks that summarize (and validate) the
calculation of $\langle \deltarsd \deltarsd \rangle$ in SPT,
and software tools to compute the loop integrals needed for
numerical evaluation.

\subsection{Mathematica notebooks}
\begin{grouppanel}
{
\renewcommand{\arraystretch}{1.1}
\begin{tabular}{ll}
    \semibold{License} & \href{https://creativecommons.org/licenses/by/4.0/}{Creative Commons Attribution 4.0 International} \;
    \includegraphics[scale=0.12]{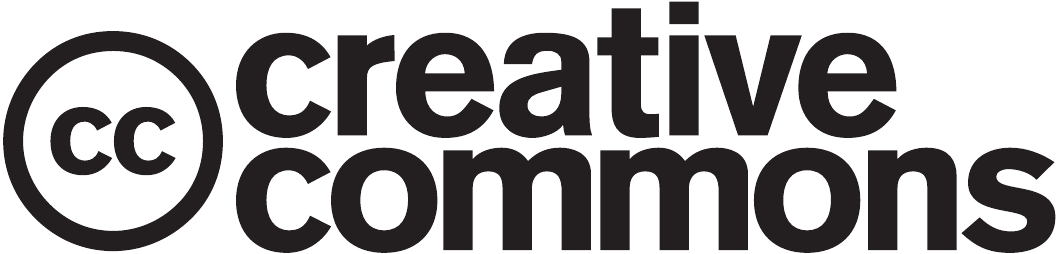}
    \includegraphics[scale=0.15]{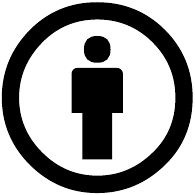} \\
    \semibold{Author} & $\copyright$ University of Sussex 2017. Contributed by David Seery \\
    \semibold{DOI} & \href{https://doi.org/10.5281/zenodo.495795}{\raisebox{-1mm}{\includegraphics[scale=0.6]{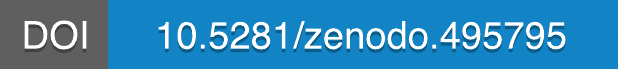}}} \\
    \semibold{Attribution} & Please cite \texttt{zenodo.org} DOI and this paper \\
    \semibold{Download} & {\small \url{https://zenodo.org/record/495795}}
\end{tabular}
}
\end{grouppanel}

\par\noindent
This deposit contains two Mathematica notebooks:
\begin{itemize}
\item {\packagefont FabrikantIntegrals.nb} \\ This notebook
implements Fabrikant's method for evaluation of
the three-Bessel integrals, as described in Appendix~\ref{appendix:fabrikant}.
It will automatically test the resulting analytic formulae against
numerical results obtained using Mathematica's built-in integration strategies.

\item {\packagefont SPTPowerSpectrum.nb} \\ This notebook
summarizes our analytic calculation of the redshift-space
power spectrum up to one-loop. It also validates the
result against Matsubara's result for the redshift-space
power spectrum using the Einstein--de Sitter
approximation~\cite{Matsubara:2007wj},
and with the formulae for the velocity
power spectrum given by Makino et al.~\cite{Makino:1991rp}.
\end{itemize}

\subsection{One-loop SPT integrals in redshift space}
\subsubsection{Pipeline A}\label{software:david-pipeline}
\begin{grouppanel}
{
\renewcommand{\arraystretch}{1.1}
\begin{tabular}{ll}
    \semibold{License} & \href{https://www.gnu.org/licenses/old-licenses/gpl-2.0.en.html}{GNU GPL v2.0 or a later version} \\
    \semibold{Author} & $\copyright$ University of Sussex 2017. Contributed by David Seery \\
    \semibold{DOI} & \href{https://doi.org/10.5281/zenodo.546725}{\raisebox{-1mm}{\includegraphics[scale=0.6]{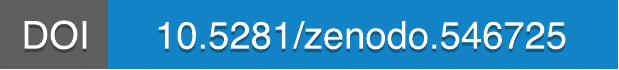}}} \\
    \semibold{GitHub} & {\small\url{https://github.com/ds283/LSSEFT}} \\
    \semibold{Git clone} & \texttt{\small git clone https://github.com/ds283/LSSEFT.git}
\end{tabular}
}
\end{grouppanel}

\vspace{2mm}\par\noindent This is a {\CC} pipeline for computation of the growth factors and
loop integrals needed to construct the renormalized redshift-space power spectrum and its
multipoles. It implements the Vlah et al. resummation scheme described in \S\S\ref{sec:resummation-methods}
and~\ref{sec:rsd-resummation}.

The implementation is parallelized using MPI and uses adaptive load balancing
to spread work over available cores.
The \href{http://www.feynarts.de/cuba/}{Cuba library} is used
to perform the multidimensional integrations
that are required~\cite{Hahn:2004fe},
and the \href{https://github.com/bgrimstad/splinter}{SPLINTER library}
is used to construct B-splines~\cite{SPLINTER}.
Results are stored as
\href{https://www.sqlite.org}{{\SQLite}} databases.
The counterterm fits are performed using the
\href{https://bitbucket.org/joezuntz/cosmosis/wiki/Home}{CosmoSIS} parameter-estimation
framework and the 
\href{http://dan.iel.fm/emcee/current}{emcee} sampler,
for which a Python module is supplied~\cite{Zuntz:2014csq}.
The power spectra presented in this paper were computed using
{\git} revision
\GitRevision{977e5b03}{https://github.com/ds283/LSSEFT/commit/977e5b038c79d0620f859276b81d4f67eae7b132}.

This pipeline shares some code with the {\CppTransport}
platform for computing correlation functions of
inflationary density perturbations~\cite{Dias:2016rjq}.

\subsubsection{Pipeline B}\label{software:donough-pipeline}
\begin{grouppanel}
{
\renewcommand{\arraystretch}{1.1}
\begin{tabular}{ll}
    \semibold{License} & \href{https://www.gnu.org/licenses/old-licenses/gpl-2.0.en.html}{GNU GPL v2.0 or a later version} \\
    \semibold{Author} & $\copyright$ University of Sussex 2017. Contributed by Donough Regan \\
    \semibold{DOI} & \href{https://doi.org/10.5281/zenodo.495113}{\raisebox{-1mm}{\includegraphics[scale=0.65]{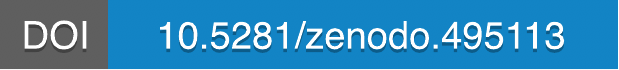}}} \\
    \semibold{GitHub} & {\small\url{https://github.com/DonRegan/PowSpec\_EFTofLSS}} \\
    \semibold{Git clone} & \texttt{\small git clone https://github.com/DonRegan/PowSpec\_EFTofLSS.git}
\end{tabular}
}
\end{grouppanel}

This is a second, independent
pipeline that duplicates the functionality of pipeline A,
but with slightly different implementation choices.
The multidimensional integrations and splines are evaluated using
the \href{http://www.gnu.org/software/gsl/}{GNU Scientific Library}~\cite{GSLGNU}.
The results presented in this paper have been computed using both pipelines A and B,
and a third C pipeline implemented using Mathematica.
The plots and numerical results are those belonging to pipeline A.
The pipeline B results were obtained using
{\git} revision
\GitRevision{1126b040}{https://github.com/DonRegan/PowSpec_EFTofLSS/commit/1126b040663674da5bfc00b862e571fd28a9b714}.
We find very good agreement between all pipelines,
indicating that our numerics are
robust to changes in integration strategy, filtering methods
and the counterterm fitting procedure.

\subsection{Supporting dataset}
\begin{grouppanel}
{
\renewcommand{\arraystretch}{1.1}
\begin{tabular}{ll}
    \semibold{License} & \href{https://creativecommons.org/licenses/by/4.0/}{Creative Commons Attribution 4.0 International} \;
    \includegraphics[scale=0.12]{LicenseLogos/CCLogo}
    \includegraphics[scale=0.15]{LicenseLogos/CCBy} \\
    \semibold{Author} & Contributed by David Seery and Shaun Hotchkiss \\
    \semibold{DOI} & \href{https://doi.org/10.5281/zenodo.546734}{\raisebox{-1mm}{\includegraphics[scale=0.65]{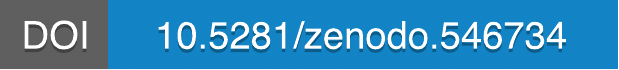}}} \\ 
    \semibold{Attribution} & Please cite \texttt{zenodo.org} DOI and this paper \\
    \semibold{Download} & {\small \url{https://zenodo.org/record/546734}}
\end{tabular}
}
\end{grouppanel}

This dataset includes the components necessary to reproduce
our numerical results. It comprises:
\begin{itemize}
	\item {\SQLite} databases
	containing the output of the pipeline
	described in~\ref{software:david-pipeline}
	for the Planck2015~\cite{Ade:2015xua}
	and MultiDark MDR1 cosmologies~\cite{Prada:2011jf}.
	These were used to construct the renormalized
	real-space power spectrum in~\S\ref{sec:one-loop-real}
	and the redshift-space power spectrum in~\S\ref{sec:one-loop-rsd},
	respectively.
	
	\item A settings file for
	the \href{https://github.com/gevolution-code/gevolution-1.1}{\gevolution}
	numerical relativity code, which was used to perform the
	custom $N$-body simulations described in~\S\ref{sec:simulations}.
	Our results used version 1.1 of the {\gevolution} framework.
	The initial conditions are generated dynamically from the
	settings file.
	
	\item \href{http://camb.info}{{\CAMB}} parameter files for the
	linear power spectra
	used to construct our one-loop results, for both the
	Planck2015 and MDR1 cosmologies.
	For the Planck2015 model we also include a {\CAMB} parameter file
	to compute the final non-linear power spectrum using {\Halofit}.
	These power spectra are also emebedded in the {\SQLite}
	databases containing our numerical results.
\end{itemize}
	
\bibliographystyle{JHEP}
\bibliography{paper}

\providecommand{\href}[2]{#2}\begingroup\raggedright\begin{thebibliography}{100}

\bibitem{1981MNRAS.197..931J}
R.~{Juszkiewicz}, {\it {On the evolution of cosmological adiabatic
  perturbations in the weakly non-linear regime}},  {\em Mon. Not. Roy. Astron.
  Soc.} {\bf 197} (1981) 931--940.

\bibitem{1983MNRAS.203..345V}
E.~T. {Vishniac}, {\it {Why weakly non-linear effects are small in a
  zero-pressure cosmology}},  {\em Mon. Not. Roy. Astron. Soc.} {\bf 203}
  (1983) 345--349.

\bibitem{Fry:1983cj}
J.~N. Fry, {\it {The Galaxy correlation hierarchy in perturbation theory}},
  {\em Astrophys. J.} {\bf 279} (1984) 499--510.

\bibitem{Goroff:1986ep}
M.~H. Goroff, B.~Grinstein, S.~J. Rey, and M.~B. Wise, {\it {Coupling of Modes
  of Cosmological Mass Density Fluctuations}},  {\em Astrophys. J.} {\bf 311}
  (1986) 6--14.

\bibitem{Bertschinger:1993zv}
E.~Bertschinger and B.~Jain, {\it {Gravitational instability of cold matter}},
  {\em Astrophys. J.} {\bf 431} (1994) 486,
  [\href{http://xxx.lanl.gov/abs/astro-ph/9307033}{{\tt astro-ph/9307033}}].

\bibitem{Suto:1990wf}
Y.~Suto and M.~Sasaki, {\it {Quasi nonlinear theory of cosmological
  selfgravitating systems}},  {\em Phys. Rev. Lett.} {\bf 66} (1991) 264--267.

\bibitem{Makino:1991rp}
N.~Makino, M.~Sasaki, and Y.~Suto, {\it {Analytic approach to the perturbative
  expansion of nonlinear gravitational fluctuations in cosmological density and
  velocity fields}},  {\em Phys. Rev.} {\bf D46} (1992) 585--602.

\bibitem{Bernardeau:2001qr}
F.~Bernardeau, S.~Colombi, E.~Gaztanaga, and R.~Scoccimarro, {\it {Large scale
  structure of the universe and cosmological perturbation theory}},  {\em Phys.
  Rept.} {\bf 367} (2002) 1--248,
  [\href{http://xxx.lanl.gov/abs/astro-ph/0112551}{{\tt astro-ph/0112551}}].

\bibitem{Baumann:2010tm}
D.~Baumann, A.~Nicolis, L.~Senatore, and M.~Zaldarriaga, {\it {Cosmological
  Non-Linearities as an Effective Fluid}},  {\em JCAP} {\bf 1207} (2012) 051,
  [\href{http://xxx.lanl.gov/abs/1004.2488}{{\tt 1004.2488}}].

\bibitem{Carrasco:2012cv}
J.~J.~M. Carrasco, M.~P. Hertzberg, and L.~Senatore, {\it {The Effective Field
  Theory of Cosmological Large Scale Structures}},  {\em JHEP} {\bf 09} (2012)
  082, [\href{http://xxx.lanl.gov/abs/1206.2926}{{\tt 1206.2926}}].

\bibitem{Carrasco:2013mua}
J.~J.~M. Carrasco, S.~Foreman, D.~Green, and L.~Senatore, {\it {The Effective
  Field Theory of Large Scale Structures at Two Loops}},  {\em JCAP} {\bf 1407}
  (2014) 057, [\href{http://xxx.lanl.gov/abs/1310.0464}{{\tt 1310.0464}}].

\bibitem{Porto:2013qua}
R.~A. Porto, L.~Senatore, and M.~Zaldarriaga, {\it {The Lagrangian-space
  Effective Field Theory of Large Scale Structures}},  {\em JCAP} {\bf 1405}
  (2014) 022, [\href{http://xxx.lanl.gov/abs/1311.2168}{{\tt 1311.2168}}].

\bibitem{Senatore:2014via}
L.~Senatore and M.~Zaldarriaga, {\it {The IR-resummed Effective Field Theory of
  Large Scale Structures}},  {\em JCAP} {\bf 1502} (2015), no.~02 013,
  [\href{http://xxx.lanl.gov/abs/1404.5954}{{\tt 1404.5954}}].

\bibitem{Senatore:2014vja}
L.~Senatore and M.~Zaldarriaga, {\it {Redshift Space Distortions in the
  Effective Field Theory of Large Scale Structures}},
  \href{http://xxx.lanl.gov/abs/1409.1225}{{\tt 1409.1225}}.

\bibitem{Vlah:2015sea}
Z.~Vlah, M.~White, and A.~Aviles, {\it {A Lagrangian effective field theory}},
  {\em JCAP} {\bf 1509} (2015), no.~09 014,
  [\href{http://xxx.lanl.gov/abs/1506.05264}{{\tt 1506.05264}}].

\bibitem{Vlah:2015zda}
Z.~Vlah, U.~Seljak, M.~Y. Chu, and Y.~Feng, {\it {Perturbation theory,
  effective field theory, and oscillations in the power spectrum}},  {\em JCAP}
  {\bf 1603} (2016), no.~03 057,
  [\href{http://xxx.lanl.gov/abs/1509.02120}{{\tt 1509.02120}}].

\bibitem{Jackson:2008yv}
J.~C. Jackson, {\it {Fingers of God: A critique of Rees' theory of primoridal
  gravitational radiation}},  {\em Mon. Not. Roy. Astron. Soc.} {\bf 156}
  (1972) 1P--5P, [\href{http://xxx.lanl.gov/abs/0810.3908}{{\tt 0810.3908}}].

\bibitem{Lewandowski:2015ziq}
M.~Lewandowski, L.~Senatore, F.~Prada, C.~Zhao, and C.-H. Chuang, {\it {On the
  EFT of Large Scale Structures in Redshift Space}},
  \href{http://xxx.lanl.gov/abs/1512.06831}{{\tt 1512.06831}}.

\bibitem{Perko:2016puo}
A.~Perko, L.~Senatore, E.~Jennings, and R.~H. Wechsler, {\it {Biased Tracers in
  Redshift Space in the EFT of Large-Scale Structure}},
  \href{http://xxx.lanl.gov/abs/1610.09321}{{\tt 1610.09321}}.

\bibitem{Assassi:2014fva}
V.~Assassi, D.~Baumann, D.~Green, and M.~Zaldarriaga, {\it {Renormalized Halo
  Bias}},  {\em JCAP} {\bf 1408} (2014) 056,
  [\href{http://xxx.lanl.gov/abs/1402.5916}{{\tt 1402.5916}}].

\bibitem{Gleyzes:2016tdh}
{Gleyzes, J\'{e}r\^{o}me and de Putter, Roland and Green, Daniel and Dor\'{e},
  Olivier}, {\it {Biasing and the search for primordial non-Gaussianity beyond
  the local type}},  \href{http://xxx.lanl.gov/abs/1612.06366}{{\tt
  1612.06366}}.

\bibitem{Bose:2016qun}
B.~Bose and K.~Koyama, {\it {A Perturbative Approach to the Redshift Space
  Power Spectrum: Beyond the Standard Model}},  {\em JCAP} {\bf 1608} (2016),
  no.~08 032, [\href{http://xxx.lanl.gov/abs/1606.02520}{{\tt 1606.02520}}].

\bibitem{Bose:2017dtl}
B.~Bose and K.~Koyama, {\it {A Perturbative Approach to the Redshift Space
  Correlation Function: Beyond the Standard Model}},  {\em JCAP} {\bf 1708}
  (2017), no.~08 029, [\href{http://xxx.lanl.gov/abs/1705.09181}{{\tt
  1705.09181}}].

\bibitem{Fasiello:2016qpn}
M.~Fasiello and Z.~Vlah, {\it {Nonlinear fields in generalized cosmologies}},
  {\em Phys. Rev.} {\bf D94} (2016), no.~6 063516,
  [\href{http://xxx.lanl.gov/abs/1604.04612}{{\tt 1604.04612}}].

\bibitem{Lewandowski:2017kes}
M.~Lewandowski and L.~Senatore, {\it {IR-safe and UV-safe integrands in the
  EFTofLSS with exact time dependence}},
  \href{http://xxx.lanl.gov/abs/1701.07012}{{\tt 1701.07012}}.

\bibitem{Matsubara:2007wj}
T.~Matsubara, {\it {Resumming Cosmological Perturbations via the Lagrangian
  Picture: One-loop Results in Real Space and in Redshift Space}},  {\em Phys.
  Rev.} {\bf D77} (2008) 063530, [\href{http://xxx.lanl.gov/abs/0711.2521}{{\tt
  0711.2521}}].

\bibitem{doi:10.1137/0520067}
A.~Gervois and H.~Navelet, {\it Infinite integrals involving three spherical
  {Bessel} functions},  {\em SIAM Journal on Mathematical Analysis} {\bf 20}
  (1989), no.~4 1006--1018,
  [\href{http://xxx.lanl.gov/abs/http://dx.doi.org/10.1137/0520067}{{\tt
  http://dx.doi.org/10.1137/0520067}}].

\bibitem{fabrikant2013elementary}
V.~Fabrikant, {\it Elementary exact evaluation of infinite integrals of the
  product of several spherical {Bessel} functions, power and exponential},
  {\em Quarterly of Applied Mathematics} {\bf 71} (2013), no.~3 573--581.

\bibitem{bailey1936some}
W.~N. Bailey, {\it Some infinite integrals involving {Bessel} functions},  {\em
  Proceedings of the London Mathematical Society} {\bf s2-40} (1936), no.~1
  37--48.

\bibitem{Mercolli:2013bsa}
L.~Mercolli and E.~Pajer, {\it {On the velocity in the Effective Field Theory
  of Large Scale Structures}},  {\em JCAP} {\bf 1403} (2014) 006,
  [\href{http://xxx.lanl.gov/abs/1307.3220}{{\tt 1307.3220}}].

\bibitem{Meszaros:1974tb}
P.~M\'{e}sz\'{a}ros, {\it {The behaviour of point masses in an expanding
  cosmological substratum}},  {\em Astron. Astrophys.} {\bf 37} (1974)
  225--228.

\bibitem{1975A&A....41..143G}
E.~J. {Groth} and P.~J.~E. {Peebles}, {\it {Closed-form solutions for the
  evolution of density perturbations in some cosmological models}},  {\em
  Astron. Astrophys.} {\bf 41} (June, 1975) 143--145.

\bibitem{Linder:2005in}
E.~V. Linder, {\it {Cosmic growth history and expansion history}},  {\em Phys.
  Rev.} {\bf D72} (2005) 043529,
  [\href{http://xxx.lanl.gov/abs/astro-ph/0507263}{{\tt astro-ph/0507263}}].

\bibitem{Scoccimarro:1997st}
R.~Scoccimarro, S.~Colombi, J.~N. Fry, J.~A. Frieman, E.~Hivon, and A.~Melott,
  {\it {Nonlinear evolution of the bispectrum of cosmological perturbations}},
  {\em Astrophys. J.} {\bf 496} (1998) 586,
  [\href{http://xxx.lanl.gov/abs/astro-ph/9704075}{{\tt astro-ph/9704075}}].

\bibitem{Wise:1988kua}
M.~B. Wise, {\it {Non-Gaussian Fluctuations}},  {\em NATO Sci. Ser. C.} {\bf
  219} (1988) 215--238.

\bibitem{Scoccimarro:1995if}
R.~Scoccimarro and J.~Frieman, {\it {Loop corrections in nonlinear cosmological
  perturbation theory}},  {\em Astrophys. J. Suppl.} {\bf 105} (1996) 37,
  [\href{http://xxx.lanl.gov/abs/astro-ph/9509047}{{\tt astro-ph/9509047}}].

\bibitem{Scoccimarro:1996se}
R.~Scoccimarro and J.~Frieman, {\it {Loop corrections in nonlinear cosmological
  perturbation theory 2. Two point statistics and selfsimilarity}},  {\em
  Astrophys. J.} {\bf 473} (1996) 620,
  [\href{http://xxx.lanl.gov/abs/astro-ph/9602070}{{\tt astro-ph/9602070}}].

\bibitem{Ade:2015xua}
{\bf Planck} Collaboration, P.~A.~R. Ade {\em et~al.}, {\it {Planck 2015
  results. XIII. Cosmological parameters}},  {\em Astron. Astrophys.} {\bf 594}
  (2016) A13, [\href{http://xxx.lanl.gov/abs/1502.01589}{{\tt 1502.01589}}].

\bibitem{Scoccimarro:2004tg}
R.~Scoccimarro, {\it {Redshift-space distortions, pairwise velocities and
  nonlinearities}},  {\em Phys. Rev.} {\bf D70} (2004) 083007,
  [\href{http://xxx.lanl.gov/abs/astro-ph/0407214}{{\tt astro-ph/0407214}}].

\bibitem{Sugiyama:2013gza}
N.~S. Sugiyama and D.~N. Spergel, {\it {How does non-linear dynamics affect the
  baryon acoustic oscillation?}},  {\em JCAP} {\bf 1402} (2014) 042,
  [\href{http://xxx.lanl.gov/abs/1306.6660}{{\tt 1306.6660}}].

\bibitem{Pajer:2013jj}
E.~Pajer and M.~Zaldarriaga, {\it {On the Renormalization of the Effective
  Field Theory of Large Scale Structures}},  {\em JCAP} {\bf 1308} (2013) 037,
  [\href{http://xxx.lanl.gov/abs/1301.7182}{{\tt 1301.7182}}].

\bibitem{Carrasco:2013sva}
J.~J.~M. Carrasco, S.~Foreman, D.~Green, and L.~Senatore, {\it {The 2-loop
  matter power spectrum and the IR-safe integrand}},  {\em JCAP} {\bf 1407}
  (2014) 056, [\href{http://xxx.lanl.gov/abs/1304.4946}{{\tt 1304.4946}}].

\bibitem{Donoghue:1994dn}
J.~F. Donoghue, {\it {General relativity as an effective field theory: The
  leading quantum corrections}},  {\em Phys. Rev.} {\bf D50} (1994) 3874--3888,
  [\href{http://xxx.lanl.gov/abs/gr-qc/9405057}{{\tt gr-qc/9405057}}].

\bibitem{Donoghue:1995cz}
J.~F. Donoghue, {\it {Introduction to the effective field theory description of
  gravity}},  in {\em {Advanced School on Effective Theories Almunecar, Spain,
  June 25-July 1, 1995}}, 1995.
\newblock \href{http://xxx.lanl.gov/abs/gr-qc/9512024}{{\tt gr-qc/9512024}}.

\bibitem{1974A&A....32..391P}
P.~J.~E. {Peebles}, {\it {The Effect of a Lumpy Matter Distribution on the
  Growth of Irregularities in an Expanding Universe}},  {\em Astron.
  Astrophys.} {\bf 32} (1974) 391.

\bibitem{Carroll:2013oxa}
S.~M. Carroll, S.~Leichenauer, and J.~Pollack, {\it {Consistent effective
  theory of long-wavelength cosmological perturbations}},  {\em Phys. Rev.}
  {\bf D90} (2014), no.~2 023518,
  [\href{http://xxx.lanl.gov/abs/1310.2920}{{\tt 1310.2920}}].

\bibitem{Fuhrer:2015cia}
F.~Führer and G.~Rigopoulos, {\it {Renormalizing a Viscous Fluid Model for
  Large Scale Structure Formation}},  {\em JCAP} {\bf 1602} (2016), no.~02 032,
  [\href{http://xxx.lanl.gov/abs/1509.03073}{{\tt 1509.03073}}].

\bibitem{Feynman:1963fq}
R.~P. Feynman and F.~L. Vernon, Jr., {\it {The Theory of a general quantum
  system interacting with a linear dissipative system}},  {\em Annals Phys.}
  {\bf 24} (1963) 118--173. [Annals Phys.281,547(2000)].

\bibitem{Caldeira:1981rx}
A.~O. Caldeira and A.~J. Leggett, {\it {Influence of dissipation on quantum
  tunneling in macroscopic systems}},  {\em Phys. Rev. Lett.} {\bf 46} (1981)
  211.

\bibitem{Caldeira:1982iu}
A.~O. Caldeira and A.~J. Leggett, {\it {Path integral approach to quantum
  Brownian motion}},  {\em Physica} {\bf 121A} (1983) 587--616.

\bibitem{Caldeira:1982uj}
A.~O. Caldeira and A.~J. Leggett, {\it {Quantum tunneling in a dissipative
  system}},  {\em Annals Phys.} {\bf 149} (1983) 374--456.

\bibitem{kamenev2011field}
A.~Kamenev, {\em Field theory of non-equilibrium systems}.
\newblock Cambridge University Press, 2011.

\bibitem{Calzetta:1996sy}
E.~Calzetta and B.-L. Hu, {\it {Stochastic behavior of effective field theories
  across threshold}},  {\em Phys. Rev.} {\bf D55} (1997) 3536--3551,
  [\href{http://xxx.lanl.gov/abs/hep-th/9603164}{{\tt hep-th/9603164}}].

\bibitem{Calzetta:1999xh}
E.~Calzetta and B.~L. Hu, {\it {Stochastic dynamics of correlations in quantum
  field theory: From Schwinger-Dyson to Boltzmann-Langevin equation}},  {\em
  Phys. Rev.} {\bf D61} (2000) 025012,
  [\href{http://xxx.lanl.gov/abs/hep-ph/9903291}{{\tt hep-ph/9903291}}].

\bibitem{Itzykson:1980rh}
C.~Itzykson and J.~B. Zuber, {\em {Quantum Field Theory}}.
\newblock International Series In Pure and Applied Physics. McGraw-Hill, New
  York, 1980.

\bibitem{Collins:1984xc}
J.~C. Collins, {\em {Renormalization}}, vol.~26 of {\em Cambridge Monographs on
  Mathematical Physics}.
\newblock Cambridge University Press, Cambridge, 1986.

\bibitem{Polkinghorne:1980mk}
J.~C. Polkinghorne, {\em {Models of High Energy Processes}}.
\newblock Cambridge Monographs on Mathematical Physics. Cambridge Univ. Press,
  Cambridge, UK, 2010.

\bibitem{1988grra.conf..385B}
J.~R. {Bond} and H.~M.~P. {Couchman}, {\it {$w_{gg}({\theta})$ as a probe of
  large scale structure}},  in {\em Proceedings of the 2nd Canadian Conference
  on General Relativity and Relativistic Astrophysics} (A.~{Coley}, C.~{Dyer},
  and T.~{Tupper}, eds.), pp.~385--389, 1988.

\bibitem{1993cvf..conf..585T}
A.~N. {Taylor}, {\it {Statistics of self-gravitating fluctuations}},  in {\em
  Cosmic Velocity Fields} (F.~{Bouchet} and M.~{Lachieze-Rey}, eds.), p.~585,
  1993.

\bibitem{1996MNRAS.282..767T}
A.~N. {Taylor} and A.~J.~S. {Hamilton}, {\it {Non-linear cosmological power
  spectra in real and redshift space}},  {\em Mon. Not. R. Astron. Soc.} {\bf
  282} (Oct., 1996) 767--778,
  [\href{http://xxx.lanl.gov/abs/astro-ph/9604020}{{\tt astro-ph/9604020}}].

\bibitem{Okamura:2011nu}
T.~Okamura, A.~Taruya, and T.~Matsubara, {\it {Next-to-leading resummation of
  cosmological perturbations via the Lagrangian picture: 2-loop correction in
  real and redshift spaces}},  {\em JCAP} {\bf 1108} (2011) 012,
  [\href{http://xxx.lanl.gov/abs/1105.1491}{{\tt 1105.1491}}].

\bibitem{McQuinn:2015tva}
M.~McQuinn and M.~White, {\it {Cosmological perturbation theory in 1+1
  dimensions}},  {\em JCAP} {\bf 1601} (2016), no.~01 043,
  [\href{http://xxx.lanl.gov/abs/1502.07389}{{\tt 1502.07389}}].

\bibitem{Baldauf:2015xfa}
T.~Baldauf, M.~Mirbabayi, M.~Simonovi\'{c}, and M.~Zaldarriaga, {\it
  {Equivalence Principle and the Baryon Acoustic Peak}},  {\em Phys. Rev.} {\bf
  D92} (2015), no.~4 043514, [\href{http://xxx.lanl.gov/abs/1504.04366}{{\tt
  1504.04366}}].

\bibitem{Taruya:2010mx}
A.~Taruya, T.~Nishimichi, and S.~Saito, {\it {Baryon Acoustic Oscillations in
  2D: Modeling Redshift-space Power Spectrum from Perturbation Theory}},  {\em
  Phys. Rev.} {\bf D82} (2010) 063522,
  [\href{http://xxx.lanl.gov/abs/1006.0699}{{\tt 1006.0699}}].

\bibitem{Blas:2016sfa}
D.~Blas, M.~Garny, M.~M. Ivanov, and S.~Sibiryakov, {\it {Time-Sliced
  Perturbation Theory II: Baryon Acoustic Oscillations and Infrared
  Resummation}},  {\em JCAP} {\bf 1607} (2016), no.~07 028,
  [\href{http://xxx.lanl.gov/abs/1605.02149}{{\tt 1605.02149}}].

\bibitem{Eisenstein:1997jh}
D.~J. Eisenstein and W.~Hu, {\it {Power spectra for cold dark matter and its
  variants}},  {\em Astrophys. J.} {\bf 511} (1997) 5,
  [\href{http://xxx.lanl.gov/abs/astro-ph/9710252}{{\tt astro-ph/9710252}}].

\bibitem{Eisenstein:2006nk}
D.~J. Eisenstein, H.-j. Seo, E.~Sirko, and D.~Spergel, {\it {Improving
  Cosmological Distance Measurements by Reconstruction of the Baryon Acoustic
  Peak}},  {\em Astrophys. J.} {\bf 664} (2007) 675--679,
  [\href{http://xxx.lanl.gov/abs/astro-ph/0604362}{{\tt astro-ph/0604362}}].

\bibitem{Eisenstein:2006nj}
D.~J. Eisenstein, H.-j. Seo, and M.~J. White, {\it {On the Robustness of the
  Acoustic Scale in the Low-Redshift Clustering of Matter}},  {\em Astrophys.
  J.} {\bf 664} (2007) 660--674,
  [\href{http://xxx.lanl.gov/abs/astro-ph/0604361}{{\tt astro-ph/0604361}}].

\bibitem{Crocce:2005xz}
M.~Crocce and R.~Scoccimarro, {\it {Memory of initial conditions in
  gravitational clustering}},  {\em Phys. Rev.} {\bf D73} (2006) 063520,
  [\href{http://xxx.lanl.gov/abs/astro-ph/0509419}{{\tt astro-ph/0509419}}].

\bibitem{Crocce:2007dt}
M.~Crocce and R.~Scoccimarro, {\it {Nonlinear Evolution of Baryon Acoustic
  Oscillations}},  {\em Phys. Rev.} {\bf D77} (2008) 023533,
  [\href{http://xxx.lanl.gov/abs/0704.2783}{{\tt 0704.2783}}].

\bibitem{Smith:2002dz}
{\bf VIRGO Consortium} Collaboration, R.~E. Smith, J.~A. Peacock, A.~Jenkins,
  S.~D.~M. White, C.~S. Frenk, F.~R. Pearce, P.~A. Thomas, G.~Efstathiou, and
  H.~M.~P. Couchmann, {\it {Stable clustering, the halo model and nonlinear
  cosmological power spectra}},  {\em Mon. Not. Roy. Astron. Soc.} {\bf 341}
  (2003) 1311, [\href{http://xxx.lanl.gov/abs/astro-ph/0207664}{{\tt
  astro-ph/0207664}}].

\bibitem{Takahashi:2012em}
R.~Takahashi, M.~Sato, T.~Nishimichi, A.~Taruya, and M.~Oguri, {\it {Revising
  the Halofit Model for the Nonlinear Matter Power Spectrum}},  {\em Astrophys.
  J.} {\bf 761} (2012) 152, [\href{http://xxx.lanl.gov/abs/1208.2701}{{\tt
  1208.2701}}].

\bibitem{Jennings:2012pt}
E.~Jennings, C.~M. Baugh, B.~Li, G.-B. Zhao, and K.~Koyama, {\it {Redshift
  space distortions in f(R) gravity}},  {\em Mon. Not. Roy. Astron. Soc.} {\bf
  425} (2012) 2128--2143, [\href{http://xxx.lanl.gov/abs/1205.2698}{{\tt
  1205.2698}}].

\bibitem{Burrage:2015lla}
C.~Burrage, D.~Parkinson, and D.~Seery, {\it {Beyond the growth rate of cosmic
  structure: Testing modified gravity models with an extra degree of freedom}},
   \href{http://xxx.lanl.gov/abs/1502.03710}{{\tt 1502.03710}}.

\bibitem{1987MNRAS.227....1K}
N.~{Kaiser}, {\it {Clustering in real space and in redshift space}},  {\em Mon.
  Not. Roy. Astron. Soc.} {\bf 227} (July, 1987) 1--21.

\bibitem{1989MNRAS.236..851L}
P.~B. {Lilje} and G.~{Efstathiou}, {\it {Gravitationally induced velocity
  fields in the universe. I -- Correlation functions}},  {\em Mon. Not. Roy.
  Astron. Soc.} {\bf 236} (Feb., 1989) 851--864.

\bibitem{1990MNRAS.242..428M}
C.~{McGill}, {\it {The redshift projection. I -- Caustics and correlation
  functions}},  {\em Mon. Not. Roy. Astron. Soc.} {\bf 242} (Feb., 1990)
  428--438.

\bibitem{Cole:1993kh}
S.~Cole, K.~B. Fisher, and D.~H. Weinberg, {\it {Fourier analysis of redshift
  space distortions and the determination of $\Omega$}},  {\em Mon. Not. Roy.
  Astron. Soc.} {\bf 267} (1994) 785,
  [\href{http://xxx.lanl.gov/abs/astro-ph/9308003}{{\tt astro-ph/9308003}}].

\bibitem{Heavens:1998es}
A.~F. Heavens, S.~Matarrese, and L.~Verde, {\it {The Nonlinear redshift-space
  power spectrum of galaxies}},  {\em Mon. Not. Roy. Astron. Soc.} {\bf 301}
  (1998) 797--808, [\href{http://xxx.lanl.gov/abs/astro-ph/9808016}{{\tt
  astro-ph/9808016}}].

\bibitem{Scoccimarro:1999ed}
R.~Scoccimarro, H.~M.~P. Couchman, and J.~A. Frieman, {\it {The Bispectrum as a
  Signature of Gravitational Instability in Redshift-Space}},  {\em Astrophys.
  J.} {\bf 517} (1999) 531--540,
  [\href{http://xxx.lanl.gov/abs/astro-ph/9808305}{{\tt astro-ph/9808305}}].

\bibitem{Weinzierl:2006qs}
S.~Weinzierl, {\it {The Art of computing loop integrals}},  in {\em
  {Universality and renormalization: From stochastic evolution to
  renormalization of quantum fields. Proceedings, Workshop on 'Percolation, SLE
  and related topics', Toronto, Canada, September 20-24, 2005, and Workshop on
  'Renormalization and universality in mathematical physics', Toronto, Canada,
  October 18-22, 2005}}, pp.~345--395, 2006.
\newblock \href{http://xxx.lanl.gov/abs/hep-ph/0604068}{{\tt hep-ph/0604068}}.

\bibitem{neumann}
F.~E. Neumann, {\em {Beitr\"{a}ge zur Theorie der Kugelfuntionen}}.
\newblock Leipzig, 1878.

\bibitem{adams}
J.~C. Adams, {\it {On the Expression of the Product of Any Two Legendre's
  Coefficients by Means of a Series of Legendre's Coefficients}},  {\em Proc.
  Roy. Soc.} {\bf 27} (1878).

\bibitem{bailey1933}
W.~N. Bailey, {\it {On the product of two Legendre polynomials}},  {\em
  Mathematical Proceedings of the Cambridge Philosophical Society} {\bf 29}
  (005, 1933) 173--177.

\bibitem{McEwen:2016fjn}
J.~E. McEwen, X.~Fang, C.~M. Hirata, and J.~A. Blazek, {\it {FAST-PT: a novel
  algorithm to calculate convolution integrals in cosmological perturbation
  theory}},  {\em JCAP} {\bf 1609} (2016), no.~09 015,
  [\href{http://xxx.lanl.gov/abs/1603.04826}{{\tt 1603.04826}}].

\bibitem{Fang:2016wcf}
X.~Fang, J.~A. Blazek, J.~E. McEwen, and C.~M. Hirata, {\it {FAST-PT II: an
  algorithm to calculate convolution integrals of general tensor quantities in
  cosmological perturbation theory}},  {\em JCAP} {\bf 1702} (2017), no.~02
  030, [\href{http://xxx.lanl.gov/abs/1609.05978}{{\tt 1609.05978}}].

\bibitem{Schmittfull:2016jsw}
M.~Schmittfull, Z.~Vlah, and P.~McDonald, {\it {Fast large scale structure
  perturbation theory using one-dimensional fast Fourier transforms}},  {\em
  Phys. Rev.} {\bf D93} (2016), no.~10 103528,
  [\href{http://xxx.lanl.gov/abs/1603.04405}{{\tt 1603.04405}}].

\bibitem{Schmittfull:2016yqx}
M.~Schmittfull and Z.~Vlah, {\it {FFT-PT: Reducing the two-loop large-scale
  structure power spectrum to low-dimensional radial integrals}},  {\em Phys.
  Rev.} {\bf D94} (2016), no.~10 103530,
  [\href{http://xxx.lanl.gov/abs/1609.00349}{{\tt 1609.00349}}].

\bibitem{Peacock:1993xg}
J.~A. Peacock and S.~J. Dodds, {\it {Reconstructing the linear power spectrum
  of cosmological mass fluctuations}},  {\em Mon. Not. Roy. Astron. Soc.} {\bf
  267} (1994) 1020--1034, [\href{http://xxx.lanl.gov/abs/astro-ph/9311057}{{\tt
  astro-ph/9311057}}].

\bibitem{Yamamoto:2008gr}
K.~Yamamoto, T.~Sato, and G.~Huetsi, {\it {Testing general relativity with the
  multipole spectra of the SDSS luminous red galaxies}},  {\em Prog. Theor.
  Phys.} {\bf 120} (2008) 609--614,
  [\href{http://xxx.lanl.gov/abs/0805.4789}{{\tt 0805.4789}}].

\bibitem{Blake:2011rj}
C.~Blake {\em et~al.}, {\it {The WiggleZ Dark Energy Survey: the growth rate of
  cosmic structure since redshift z=0.9}},  {\em Mon. Not. Roy. Astron. Soc.}
  {\bf 415} (2011) 2876, [\href{http://xxx.lanl.gov/abs/1104.2948}{{\tt
  1104.2948}}].

\bibitem{Oka:2013cba}
A.~Oka, S.~Saito, T.~Nishimichi, A.~Taruya, and K.~Yamamoto, {\it {Simultaneous
  constraints on the growth of structure and cosmic expansion from the
  multipole power spectra of the SDSS DR7 LRG sample}},  {\em Mon. Not. Roy.
  Astron. Soc.} {\bf 439} (2014) 2515--2530,
  [\href{http://xxx.lanl.gov/abs/1310.2820}{{\tt 1310.2820}}].

\bibitem{Beutler:2013yhm}
{\bf BOSS} Collaboration, F.~Beutler {\em et~al.}, {\it {The clustering of
  galaxies in the SDSS-III Baryon Oscillation Spectroscopic Survey: Testing
  gravity with redshift-space distortions using the power spectrum
  multipoles}},  {\em Mon. Not. Roy. Astron. Soc.} {\bf 443} (2014), no.~2
  1065--1089, [\href{http://xxx.lanl.gov/abs/1312.4611}{{\tt 1312.4611}}].

\bibitem{Prada:2011jf}
F.~Prada, A.~A. Klypin, A.~J. Cuesta, J.~E. Betancort-Rijo, and J.~Primack,
  {\it {Halo concentrations in the standard LCDM cosmology}},  {\em Mon. Not.
  Roy. Astron. Soc.} {\bf 423} (2012) 3018--3030,
  [\href{http://xxx.lanl.gov/abs/1104.5130}{{\tt 1104.5130}}].

\bibitem{Adamek:2015eda}
J.~Adamek, D.~Daverio, R.~Durrer, and M.~Kunz, {\it {General relativity and
  cosmic structure formation}},  {\em Nature Phys.} {\bf 12} (2016) 346--349,
  [\href{http://xxx.lanl.gov/abs/1509.01699}{{\tt 1509.01699}}].

\bibitem{Jeong:2008rj}
D.~Jeong and E.~Komatsu, {\it {Perturbation Theory Reloaded II: Non-linear
  Bias, Baryon Acoustic Oscillations and Millennium Simulation In Real Space}},
   {\em Astrophys. J.} {\bf 691} (2009) 569--595,
  [\href{http://xxx.lanl.gov/abs/0805.2632}{{\tt 0805.2632}}].

\bibitem{2011MNRAS.410.2081J}
E.~{Jennings}, C.~M. {Baugh}, and S.~{Pascoli}, {\it {Modelling redshift space
  distortions in hierarchical cosmologies}},  {\em Mon. Not. Roy. Astron. Soc.}
  {\bf 410} (Jan., 2011) 2081--2094,
  [\href{http://xxx.lanl.gov/abs/1003.4282}{{\tt 1003.4282}}].

\bibitem{2016MNRAS.457.4340K}
A.~{Klypin}, G.~{Yepes}, S.~{Gottl{\"o}ber}, F.~{Prada}, and S.~{He{\ss}}, {\it
  {MultiDark simulations: the story of dark matter halo concentrations and
  density profiles}},  {\em Mon. Not. Roy. Astron. Soc.} {\bf 457} (Apr., 2016)
  4340--4359, [\href{http://xxx.lanl.gov/abs/1411.4001}{{\tt 1411.4001}}].

\bibitem{Cataneo:2016suz}
M.~Cataneo, S.~Foreman, and L.~Senatore, {\it {Efficient exploration of
  cosmology dependence in the EFT of LSS}},
  \href{http://xxx.lanl.gov/abs/1606.03633}{{\tt 1606.03633}}.

\bibitem{Hertzberg:2012qn}
M.~P. Hertzberg, {\it {Effective field theory of dark matter and structure
  formation: Semianalytical results}},  {\em Phys. Rev.} {\bf D89} (2014),
  no.~4 043521, [\href{http://xxx.lanl.gov/abs/1208.0839}{{\tt 1208.0839}}].

\bibitem{Taruya:2007xy}
A.~Taruya and T.~Hiramatsu, {\it {A Closure Theory for Non-linear Evolution of
  Cosmological Power Spectra}},  {\em Astrophys. J.} {\bf 674} (2008) 617,
  [\href{http://xxx.lanl.gov/abs/0708.1367}{{\tt 0708.1367}}].

\bibitem{Taruya:2009ir}
A.~Taruya, T.~Nishimichi, S.~Saito, and T.~Hiramatsu, {\it {Non-linear
  Evolution of Baryon Acoustic Oscillations from Improved Perturbation Theory
  in Real and Redshift Spaces}},  {\em Phys. Rev.} {\bf D80} (2009) 123503,
  [\href{http://xxx.lanl.gov/abs/0906.0507}{{\tt 0906.0507}}].

\bibitem{Bertolini:2015fya}
D.~Bertolini, K.~Schutz, M.~P. Solon, J.~R. Walsh, and K.~M. Zurek, {\it
  {Non-Gaussian Covariance of the Matter Power Spectrum in the Effective Field
  Theory of Large Scale Structure}},  {\em Phys. Rev.} {\bf D93} (2016), no.~12
  123505, [\href{http://xxx.lanl.gov/abs/1512.07630}{{\tt 1512.07630}}].

\bibitem{Mohammed:2016sre}
I.~Mohammed, U.~Seljak, and Z.~Vlah, {\it {Perturbative approach to covariance
  matrix of the matter power spectrum}},  {\em Mon. Not. Roy. Astron. Soc.}
  (2016) [\href{http://xxx.lanl.gov/abs/1607.00043}{{\tt 1607.00043}}].

\bibitem{fabrikant2001elementary}
V.~Fabrikant and G.~D{\^o}me, {\it Elementary evaluation of certain infinite
  integrals involving {Bessel} functions},  {\em Quarterly of Applied
  Mathematics} {\bf 59} (2001), no.~1 1--24.

\bibitem{fabrikant2003computation}
V.~I. Fabrikant, {\it Computation of infinite integrals involving three
  {Bessel} functions by introduction of new formalism},  {\em ZAMM-Journal of
  Applied Mathematics and Mechanics/Zeitschrift f{\"u}r Angewandte Mathematik
  und Mechanik} {\bf 83} (2003), no.~6 363--374.

\bibitem{mehrem2011plane}
R.~Mehrem, {\it The plane wave expansion, infinite integrals and identities
  involving spherical {Bessel} functions},  {\em Applied Mathematics and
  Computation} {\bf 217} (2011), no.~12 5360--5365,
  [\href{http://xxx.lanl.gov/abs/0909.0494}{{\tt 0909.0494}}].

\bibitem{Mehrem:2010qk}
R.~Mehrem and A.~Hohenegger, {\it {A Generalisation For The Infinite Integral
  Over Three Spherical {Bessel} Functions}},  {\em J. Phys.} {\bf A43} (2010),
  no.~45 9, [\href{http://xxx.lanl.gov/abs/1006.2108}{{\tt 1006.2108}}].

\bibitem{Gervois:1984ck}
A.~Gervois and H.~Navelet, {\it Some integrals involving three {Bessel}
  functions when their arguments satisfy the triangle inequalities},  {\em
  Journal of Mathematical Physics} {\bf 25} (1984), no.~11 3350--3356,
  [\href{http://xxx.lanl.gov/abs/http://dx.doi.org/10.1063/1.526062}{{\tt
  http://dx.doi.org/10.1063/1.526062}}].

\bibitem{Hahn:2004fe}
T.~Hahn, {\it {CUBA: A Library for multidimensional numerical integration}},
  {\em Comput. Phys. Commun.} {\bf 168} (2005) 78--95,
  [\href{http://xxx.lanl.gov/abs/hep-ph/0404043}{{\tt hep-ph/0404043}}].

\bibitem{SPLINTER}
B.~Grimstad {\em et~al.}, ``{SPLINTER: a library for multivariate function
  approximation with splines}.'' \url{http://github.com/bgrimstad/splinter},
  2015.
\newblock Accessed: 2016-03-20.

\bibitem{Zuntz:2014csq}
J.~Zuntz, M.~Paterno, E.~Jennings, D.~Rudd, A.~Manzotti, S.~Dodelson,
  S.~Bridle, S.~Sehrish, and J.~Kowalkowski, {\it {CosmoSIS: Modular
  Cosmological Parameter Estimation}},  {\em Astron. Comput.} {\bf 12} (2015)
  45--59, [\href{http://xxx.lanl.gov/abs/1409.3409}{{\tt 1409.3409}}].

\bibitem{Dias:2016rjq}
M.~Dias, J.~Frazer, D.~J. Mulryne, and D.~Seery, {\it {Numerical evaluation of
  the bispectrum in multiple field inflation---the transport approach with
  code}},  {\em JCAP} {\bf 1612} (2016), no.~12 033,
  [\href{http://xxx.lanl.gov/abs/1609.00379}{{\tt 1609.00379}}].

\bibitem{GSLGNU}
M.~Galassi {\em et~al.}, ``{GNU Scientific Library Reference Manual (3rd
  Ed.)}.'' \url{http://savannah.gnu.org/projects/gsl}.

\end{thebibliography}\endgroup

\end{document}